\documentclass[moor]{informs1} 

\usepackage[utf8]{inputenc}
\usepackage[english]{babel}
\usepackage{textcomp}
\usepackage{mathtools}
\usepackage{enumerate}
\usepackage{accents}

\usepackage{natbib}
 \NatBibNumeric
 \def\bibfont{\small}%
 \def\newblock{\ }%
 \bibpunct[, ]{[}{]}{,}{n}{}{,}%

\usepackage[colorlinks=true,breaklinks=true,bookmarks=true,urlcolor=blue,
     citecolor=blue,linkcolor=blue,bookmarksopen=false,draft=false]{hyperref}

\def\EMAIL#1{\href{mailto:#1}{#1}}

\TheoremsNumberedThrough     

\EquationsNumberedThrough    

\newcommand{\w}{\omega}

\author{FERHOUNE Massinissa}
\begin{document}

\TITLE{Discrete time optimal investment under model uncertainty}
\ARTICLEAUTHORS{%
\AUTHOR{Laurence Carassus}
\AFF{Laboratoire de Mathématiques de Reims, UMR9008 CNRS and Université de Reims
Champagne-Ardenne, France \EMAIL{laurence.carassus@univ-reims.fr}}
\AUTHOR{Massinissa Ferhoune}
\AFF{Laboratoire de Mathématiques de Reims, UMR9008 CNRS and Université de Reims
Champagne-Ardenne, France, \EMAIL{fmassinissa@free.fr}}
} 
\ABSTRACT{We study a robust utility maximization problem in a general  discrete-time frictionless market under quasi-sure no-arbitrage. The investor is assumed to have a random and concave utility function defined on the whole real-line. She also faces model  ambiguity on her beliefs  about the market, which is modeled through a set of priors.  We prove the existence of an optimal investment strategy using only primal methods. 
For that we assume classical assumptions on the market and on the random utility function as asymptotic elasticity constraints. Most of our other assumptions 
 are stated on a prior-by-prior basis and correspond  to generally accepted assumptions in the literature on markets without ambiguity.  
 We also propose a general setting including utility functions with benchmark  for which our assumptions are easily checked.  }

\MSCCLASS{Primary: 93E20, 91B28; secondary: 91B16, 28B20}
\ORMSCLASS{Primary: utility/preference: theory, dynamic programming/optimal control; secondary: finance: portfolio}

\maketitle




\section{Introduction}

In this article, we are interested in the following question: In a frictionless discrete-time market, does an investor who tolerates negative wealth and faces model ambiguity have an optimal investment strategy? Model uncertainty refers to random phenomena that have an uncertain probability of occurrence. For example, in the Ellsberg experiments, people have to choose between a risky urn which composition is known and an uncertain urn which composition is unknown.  
As the world becomes more and more unpredictable, model ambiguity has become a major issue especially in financial economics. This notion of model ambiguity, sometimes referred to as ``unknown unknown",  goes back to Knight, see \cite{refkni}, and is therefore called Knightian uncertainty. To address our research question, we will model the preference of the investor with a utility function. The classical utility maximization problem i.e. without Knightian uncertainty, 
has a rich literature and we refer to \citep{refFS} and the references therein for a detailed survey. In a general semimartingale setting for concave utility functions defined on $(0,\infty)$, \citep{refAE1} uses dual methods to show that Asymptotic Elasticity (AE) constraint on $+\infty$ is necessary in order to obtain the existence of an optimal investment strategy. Roughly speaking, the AE constraint at $+\infty$ states that for positive  large enough  wealth, the marginal utility is small  compared to the average  utility. 
When the utility function is defined on $\mathbb{R},$ \citep{refAE2} shows that the AE constraints  on both $-\infty$ and $+\infty$ are necessary to obtain existence. In discrete time, \citep{ref2} proves similar results using primal methods and postulating the AE constraint only on one side. Later, \citep{ref7} extends these results to non-concave utility functions, see also \citep{ref10}. 
As a result, constraints on AE have become classic assumptions for solving utility maximization problems.

Going back to model uncertainty,  a set $\mathcal{Q}$ of probability measures, also called priors, usually models all the investor's beliefs about the market. The earliest literature assumed that $\mathcal{Q}$ is dominated. We refer to \citep{refdoml} for a comprehensive survey of the dominated case. Unfortunately, this setting excludes volatility uncertainty and  Bouchard and Nutz  introduce the so-called quasi-sure uncertainty, see \citep{refbn1}. Random sets of ``local" priors are first given. These probability measures are ``local" in the sense that they represent the investor's belief between two successive moments. The cornerstone assumption of \citep{refbn1} is that the graphs of these random sets are analytic sets, see \cite{ref1} for a comprehensive presentation of analytic sets. This allows measurable selection techniques  and 
the set of priors for the whole market is then constructed by taking the Fubini product of measurable versions of the ``local" priors.   Note that the set of probability measures $\mathcal{Q}$ is neither assumed to be compact nor to be dominated by any particular measure. In the case of model ambiguity, Gilboa and Schmeidler \citep{refgs} characterize the preference of the agents with the following numerical representation $X\mapsto \inf_{Q\in\mathcal{Q}} \mathbb{E}_Q U(X)$. This is coherent with  Ellsberg experiments: participants   strictly prefer the risky urn to the uncertain one, showing  uncertainty aversion. 
Thus, optimal investment amounts to show existence in a maxmin expected utility problem.

In the discrete time quasi-sure setting, \citep{ref9} solves this maxmin problem for concave and bounded from above utility functions, when only positive wealth is admissible, i.e. the utility function is defined on the positive axis. The same result is proved in \citep{ref4} for unbounded functions.  The literature dealing with  potentially negative wealth in the discrete time nondominated quasi-sure setting is quite recent.  The existence of an optimal investment strategy is proved in \cite{ref11} for a one-period market. Assuming that the strategies are discrete (in $\omega$), \citep{refns} obtains existence for non-concave bounded from above utility functions. Using a dual approach, \citep{refbar1} solves the utility maximization problem for an exponential utility function assuming a stronger no-arbitrage condition. 
Some of these results have been extended in \citep{refbar3} always assuming a boundedness condition on the utility function but also under the  assumption that medial limits exist. 
Note also that \citep{ref8} have derived existence results in a different framework where the uncertainty is represented by a set of stochastic processes. There seems to be no  obvious connection between their framework and the quasi-sure framework of \citep{refbn1}. 

To the best of our knowledge, there are no general results on the existence of an optimal investment strategy in the discrete-time nondominated quasi-sure setting of \citep{refbn1}, when the wealth of the investor can be negative. In this article, 
we show that under expected conditions (involving no medial limits) on the market and on the utility function, an optimal investment strategy exists.  
Since beliefs are uncertain, we consider random utility functions. We assume that the quasi-sure no-arbitrage condition of \citep{refbn1} called $NA(\mathcal{Q})$ holds true. 
To solve our optimisation problem, we first consider a one-period case with strategy in $\mathbb{R}^d$. We then ``glue" together the solutions found in the one-period case using dynamic programming together with measurable selection techniques. This is where the analycity of the graph of priors is required. 
We now comment on our other assumptions.  The first one is the classical AE constraint. The second one requires that $U$ is ``negative enough"  and is automatically satisfied for deterministic non-constant concave utility functions. We provide an example where there is no optimal strategy when this assumption fails. The third one states that some simple strategies are admissible and is the only assumption that is postulated in a strong sense, i.e. considering a supremum over all priors. This assumption ensures a control from below for the value functions. These first three assumptions can be checked directly when the utility function and the market are specified. This is not the case for the last one which requires that the value function $U_0^{P}$ at time 0 is finite for every prior $P\in \mathcal{H}$. 
The set $\mathcal{H}$ is the the set of all priors $P$ such that the $P$ no-arbitrage condition holds true in a quasi-sure sense. The existence of such priors has been proved in \citep{ref4} under $NA(\mathcal{Q})$. 
Note that this last assumption is stated on a prior-by-prior basis, not for the supremum over all priors. So, it can be verified as in the setting without ambiguity and this assumption is indeed required in  \citep{ref2}.  It provides a control from above for the value functions.
 We will provide an example where there is some prior $P^*\in \mathcal{H}$ such that $U_0^{P^*}=+\infty$ and there is no optimal strategy even if the value function is finite. 
The  one-period assumptions and the dynamic programming procedure of \citep{ref9} and \citep{ref3} are inspired from \citep{ref2} and \citep{ref2bis}. 
Here  we follow a different path and adapt to the quasi-sure setting the approach of \cite{ref11} that was initially formulated for non-concave utility functions.  
We also use the set $\mathcal{H}$ and formulate our  one-period assumptions for a given prior in $\mathcal{H}$ and not for all $P \in \mathcal{Q}$ or for $\sup_{P \in \mathcal{Q}}$  as \cite{ref9, ref3, ref11}. 
Thus, our methods for finding  optimal strategies are different from those used before in the quasi-sure literature. Here are some further examples. First, we introduce the random variables $N^*_{t}$ and $N^P_{t}$ that represent the cash positions for which the value functions $U_{t+1}$ and $U^P_{t+1}$ are below a certain threshold.
These random variables are unnecessary for utility functions defined only on the positive half-line, as they take the value $-\infty$ on the negative half-line, and this condition propagates through dynamic programming, eliminating the need to check whether value functions are ``sufficiently negative".
The $N^*_{t}$ and $N^P_{t}$ are important for converting some of our assumptions into integrability conditions. We provide both finiteness and integrability results for them. These results are new and nontrivial.  Secondly, we introduce  the AE constraints in the context of uncertainty, which is new to the best of knowledge. We define the random variables $C_t$ and show that they are quasi-surely finite and that their finiteness ensures that the dynamic AE constraints on the value functions are satisfied. 
Finally, from a more technical side, we show that some results of \citep{refbn1} which are proved assuming that $-\infty +\infty=+\infty$ remain true under the opposite convention.  The convention $-\infty +\infty=-\infty$ is crucial in order to propagate the concavity property through the dynamic programming procedure.
Indeed unlike \citep{ref7}, we have to assume  in our quasi-sure setting that the utility function is concave in order to obtain the regularity of the value functions $U_t$. \\
From all our assumptions,  we prove in Theorem \ref{one_step_strategy} that for any initial wealth $x$ the optimal one-step strategies $\Phi_{t+1}^{*,x}$ exist and that if these strategy are admissible, then they are optimal. 
The fact that the $\Phi_{t+1}^{*,x}$ are not automatically admissible is not due to the quasi-sure uncertainty. Already,  in \citep{ref2} (see \citep[Remark 15]{ref11}), one need to assume that $E(V_T^{x,\Phi^{*,x}})<\infty$ to get that $\Phi^{*,x}$ is admissible. Thus, as in Rasonyi and Stettner, we propose in Theorem \ref{optimality_M_typeA} a general setting where 
not only the one-step optimal strategies are admissible, but also $U_0^{P}<\infty$  for every prior $P\in \mathcal{H}$. 
 For this purpose, we introduce a new kind of random utility functions called utility functions of type (A). They  satisfy the AE constraints, are negative enough but bounded from below by some power greater than 1, all of this with integrability conditions. We show that random utility functions with benchmark introduced by \cite{refkahn} are of type (A).  For that we use nontrivial and new results on AE constraints in the context of uncertainty. The proof of Theorem \ref{optimality_M_typeA} uses integrability results for  $N^*_{t}$ and $N^P_{t}$. 

The remainder of this article is organized as follows. In Section \ref{sett_main}, we introduce the financial model, the assumptions as well as the main results. In Section \ref{op}, we solve the utility maximization problem in a one-period market, while in Section \ref{muti_per}, we prepare the dynamic programming procedure. Section \ref{proof_th1} presents the proof of Theorem \ref{one_step_strategy}, while Section \ref{proof_optimality_M} presents the one of Theorem \ref{optimality_M_typeA}. The appendix contains further proofs and results, in particular on AE and on a specific integral that we introduce in order to preserve concavity through the dynamic programming procedure.

\section{Setting and main result}
\label{sett_main}
\subsection{Setting}
\label{sett}
We fix a time horizon $T$ and introduce a family of Polish spaces $(\Omega_t)_{1\leq t \leq T}$. For some $0\leq t\leq T$, let $\Omega^t:=\Omega_1 \times \cdot\cdot\cdot \times \Omega_t$ with the convention that $\Omega^0$ is a singleton. Let $X$ be a Borel space (see \citep[Definition 7.7]{ref1}). We denote by $\mathfrak P(X)$, the set of probability measures defined on the measurable space $(X,\mathcal{B}(X))$, where $\mathcal{B}(X)$ is the usual Borel sigma-algebra on $X$. We denote by $\mathcal{B}_c(X)$ the completion of $(X,\mathcal{B}(X))$ with respect to all $P\in\mathfrak{P}(X)$. 
Let $S:=(S_t)_ {0\leq t\leq T}$ be a $\mathbb{R}^d$-valued process representing the discounted price of $d$ risky assets over time. We first assume that $S$ is Borel. This assumption is usual in the quasi-sure financial literature (see \citep{refbn1}) and aims at solving measurability issues.
\begin{assumption}
For all $0\leq t \leq T$, $S_t$ is $\mathcal{B}(\Omega^t)$-measurable.
\label{S_borel}
\end{assumption}
We consider a random utility function defined on the whole real line which models the investor's preference on the market in the case of possible negative wealth. 

\begin{definition}
A random utility $U$ : $\Omega^T \times \mathbb{R} \to \mathbb{R}\cup \{-\infty,+\infty\}$ is a function such that for all $x \in \mathbb{R}$,  $U(\cdot,x)$ is $\mathcal{B}(\Omega^T)$-measurable$\;$ and $U(\w^T,\cdot)$ is nondecreasing, concave and upper-semicontinuous (usc) for all $\w^T \in \Omega^T$. 
\label{U_hp}
\end{definition}

In presence of uncertainty about the distribution of the states of nature, we expect that any characterization of the investor\rq{}s behavior should also be uncertain. Hence, the choice of a random utility. The concavity and monotonicity of $U$ ensure that the investor is risk-averse and always seeks for more money. Remark that as $U$ is not necessarily finite, the usc condition is not vacuous. 
\begin{remark}
Using \citep[Lemma 5.10]{ref10}, $U(\w,\cdot)$ is right-continuous  and thus  $U$ is $\mathcal{B}(\Omega^T) \otimes \mathcal{B}(\mathbb{R})$-measurable (see \citep[Lemma 5.13]{ref10}). 
\label{joint_mes_U}
\end{remark}

We now construct the set $\mathcal{Q}^T$ of all priors on the market. The set $\mathcal{Q}^T$ captures all the investor's beliefs about the distribution of the future states of nature. For all $0\leq t \leq T-1$, let\footnote{The notation $\twoheadrightarrow$ stands for set-valued mapping.} $\mathcal{Q}_{t+1} : \Omega^t \twoheadrightarrow \mathfrak P(\Omega_{t+1})$ where $\mathcal{Q}_{t+1}(\w^t)$ can be seen as the set of all possible priors for the $(t+1)$-th period given the state $\w^t$ at time $t$. The following assumption is now classical in the quasi-sure financial literature (see \citep{ref9}, \citep{refbn1}, \citep{ref3}, \citep{ref4}, \citep{refns} and \citep{refbar1}) and allows to do measurable selection.
\begin{assumption}
The set $\mathcal{Q}_1$ is nonempty and convex. For all $\;1\leq t \leq T-1$, $\mathcal{Q}_{t+1}$ is a nonempty and convex-valued random set such that $graph(\mathcal{Q}_{t+1}):=\{(\w^t,p)\in \Omega^t \times \mathfrak P(\Omega_{t+1}), p\in\mathcal{Q}_{t+1}(\w^t)\}$ is an analytic set.
\label{analytic_graph}
\end{assumption}
Let $X$ be a Polish space. An analytic set of $X$ is the image of a Borel subset of some other Polish space under some Borel measurable function (see \citep[Theorem 12.24, p447]{refAB}). We denote by $\mathcal{A}(X)$ the class of analytic sets of X. Then, $\mathcal{B}(X) \subset \mathcal{A}(X) \subset \mathcal{B}_c(X)$ (see \citep[p171]{ref1}).

Explicit examples of nondominated financial markets satisfying Assumption \ref{analytic_graph} can be found in \citep{ref4} and \citep{refbar1}. Among them is a robust discrete time Black-Scholes model and a robust binomial model where the uncertainty affects  the probability of jumps and  their size. 
We also propose in Example \ref{example1} a diffusion discretisation where the driving process have an unknown centered distribution, with variance equal to 1 and bounded exponential moment.

For all $0\leq t\leq T-1$, let $SK_{t+1}$ be the set of universally measurable stochastic kernels on $\Omega_{t+1}$ given $\Omega^t$. For $q_{t+1}(\cdot|\cdot) : \mathcal{B}(\Omega_{t+1})\times \Omega^t \to \mathbb{R}$, we have that $q_{t+1}\in SK_{t+1}$ if for all $\w^t\in\Omega^t$, $q_{t+1}(\cdot|\w^t) \in \mathfrak{P}(\Omega_{t+1})$ and for all $B\in\mathcal{B}(\Omega_{t+1})$, $q_{t+1}(B|\cdot)$ is $\mathcal{B}_c(\Omega^t)$-measurable (see \citep[Definition 7.12, p134]{ref1} and \citep[Lemma 7.28, p174]{ref1}). 
Then, from the Jankov-von Neumann theorem (see \citep[Proposition 7.49 p182]{ref1}) and Assumption \ref{analytic_graph}, there exists $q_{t+1}\in SK_{t+1}$ such that for all $\w^t\in \Omega^t$, $q_{t+1}(\cdot|\w^t)\in \mathcal{Q}_{t+1}(\w^t).$ For all $1\leq t\leq T$, let $\mathcal{Q}^t \subset \mathfrak P(\Omega^t)$ be defined by 
\begin{eqnarray}
\mathcal{Q}^t &:=& \{q_1 \otimes q_2 \otimes \cdot\cdot\cdot \otimes q_t, q_1\in \mathcal{Q}_1,\; q_{s+1}\in SK_{s+1}, q_{s+1}(\cdot|\w^s)\in \mathcal{Q}_{s+1}(\w^s),\; \forall \w_s\in \Omega^s,\forall 1\leq s \leq t-1\}, \label{def_set_Q}
\end{eqnarray}
where $q_1 \otimes q_2 \otimes \cdots \otimes  q_t$ denotes the t-fold application of Fubini's theorem, see
\citep[Proposition 7.45 p175]{ref1}, and belongs to $\mathfrak{P}(\Omega^t)$. From now, we will use Fubini's theorem without further reference. We also set $\mathcal{Q}^0:= \{ \delta_{\w_0}\}$, where $\delta_{\w_0}$ is the Dirac measure on the single element $\w_0$ of $\Omega^0$.
If $P:= q_1 \otimes q_2 \otimes \cdot\cdot\cdot \otimes q_T \in\mathcal{Q}^T$, we write for any $1\leq t\leq T$, $P^t:= q_1 \otimes q_2 \otimes \cdot\cdot\cdot \otimes q_t$ and $ P^t\in\mathcal{Q}^t$. In this paper, most of the time, we work directly on the disintegration of $P$ rather than $P$. So, from now, we will specify the fixed disintegration for which the required result holds true. Usually the letter $q$ will designate a stochastic kernel while the letter $p$ will be used for a probability measure.

Trading strategies are represented by $d$-dimensional, $(\mathcal{B}_c(\Omega^{t-1}))_{1\leq t \leq T}$-adapted processes $\phi:=\{\phi_t,\; 1\leq t\leq T\}$ representing the investor's holdings in each of the $d$ risky assets over time. The set of all such trading strategies is denoted by $\Phi$. Trading is assumed to be self-financing and the value at time $t$ of a portfolio $\phi\in\Phi$ starting from initial capital $x\in \mathbb{R}$ is thus given by $V_t^{x,\phi}=x+\sum_{s=1}^{t} \phi_s \Delta S_s.$
Note that if $x,y\in \mathbb{R}^d$ then the concatenation $xy$ stands for their scalar product. The symbol $|\,.\,|$ refers to the Euclidean norm on $\mathbb{R}^d$ (or on $\mathbb{R}$) and $|\,.\,|_1$ is the norm on $\mathbb{R}^d$ defined by $|\,x\,|_1 :=\sum_{i=1}^d |x_i|$ for all $x\in \mathbb{R}^d$.\\

We use the quasi-sure no-arbitrage condition $NA(\mathcal{Q}^T)$ introduced in \citep[Definition 1.1]{refbn1}. Recall that a set $A\subset \Omega^T$ is a $\mathcal{Q}^T$-polar set if there exists $N\in \mathcal{B}(\Omega^T)$ such that $A \subset N$ and $P(N)=0$ for all $P\in \mathcal{Q}^T$. A property holds true $\mathcal{Q}^T$-quasi-surely (q.s.) if it holds true outside of a $\mathcal{Q}^T$-polar set. The complement of a $\mathcal{Q}^T$-polar set is called a $\mathcal{Q}^T$-full-measure set. Of course, any $\mathcal{Q}^T$-full-measure set is a $P$-full-measure set for all $P\in\mathcal{Q}^T$.\\

\begin{definition}
\label{defNA}
The $NA(\mathcal{Q}^T)$ condition holds true if $V_T^{0,\phi}\geq 0$ $\mathcal{Q}^T$-$\mbox{q.s.}$ for some $\phi\in \Phi$ implies that $V_T^{0,\phi}= 0$ $\mathcal{Q}^T\mbox{-q.s.}$\\ 
\end{definition}


In this paper, we will use an adhoc integral called $\int_{-}$ in order to preserve the concavity of the value functions. When this integral is finite, it is just the usual integral. Else, it is computed using the convention:
\begin{eqnarray}
\label{cvt_inf}
-\infty+\infty=+\infty-\infty=-\infty.
\end{eqnarray}
Unfortunately, \citep{ref1} uses the opposite convention. Nevertheless, we will show in Appendix \ref{inf_cvt_sub} that important results of \citep{ref1} still hold true for $\int_{-}$. From now, we will simply write $\int$ (or $\mathbb{E}$) instead of $\int_{-}$ except when we need to clarify the difference between both integrals. We now introduce the set of admissible strategies for the utility maximization problem. 

\begin{definition}
Let $U$ be a random utility function as in Definition \ref{U_hp}. Let $P\in\mathfrak{P}(\Omega^T)$ and $x\in\mathbb{R}$. 
\begin{eqnarray*}
\Phi(x,U,P) &:=& \{\phi\in\Phi,\; \mathbb{E}_P U^-(\cdot,V_T^{x,\phi}(\cdot))<+\infty\}\\
\Phi(x,U,\mathcal{Q}^T)&:=&\bigcap_{P\in \mathcal{Q}^T}\Phi(x,U,P).
\end{eqnarray*}
\label{admissibility_def}
\end{definition} 
We want to prove the existence of an optimal solution for  the following max-min utility problem  when the uncertainty about the true probability  is modeled by $\mathcal{Q}^T$:
\begin{eqnarray}
u(x):=\sup_{\phi\in \Phi(x,U,\mathcal{Q}^T)}\inf_{P\in \mathcal{Q}^T}\mathbb{E}_P U(\cdot,V_T^{x,\phi}(\cdot)).
\label{RUMP}
\end{eqnarray} 
Note that $u(x)=\sup_{\phi\in \Phi}\inf_{P\in \mathcal{Q}^T}\mathbb{E}_P U(\cdot,V_T^{x,\phi}(\cdot)).$ Indeed, if $\phi\in \Phi\setminus \Phi(x,U,\mathcal{Q}^T)$, then $\inf_{P\in\mathcal{Q}^T}\mathbb{E}_P U(\cdot,V_T^{x,\phi}(\cdot))=-\infty$ thanks to convention \eqref{cvt_inf}. 
We now introduce the different assumptions needed for the existence of an optimal strategy in \eqref{RUMP}.

\subsection{Direct Assumptions on $U$} 

We start with the assumptions which can directly and easily be checked. The first one is on the asymptotic behavior of the random utility $U$. In the uni-prior setting, this condition already appeared in \citep[Proposition 4]{ref7} for a finite nondecreasing continuous  and non necessarily concave function. See also \citep[Proposition 5.1]{ref2} and \citep[Proposition 3.24]{ref3}. 
\begin{assumption}
There exist $\gamma>0$ such that $\gamma \neq 1$ and a $\mathcal{B}(\Omega^T)$-measurable random variable $C : \Omega^T \to \mathbb{R}^+ \cup\{+\infty\}$ such that $\sup_{P\in\mathcal{Q}^T} \mathbb{E}_P \, C<+\infty$ and such that for all $\w^T\in\Omega^T$ satisfying $C(\w^T)<+\infty$, for all $\lambda\geq 1$ and $x\in\mathbb{R}$,
\begin{eqnarray}
U(\w^T,\lambda x)&\leq& \lambda^{\gamma}(U(\w^T,x)+C(\w^T)).\label{elas_gammaf}
\end{eqnarray}
\label{AE}
\end{assumption}
A consequence of Assumption \ref{AE} is that a one-step strategy must be bounded in order to be optimal (see Proposition \ref{sub_optimal}) and this compactness result will allow us to prove the existence of an optimal strategy. 
Assumption \ref{AE} is related to the notion of Reasonnable Asymptotic Elasticity (RAE) introduced in  \cite{refAE1} and \cite{refAE2}. To see that, assume for a moment that $U : \mathbb{R}\to \mathbb{R}$ is a non-random, non-constant and continuously differentiable function such that $\lim_{x\rightarrow +\infty}U(x)>0$ and define
\begin{eqnarray}
\mbox{AE}_{+\infty}(U):=\underset{x\rightarrow +\infty}{\lim \sup} \frac{xU'(x)}{U(x)} \quad \mbox{and}  \quad 
\mbox{AE}_{-\infty}(U):=\underset{x\rightarrow -\infty}{\lim \inf} \frac{xU'(x)}{U(x)}.\label{AE_brut_-inf}\label{AE_brut_inf}
\end{eqnarray}
$\mbox{AE}_{+\infty}(U)$ (resp. $\mbox{AE}_{-\infty}(U)$) is well-defined and is called asymptotic elasticity of $U$ at $+\infty$ (resp. $-\infty$). Note that $(x U'(x))/U(x)$ can be seen as the ratio of the marginal utility $U'(x)$ and the average utility $U(x)/x$. The reader may refer to \cite{refAE1}, \cite{refAE2} and \cite{refGI} for details on asymptotic elasticity. If  $U$ is also nondecreasing and concave, then $\mbox{AE}_{+\infty}(U)\leq 1$ and $\mbox{AE}_{-\infty}(U)
\geq 1$ (see Proposition \ref{annex_AE1}). A utility function is said to have RAE if $\mbox{AE}_{+\infty}(U)< 1$ and $\mbox{AE}_{-\infty}(U)>1$. In the uni-prior and continuous time setting, RAE is a necessary condition to get existence of an optimal solution for the utility maximization problem (see \citep{refAE2}). However, \citep{ref2} shows that in discrete time, it is enough to have either $\mbox{AE}_{+\infty}(U)< 1$ or $\mbox{AE}_{-\infty}(U)>1$ and this condition will be called RAE in discrete time. We show in Proposition \ref{annex_AE2} that RAE in discrete time implies \eqref{elas_gammaf}. Note that \citep[Example 7.3]{ref2} gives an example of a market situation where $\mbox{AE}_{+\infty}(U)=\mbox{AE}_{-\infty}(U)=1$ and the utility maximization problem has no solution. 
The next assumption ensures that $U$ takes negative values.
\begin{assumption}
There exists a $\mathcal{B}_c(\Omega^T)$-measurable random variable $\underline{X}: \Omega^T\to \mathbb{R}$ such that for all $\w^T\in\Omega^T$, $\underline{X}(\w^T)<0$ and $U(\cdot,\underline{X}(\cdot))<-C(\cdot)$ $\mathcal{Q}^T \mbox{q.s}$, where $C(\cdot)\geq 0$ has been introduced in Assumption \ref{AE}.
\label{nncst}
\end{assumption}
Assumption \ref{nncst} is of course satisfied if $U$ is deterministic, nondecreasing, non-constant and concave as $\lim_{x\to -\infty} U(x) =-\infty$. Utility functions with benchmark (see Definition \ref{utility_bench_def}) are examples of random utility functions satisfying Assumptions \ref{AE} and \ref{nncst} (see Proposition \ref{utility_bench_is_typeA}). 
If Assumption \ref{nncst} is not satisfied, then \eqref{RUMP} may have no solution, see Remark \ref{rm_mino} in the one-period case Section. 

The following assumption requires admissibility of simple strategies uniformly in $P$ and is discussed in the one-period setting after Assumption \ref{integ_V+}. 
\begin{assumption}
For all $1\leq t\leq T$, $x\in \mathbb{Q}$, $h \in \mathbb{Q}^d$, 
$\sup_{P\in\mathcal{Q}^T}\mathbb{E}_P U^-(\cdot,x+h\Delta S_t(\cdot))<+\infty.$ 
\label{simple_adm}
\end{assumption}

\subsection{Value functions and Assumption on $U_0^P$}

We need a last assumption which will provide a control from above on the value functions. This assumption stated on the value function at time $0$ relative to some priors $P$ is similar to \citep[Assumption 2.3 (1)]{ref2} and is thus well-accepted in the uni-prior setting. What is nice here is that the assumption is postulated prior by prior and not uniformly on all of them. Nevertheless, it is obviously not easy to verify. We will propose a still general context where it holds automatically true (see Theorem \ref{optimality_M_typeA}). \\
Before stating Assumption \ref{U0}, we need some further notation.
We introduce the conditional support of the price increments. Let $0\leq t\leq T-1$ and $P\in\mathfrak{P}(\Omega^T)$ with the fixed disintegration $P:=q_1^P \otimes \cdot\cdot\cdot \otimes q_T^P$, the multiple-priors conditional support $D^{t+1} : \Omega^t \twoheadrightarrow \mathbb{R}^d$  and the conditional support relatively to $P$, $D^{t+1}_P : \Omega^t \twoheadrightarrow \mathbb{R}^d$  are defined by
\begin{eqnarray*}
D^{t+1}(\w^t)&:=&\bigcap \{A\subset \mathbb{R}^d,\; \mbox{closed},\; p(\Delta S_{t+1}(\w^t,\cdot)\in A)=1,\;\forall p \in \mathcal{Q}_{t+1}(\w^t)\}\\
D^{t+1}_P(\w^t)&:=&\bigcap \{A\subset \mathbb{R}^d,\; \mbox{closed},\; q_{t+1}^P(\Delta S_{t+1}(\w^t,\cdot)\in A|\w^t)=1\}.
\end{eqnarray*}
Additionally, for some $R\subset \mathbb{R}^d$, let
\begin{eqnarray*}
\mbox{Aff}(R)&:=&\bigcap \{A\subset \mathbb{R}^d,\; \mbox{affine subspace},\; R\subset A\}\quad
\mbox{Conv}(R):=\bigcap \{C\subset \mathbb{R}^d,\; \mbox{convex},\; R\subset C\}
\end{eqnarray*}
and if $R$ is convex, $\mbox{ri}(R)$ is the interior of $R$ relatively to $\mbox{Aff}(R)$.

The closure of a random function $F$ : $\Omega^t\times \mathbb{R} \to \mathbb{R}\cup\{-\infty,+\infty\}$ is defined as follows. Fix $\w^t\in\Omega^t$. Then, $x\mapsto F_{\w^t}(x):=F(\w^t,x)$ is an extended real-valued function and its closure, denoted by $\textup{Cl}(F_{\w^t})$, is the smallest upper-semicontinuous (usc) function $w$ : $\mathbb{R}\to \mathbb{R}\cup\{-\infty,+\infty\}$ such that $F_{\w^t}\leq w$. Now $\textup{Cl}(F)$ : $\Omega^t\times \mathbb{R} \to \mathbb{R}\cup\{-\infty,+\infty\}$ is defined by $\textup{Cl}(F)(\w^t,x):=\textup{Cl}(F_{\w^t})(x)$.

We now introduce the dynamic programming procedure and the associated value functions. We do this for the multiple-priors utility maximization problem \eqref{RUMP} with the value functions $U_t$ and also for the utility problem \eqref{RUMP_one_prior} related to a given prior $P$ with the value functions $U_t^P$. Fix $P\in\mathcal{Q}^T$ with the fixed disintegration $q_1^P\otimes \cdots \otimes q_T^P$. Let  
\begin{eqnarray}
u_P(x):=\sup_{\phi\in\Phi(x,U,P)}\mathbb{E}_P U(\cdot, V_T^{x,\phi}(\cdot)). \label{RUMP_one_prior}
\end{eqnarray}
For all $0\leq t\leq T-1$, we define
\begin{eqnarray}
&&\left\{
    \begin{array}{ll}
    \mathcal{U}_T^P(\w^T,x)&:=U(\w^T,x)\\
	\mathcal{U}_t^P(\w^t,x)&:=
	\sup_{h\in\mathbb{Q}^d}\mathbb{E}_{q_{t+1}^P(\cdot|\w^t)} U_{t+1}^P(\w^t,\cdot,x+h\Delta S_{t+1}(\w^t,\cdot)) \\
	U_t^P(\w^t,x)&:=\textup{Cl}(\mathcal{U}_t^P)(\w^t,x).
    \end{array} \label{state_val_t_p}
\right.\\
&&\left\{
	\begin{array}{ll}
    \mathcal{U}_T(\w^T,x)&:=U(\w^T,x)\\
	\mathcal{U}_t(\w^t,x)&:=
	\sup_{h\in\mathbb{Q}^d}\inf_{p\in\mathcal{Q}_{t+1}(\w^t)} \mathbb{E}_p U_{t+1}(\w^t,\cdot,x+h\Delta S_{t+1}(\w^t,\cdot))\\
	U_t(\w^t,x)&:=\textup{Cl}(\mathcal{U}_t)(\w^t,x).
    \end{array}
\right.
\label{state_val_t_rob}
\end{eqnarray}
The existence and the measurability of $U_t^P$ and $U_t$ are not trivial. The purpose of the closure is to ensure that these value functions are usc for all $\w^t\in\Omega^t$ and not only on a $\mathcal{Q}^t$-full-measure set. This is crucial for the dynamic programming procedure to hold true.

Assumption \ref{U0} will ensure a control from above on each $U_t^P$ for $P$ in the set $\mathcal{H}^T$ defined by
\begin{eqnarray}
\label{H_def}
\mathcal{H}^T &:= &\{P\in \mathcal{Q}^T,\; 0\in \textup{ri}(\textup{conv}(D_P^{s+1}))(\cdot)\; \mathcal{Q}^s \textup{-q.s.},\;\textup{Aff}(D_P^{s+1})(\cdot)=\textup{Aff}(D^{s+1})(\cdot)\; \mathcal{Q}^s \textup{-q.s.} \nonumber \quad \forall 0\leq s \leq T-1\}. 
\end{eqnarray}
We first comment on the set $\mathcal{H}^T$ and the link with the $\mbox{NA}(\mathcal{Q}^T)$ condition, see Definition \ref{defNA}.
\begin{lemma}
\label{lemmaH_nonempty}
We have that $\mathcal{H}^T\subset \mathcal{Q}^T$. Assume that Assumptions \ref{S_borel} and \ref{analytic_graph} hold true. If $\mbox{NA}(\mathcal{Q}^T)$ holds true then $\mathcal{H}^T\neq \emptyset$, $\mathcal{H}^T$ and $\mathcal{Q}^T$ have the same polar sets and $\mbox{NA}(P)$ holds true for all $P\in\mathcal{H}^T$. Conversely, if $\mathcal{H}^T\neq \emptyset$ then $\mbox{NA}(\mathcal{Q}^T)$ holds true.
\end{lemma}
Thus,  $\mathcal{H}^T$ has the same properties as the set $\mathcal{P}^T$ introduced in \citep[Theorem 3.6]{ref4}. This set $\mathcal{P}^T$ is defined as follows
\begin{eqnarray}
\mathcal{P}^T := \{(\lambda_1 q_1^* + (1-\lambda_1)q_1^Q) \otimes \cdots \otimes (\lambda_T q_T^* + (1-\lambda_T)q_T^Q),\; 0< \lambda_i \leq 1,\; Q\in\mathcal{Q}^T\} \subset \mathcal{H}^T,
\label{setPT}
\end{eqnarray}
for a fixed $P^*:= q_1^* \otimes \cdots \otimes q_T^*\in\mathcal{H}^T$, see \citep[Remarks 3.7 and 3.31]{ref4}. Note that there exist infinitely many sets  satisfying the properties of \citep[Theorem 3.6]{ref4} and $\mathcal{H}^T$ is not the smallest one for the inclusion. Indeed, let $T=1$, $d=1$, $\overline{\Omega}=\mathbb{R}$, $\Delta S_1 : \w \mapsto \w$  
$$\mathcal{Q}^1= \left\{\sum_{k=1}^n \alpha_k \delta_{a_k}, \;\; n\geq 1,\; a_k\in\mathbb{R},\; \alpha_k\geq 0\; \mbox{and}\;\sum_{k=1}^n \alpha_k =1\right\},$$
where for all $a\in \mathbb{R}$, $\delta_a$ is the Dirac measure at $a$. Set $P^*:= \frac{1}{3}\delta_{-1}+\frac{1}{3}\delta_{0}+\frac{1}{3}\delta_{1}\in\mathcal{Q}^1$ and $\overline{P}:=\frac{1}{2}\delta_{-1}+\frac{1}{2}\delta_{1}\in\mathcal{Q}^1$. Then, we trivially have that $\textup{Aff}(D^1_{\overline{P}})=\textup{Aff}(D^1_{P^*})=\mathbb{R}=\textup{Aff}(D^1)$ and that $0$ belongs to $\textup{ri}(\textup{conv}(D_{\overline{P}}^{1}))$ and to $\textup{ri}(\textup{conv}(D_{P^*}^{1}))$. Thus, $P^*$ and $\overline{P}$ belong to $\mathcal{H}^1$. Nevertheless, as $D_{P^*}^1 = \{-1,0,1\}$ and  recalling \eqref{setPT}, each $P\in \mathcal{P}^1$ must have a support containing at least three points and this is not the case for $\overline{P}$. So, $\overline{P}\notin\mathcal{P}^1$ but $\overline{P}\in\mathcal{H}^1$ and the inclusion  
$\mathcal{P}^1 \subset \mathcal{H}^1$ is strict. 

\proof{Proof of Lemma \ref{lemmaH_nonempty}}
Clearly, $\mathcal{H}^T\subset \mathcal{Q}^T$. The fact that $\mathcal{H}^T\neq \emptyset$ is equivalent to $\mbox{NA}(\mathcal{Q}^T)$ is proved in \citep[Theorem 3.29]{ref4}. Assume that $\mbox{NA}(\mathcal{Q}^T)$ holds true. Then, for all $P\in\mathcal{H}^T$, $\mbox{NA}(P)$ holds true as $0\in \textup{ri}(\textup{conv}(D_P^{s+1}))(\cdot)\; P^s \textup{-a.s}$ for all $0\leq s \leq T-1$, see \citep[Proposition 3.25]{ref4}. As $\mathcal{H}^T\subset \mathcal{Q}^T$, it is clear that a $\mathcal{Q}^T$ polar set is also a $\mathcal{H}^T$ polar set. Now, let $Q:=q_1^Q \otimes \cdots \otimes q_T^Q\in\mathcal{Q}^T$ and $P^*:= q_1^*\otimes \cdots \otimes q_T^*\in\mathcal{H}^T$. Set $R:= \frac{q_1^Q + q_1^*}{2} \otimes \cdots \otimes \frac{q_T^Q + q_T^*}{2}.$ Then, Proposition \ref{domin_lcl} shows that $Q\ll R$ and \eqref{setPT} that $R\in\mathcal{P}^T\subset \mathcal{H}^T$. So, a $\mathcal{H}^T$-polar set is also a $\mathcal{Q}^T$-polar set. Thus, $\mathcal{H}^T$ and $\mathcal{Q}^T$ have the same polar sets.
\Halmos \endproof
\begin{assumption}
For all $P\in \mathcal{H}^T$, $U_0^P(1)<+\infty$.
\label{U0}
\end{assumption}
Before commenting on Assumption \ref{U0}, we first link it to the existence of a kind of admissible one-step optimal strategies for all priors in $\mathcal{H}^T$ and to the finiteness of the uni-prior value functions.
\begin{lemma}
Let $U$ be a random utility function (see Definition \ref{U_hp}) and $\mathcal{Q}^T$ as in \eqref{def_set_Q}. Assume that $\mathcal{H}^T\neq \emptyset$, that for all $P:=q_1^P\otimes \cdots \otimes q_T^P\in\mathcal{H}^T$, $u_P(1)<+\infty$ (see \eqref{RUMP_one_prior}) 
 and that there exists $\phi^{*,P}\in \Phi(1,U,P)$ such that for all $0\leq t \leq T-1$, for all $\w^t$ in a $P^t$-full-measure set, 
\begin{eqnarray}
U_{t}^P(\w^t,V_t^{1,\phi^{*,P}}(\w^t))&=&\mathbb{E}_{q_{t+1}^P(\cdot|\w^t)} U^P_{t+1}\left(\w^t,\cdot,V_t^{1,\phi^{*,P}}(\w^t)+ \phi^{*,P}_{t+1}(\w^t)\Delta S_{t+1}(\w^t,\cdot)\right).\label{well_prepared}
\end{eqnarray}
Then, Assumption \ref{U0} holds true.
\end{lemma}
\proof{Proof.}
Let $P:=q_1^P\otimes \cdots \otimes q_T^P\in\mathcal{H}^T$. First, we apply  recursively \eqref{well_prepared} from $t=0$ to $t=T-1$, as $U_T^P=U$, 
and then, we use Fubini's theorem as $\phi^{*,P}\in\Phi(1,U,P)$
\begin{eqnarray*}
U_0^P(1) & = & \int_{\Omega_1}\cdots \int_{\Omega_T} U(\w^T, V_T^{1,\phi^{*,P}}(\w^T))\; q_T^P(d\w_T|\w^{T-1}) \cdots q_1^P(d\w_1)\\
  & = & \int_{\Omega^T} U(\w^T, V_T^{1,\phi^{*,P}}(\w^T))\; P(d\w^T) = \mathbb{E}_P U(\cdot,V_T^{1,\phi^{*,P}}(\cdot)) \leq u_P(1),
\end{eqnarray*}
 where the last inequality follows again from $\phi^{*,P}\in\Phi(1,U,P)$. As $u_P(1)<+\infty$, Assumption \ref{U0} holds true. 
\Halmos \endproof
Assumption \ref{U0} is similar to \citep[Assumption 2.3 (1)]{ref2} or \citep[Assumption 2 (11)]{ref7} postulated for all priors of $\mathcal{H}^T$. In the case of a utility function defined on the positive axis, \citep[Assumption 3.5]{ref3} also allows to control from above the value function. It should be stressed however that \citep[Assumptions 3.1 and 3.5]{ref3} ensure that the set of admissible strategies is exactly the set of trading strategies with positive terminal outcomes (see \citep[Proposition 3.25]{ref3}) and thus inadequate for our problem where outcomes can be negative. 
Assumption \ref{U0} allows to show that the one-period Assumption \ref{integ_V+} below holds on each one-period market quasi-surely. One may wonder if we can only require that $U_0^{P^*}(1)<+\infty$ for a given $P^*\in \mathcal{H}^T.$ In this case, Assumption \ref{integ_V+} will only be true on a $P^*-$related full-measure set which is not enough to solve the multiple-priors utility maximization problem for a nondominated set of priors $\mathcal{Q}^T$. 
One may also wonder if we could assume instead of Assumption \ref{U0} that $U_0(1)<+\infty.$ The answer is no, see Remark \ref{remark_CE1}.
\begin{remark}
The existence of an optimal strategy may fail when $U_0(1)<+\infty$ but if there exists some $P^*\in\mathcal{H}^T$ such that $U_0^{P^*}(1)=+\infty$. Set $T=1$, $d=1$, $\Omega^1=\mathbb{R}$, $\Delta S_1 : \w \mapsto \w$ and for all $\w\in\Omega^1$ and $x\in\mathbb{R}$, $U(\w,x):=\widetilde{U}(x)+\exp(\exp(\w))$ where $\widetilde{U}(x):=1-\exp(-x)$. The function $U$ is a random utility in the sense of Definition \ref{U_hp}. Let $P^* \in \mathfrak{P}(\Omega^1)$ be the standard normal law. Set $\mathcal{Q}^1:=\textup{Conv}(P^*,\delta_1)$. Assumptions \ref{S_borel} and \ref{analytic_graph} trivially hold true and $D_{P^*}^1 = D^1 = \mathbb{R}$. So, $P^*\in\mathcal{H}^1$ and $\textup{NA}(\mathcal{Q}^1)$ holds true, see Lemma \ref{lemmaH_nonempty}. Noting that $\mbox{AE}_{-\infty}(\widetilde{U})=+\infty>1$ and using Proposition \ref{annex_AE2} in Appendix \ref{AE_app}, Assumption \ref{AE} holds true for $\widetilde{U}$ with $\gamma \in (1, \mbox{AE}_{-\infty}(\widetilde{U}))$ and some $\widetilde{C}>0$. So, Assumption \ref{AE} holds true also for $U$ with the same $\gamma$ and $\widetilde{C}$ as $\exp(\exp(\w))\geq 0$. Assumption \ref{nncst} immediately holds true with $\underline{X}(\w):= \widetilde{U}^{-1}(-\widetilde{C}-1-\exp(\exp(\w)))$ where $\widetilde{U}^{-1}: (-\infty,1)\ni x \mapsto -\ln(1-x)$ is the (continuous) inverse function of $\widetilde{U}$. For all $(x,h)\in\mathbb{R}\times \mathbb{R}^d$, $U^-(\w,x)\leq \widetilde{U}^-(x)\leq \exp(-x)$ and we have that $\mathbb{E}_{P^*} U^-(\cdot,x+h\Delta S_1(\cdot)) \leq \exp(\frac{h^2}{2}-x) <+\infty$, $\mathbb{E}_{\delta_1} U^-(\cdot,x+h\Delta S_1(\cdot)) \leq \exp(-(x+h))<+\infty$  and also, $\sup_{P\in\mathcal{Q}^1} \mathbb{E}_{P} U^-(\cdot,x+h\Delta S_1(\cdot)) <+\infty$ : Assumption \ref{simple_adm} holds true. We  now show that $U_0^{P^*}(1)=+\infty$, $U_0(1)<+\infty$ and that the multiple-priors utility maximization problem has no solution. Note first that $\mathbb{E}_{P^*} \exp(\exp(\cdot))=+\infty$ and that $\mathbb{E}_{P^*} \widetilde{U}(1+h\Delta S_1(\cdot)) = 1-\exp(\frac{h^2}{2}-1)$ for all $h\in\mathbb{R}$. As a result, $\mathbb{E}_{P^*} U(\cdot,1+h\Delta S_{1}(\cdot)) =+\infty.$
So, $U_0^{P^*}(1)\geq \mathcal{U}_0^{P^*}(1)\geq \mathbb{E}_{P^*} U(\cdot,1)=+\infty$ (see \eqref{state_val_t_p}). The function $\Psi : h\mapsto \inf_{P\in\mathcal{Q}^1}\mathbb{E}_P U(\cdot,1+h\Delta S_1(\cdot))=\mathbb{E}_{\delta_1} U(\cdot,1+h\Delta S_1(\cdot))=1-\exp(-(1+h)) + \exp(\exp(1))$ doesn't admit a maximizer and \eqref{RUMP} has no solution. Nevertheless, as $\Psi(h)\leq 1+\exp(\exp(1))$, we get that $U_0(1)\leq 1+\exp(\exp(1))<+\infty$ (see \eqref{state_val_t_rob}).


\label{remark_CE1}
\end{remark}


\subsection{Main result}
\label{lab_main_res}
We are now in position to state our first main result.

\begin{theorem}
Let $U$ be a random utility (see Definition \ref{U_hp}). Assume that the $NA(\mathcal{Q}^T)$ condition as well as Assumptions \ref{S_borel}, \ref{analytic_graph}, \ref{AE}, \ref{nncst}, \ref{simple_adm} and designatehold true. 
Then there exist $(\phi^{*,x})_{x\in\mathbb{R}}\subset \Phi$ and for all $0\leq t \leq T-1$, some $\mathcal{Q}^t$-full-measure set $\widehat{\Omega}^t \in \mathcal{B}_c(\Omega^t)$ such that for all $x\in\mathbb{R}$ and $\w^t\in\widehat{\Omega}^t$, we have that $\phi^{*,x}_{t+1}(\w^t)\in \textup{Aff}(D_{t+1})(\w^t)$ and that
\begin{eqnarray}
U_{t}(\w^t,V_t^{x,\phi^{*,x}}(\w^t))&=&\sup_{h\in\mathbb{R}^d}\inf_{p\in\mathcal{Q}_{t+1}(\w^t)}\mathbb{E}_p U_{t+1}\left(\w^t,\cdot,V_{t}^{x,\phi^{*,x}}(\w^t)+h \Delta S_{t+1}(\w^t,\cdot)\right)\label{onevc v c _step_strategy_temp0}\\ 
&=&\inf_{p\in\mathcal{Q}_{t+1}(\w^t)}\mathbb{E}_p U_{t+1}\left(\w^t,\cdot,V_{t}^{x,\phi^{*,x}}(\w^t)+\phi^{*,x}_{t+1}(\w^t) \Delta S_{t+1}(\w^t,\cdot)\right).
\label{one_step_strategy_temp2}
\end{eqnarray}
Moreover, if $\phi^{*,x}\in \Phi(x,U,\mathcal{Q}^T)$, then 
\begin{eqnarray}
u(x)&=&U_0(x)=\inf_{P\in \mathcal{Q}^T}\mathbb{E}_P U(\cdot,V_T^{x,\phi^{*,x}}(\cdot)).
\label{eg_fdmt}
\end{eqnarray}
\label{optimality_pp}
\label{one_step_strategy}
\end{theorem}
The proof of Theorem \ref{optimality_pp} is quite involved and is delayed to Section \ref{proof_th1}. To construct $\phi^{*,x}$, one glues all the one-step optimal strategies constructed in Section \ref{op} together. Indeed, the one-step strategy $\phi^{*,x}_{t+1}$ will be the optimal strategy to adopt between time $t$ and $t+1$ starting from an initial wealth equal to $\sum_{s=1}^t \phi^{*,x}_s \Delta S_s$ i.e. if one has followed the strategies $(\phi^{*,x}_{1},\cdot\cdot\cdot,\phi^{*,x}_{t})$ until time $t$. Note that the strategy $\phi^{*,x}$ belongs to $\Phi$ but may fail to be admissible and thus to be a solution of (\ref{RUMP}). However, when $\phi^{*,x}$ is admissible, it achieves the supremum in (\ref{RUMP}). If $\phi^{*,x}$ is not admissible, $\inf_{P\in\mathcal{Q}^T}\mathbb{E}_P U(\cdot,V_T^{x,\phi^{*,x}}(\cdot))=-\infty$ by convention \eqref{cvt_inf} and $\phi^{*,x}$ is not optimal. 
Indeed, Assumption \ref{simple_adm} implies that $u(x) >-\infty$ for all $x\in \mathbb{R}$. \\
The fact that $\phi^{*,x}$ is not automatically admissible is not specific to our quasi-sure setting. Already in \citep[Theorem 2.7]{ref2} or \citep[Theorem 1]{ref7}, one has to assume that $\phi^{*,x}$ is  admissible (in their case that $\mathbb{E}_P U(V_T^{x,\phi^{*,x}}(\cdot))$ exists) in order to be optimal (see \citep[Remark 15]{ref7}). We exhibit in Example \ref{example1} below some $\phi^{*,x}$ which  is not  admissible. However, Assumption \ref{U0} is not satisfied in this example. 

\subsection{Application}
\label{sec ilus}
The condition that $\phi^{*,x}\in \Phi(x,U,\mathcal{Q}^T)$ is obviously not easy to verify and we would like to check that Theorem \ref{optimality_pp} applies for a concrete, broad class of market models and random utility functions. For that, we first define some sets of random variables which are integrable enough.
\begin{definition}
Fix $0\leq t\leq T$ and $P\in\mathfrak{P}(\Omega^t)$.
\begin{eqnarray*}
\mathcal{W}^t &:=& \{X : \Omega^t \to \mathbb{R}\cup\{-\infty, +\infty\},\; \mathcal{B}_c(\Omega^t)\mbox{-measurable such that} \sup_{P\in\mathcal{Q}^t} \mathbb{E}_P |X|^r <+\infty,\; \forall r\geq 1\}\\
\mathcal{M}^t(P) &:=& \left\{X : \Omega^t \to \mathbb{R}\cup\{-\infty, +\infty\},\; \mathcal{B}_c(\Omega^t)\mbox{-measurable such that } \mathbb{E}_P |X|^r <+\infty,\; \forall r\geq 1\right\}\\
\mathcal{M}^t &:=& \bigcap_{P\in\mathcal{Q}^t}\mathcal{M}^t(P).
\end{eqnarray*}
\label{def_W}
\label{M_tP}
\end{definition}
Note that $\mathcal{W}^0=\mathcal{M}^0=\mathcal{M}^{0}(P)=\mathbb{R}$. It is also clear that $\mathcal{W}^t \subset \mathcal{M}^t\subset \mathcal{M}^t(P)$ for all $P\in\mathcal{Q}^t$. The set $\mathcal{W}^T$ has already been introduced in \citep[p1866]{ref3} but for Borel measurable random variables. 
\begin{definition}
Let $U$ be a random utility as in Definition \ref{U_hp}. A random utility is of type (A) if $U^+(\cdot,1)\in \mathcal{M}^T$, Assumption \ref{AE} holds true for some $C\in\mathcal{M}^T$,  Assumption \ref{nncst} holds true for some $\underline{X}\in \mathcal{M}^T$ such that $1/|U(\cdot,\underline{X}(\cdot))+C(\cdot)|\in \mathcal{M}^T,$ and if there exist $p\geq 1$ and a non-negative $\mathcal{B}(\Omega^T)$-measurable random variable $C_1\in \mathcal{W}^T$ such that $\mathcal{Q}^T$-q.s, for all $x\in\mathbb{R}$
\begin{eqnarray}
U(\cdot,x)\geq -C_1(\cdot)(1+|x|^p).
\label{ineq_b_inf_det_A}
\end{eqnarray}
\label{typeA}
\end{definition}
We propose first an example of a utility function of type (A).
\begin{definition}
Let $\widetilde{U} : \mathbb{R}\to \mathbb{R}$ be a concave, nondecreasing, non-constant, continuously differentiable function satisfying either $\mbox{AE}_{+\infty}(\widetilde{U})<1$ and $\lim_{x\to +\infty} \widetilde{U}(x)>0$ or $\mbox{AE}_{-\infty}(\widetilde{U})>1$. Assume also that there exist $p\geq 1$ and $\widetilde{b}>0$ such that for all $x\in\mathbb{R}$,
\begin{eqnarray}
\widetilde{U}(x)\geq -\widetilde{b}(1+|x|^p).
\label{ineq_b_inf_det_bb}
\end{eqnarray}
Let $Z \in \mathcal{W}^T$ be $\mathcal{B}(\Omega^T)$-measurable. A function $U : \Omega^T\times \mathbb{R} \to \mathbb{R}$ is a  utility function with random benchmark if $U(\w^T,x):=\widetilde{U}(x-Z(\w^T))$ for all $\w^T\in\Omega^T$ and $x\in\mathbb{R}$.\\
\label{utility_bench_def}
\end{definition}

A utility function with random benchmark is of type (A), see Proposition \ref{utility_bench_is_typeA}. Utility functions with random benchmark have been introduced in  Kahneman and Tversky, see \citep{refkahn}. The random variable $Z$ is the so-called benchmark or reference point. The need of a shift of reference point $Z$ arises as people usually express their decision problems in terms of net position with respect to some benchmark rather than in terms of  gains and losses, see \citep[``Shifts of Reference"]{refkahn}.

The following theorem shows the existence of an optimal strategy for random utility of type $(A)$ under some integrability conditions on the market especially on the process $\alpha^P$ that we introduce now and that is related to the ``quantitative" no-arbitrage condition, see \citep[Definition 3.19]{ref4}.
\begin{lemma}
\label{simi_qt_na}
Assume that the $NA(\mathcal{Q}^T)$ condition  as well as Assumptions \ref{S_borel} and \ref{analytic_graph} hold true. Fix $P:=q_1^P\otimes \cdots \otimes q_T^P\in \mathcal{H}^T$. For all $0\leq t \leq T-1$, there exists some $\mathcal{B}_c(\Omega^t)$-measurable random variable $\alpha_t^P(\cdot)\in (0,1]$ such that $\Omega^{t,P}_{qNA}$ is a $\mathcal{Q}^t$-full-measure set, where
\begin{eqnarray}
\Omega^{t,P}_{qNA}:=\left\{\w^t\in\Omega^t,\forall h\in\mbox{Aff}(D^{t+1})(\w^t), h\neq 0, q_{t+1}^P\left(h\Delta S_{t+1}(\w^t,\cdot)<-\alpha_t^P(\w^t)|h| | \w^t \right)\geq \alpha_t^P(\w^t)\right\}.
\label{set_qt_na_H}
\end{eqnarray}
\end{lemma}
\proof{Proof.}
See \citep[Proposition 3.35]{ref4}.
\Halmos \endproof

\begin{theorem}
Let $U$ be a random utility of type $(A)$. Assume that the $NA(\mathcal{Q}^T)$ condition as well as Assumptions \ref{S_borel} and \ref{analytic_graph} hold true. Moreover, suppose that for all $P\in \mathcal{H}^T$ and for all $0\leq t \leq T-1$, $1/\alpha_{t}^P\in \mathcal{M}^{t}$ and  $|\Delta S_{t+1}|\in\mathcal{W}^{t+1}$.
Then, for all $x\in \mathbb{R}$, there exists $\phi^{*,x}\in \Phi(x,U,\mathcal{Q}^T)$ such that 
\begin{eqnarray*}
u(x)=\sup_{\phi\in \Phi(x,U,\mathcal{Q}^T)}\inf_{P\in \mathcal{Q}^T}\mathbb{E}_P U(\cdot,V_T^{x,\phi}(\cdot))
&=& \inf_{P\in \mathcal{Q}^T}\mathbb{E}_P U(\cdot,V_T^{x,\phi^{*,x}}(\cdot)).
\end{eqnarray*}
\label{optimality_M_typeA} 
\end{theorem}
\proof{Proof.}
See Section \ref{proof_optimality_M}.
\Halmos \endproof
\begin{remark}
We will show in Remark \ref{remoptiadmi} that the integrability condition on $1/\alpha_{t}^P$ can not be removed. 
The integrability conditions of Theorem \ref{optimality_M_typeA} are quite classical in the uni-prior literature on unbounded utility functions, see for example \citep[Proposition 7.1]{ref2}, \citep[Theorem 4.16]{ref10} and \citep[Proposition 7]{ref7}. In the multiple-priors setting, the chosen no-arbitrage condition and the set of priors for which the integrability condition on $\alpha_t^P$ is postulated are crucial. For general utility functions defined on the positive axis, \citep[Theorem 3.6]{ref3} states that if for all $P\in\mathcal{Q}^T$  $NA(P)$ holds true and $1/\alpha_t^P\in\mathcal{W}^t$, then there exists an optimal strategy. In \citep[Corollary 3.16]{ref4}, the same result is obtained under $NA(\mathcal{Q}^T)$ if $1/\alpha_t^P\in \mathcal{W}^t$ for all $P\in \mathcal{H}^T$. So, Theorem \ref{optimality_M_typeA} extends \citep[Corollary 3.16]{ref4} to utility functions of type (A)  defined on $\mathbb{R}$ assuming only  that $1/\alpha_t^P\in \mathcal{M}^t$ instead of $\mathcal{W}^t$. 
Moreover, the assumption that $U^+(\cdot,1), \, C, \, \underline{X}, 1/|U(\cdot,\underline{X}(\cdot))+C(\cdot)| \in \mathcal{M}^T,$  $C_1\in \mathcal{W}^T,$ 
$1/\alpha_{t}^P\in \mathcal{M}^{t}$ and  $|\Delta S_{t+1}|\in\mathcal{W}^{t+1}$ for all $P\in \mathcal{H}^T$ and for all $0\leq t \leq T-1$ 
could be weakened to the existence 
of the $N$-th moment for $N$ large enough but this would lead to complicated book-keeping with no essential gain in generality, which we prefer to avoid. Note finally that in \citep[Theorem 3.11]{ref8}, for a utility function defined on $\mathbb{R}$, the existence of an optimal strategy is proved under similar integrability conditions as in Theorem \ref{optimality_M_typeA} but in a different setup for the uncertainty and for another kind of no-arbitrage condition.\\
\end{remark}

\begin{example}
\label{example1}
We provide an illustration of Theorem \ref{optimality_M_typeA}. Our example is similar to the one at the end of \citep[Section 4.2]{ref4}. Let $d=1$ and $\Omega^t:=\Omega$ for some Polish space $\Omega$. Let $Z$ be some Borel-measurable  random variable defined on $\Omega$. Define $F : \mathfrak{P}(\Omega) \to \mathbb{R}^3$ by 
$$F(p) := \left(\mathbb{E}_p(Z),\; \mathbb{E}_p(Z^2)-1,\; 1_{\mathbb{E}_p \exp(|Z|)>C}\right),$$ for some fixed $C\geq 2 \exp(1/2).$  Then, let $\mathcal{Q}_{t+1}(\w^t) := F^{-1}(\{0\}) =: \mathcal{Q}$ for all $0\leq t\leq T-1$ and $\w^t\in\Omega^t$. For each $\w^t\in\Omega^t$ and 
$p \in \mathcal{Q}_{t+1}(\w^t)$, the law of $Z$ under $p$ for the next period is centered, have a variance equal to 1 and a bounded exponential moment. Now, choose $S_0 \in \mathbb{R}$ and for all $0\leq t\leq T-1$ and $(\w^t, \w_{t+1})\in \Omega^t\times \Omega_{t+1}$, set
$$S_{t+1}(\w^t,\w_{t+1}) := S_t(\w^t) + r +\sigma Z(\w_{t+1}),$$
for some $r\in\mathbb{R}$ and $\sigma>0$ such that $-\beta <r/\sigma< \beta$ where $\beta$ is a ``universal" constant defined in Lemma \ref{lemma_temp_exemple} below.
Then,  \citep[Assumption 4.7]{ref4} is satisfied with $P_0 := p_0 \otimes \cdots \otimes p_0$, where 
$p_0 \in\mathfrak{P}(\Omega)$ is such that the law of $Z$ under $p_0$ is a standard Gaussian distribution. 
Thus, \citep[Lemmata 4.8 and 4.9]{ref4} show that the $NA(\mathcal{Q}^T)$ condition as well as Assumptions \ref{S_borel} and \ref{analytic_graph} hold true. The next lemma allows to construct $\alpha_{t}^P$ that satisfies \eqref{set_qt_na_H} for all $P\in\mathcal{Q}^T$. Note that in \citep[Section 4.2]{ref4}, the result of Lemma \ref{lemma_temp_exemple} is proved for $p_0$ only. 
\noindent \begin{lemma}
There exists $\beta\in (0,1]$ such that $p(Z < -\beta)\geq \beta$ and $p(-Z < -\beta)\geq  \beta$ for all $p\in\mathcal{Q}$. \label{lemma_temp_exemple}
\end{lemma}
\proof{Proof.}
See Section \ref{sec lemma_temp_exemple}.
\Halmos \endproof
Fix $P\in\mathcal{Q}^T$. Let $\alpha^P_{t+1}(\w^t) := \min(\beta,\; \sigma \beta - r,\; \sigma \beta + r)$ for all $0\leq t\leq T-1$ and $\w^t\in\Omega^t$. As $-\beta <r/\sigma< \beta$ and $\beta \in (0,1]$, $\alpha^P_{t+1}(\cdot)\in(0,1].$ Moreover, Lemma \ref{lemma_temp_exemple}  shows that  
\eqref{set_qt_na_H} is satisfied. As $\alpha_t^P$ is deterministic, it is immediate that $1/\alpha_t^P \in \mathcal{M}^t$ for all $P\in\mathcal{H}^T \subset \mathcal{Q}^T$ and $0\leq t\leq T-1$. It remains to show that $|\Delta S_{t+1}| \in\mathcal{W}^{t+1}$ for all $0\leq t\leq T-1$. 
Let $P =q_1^P \otimes \cdots \otimes q_{t+1}^P \in\mathcal{Q}^{t+1}$. Using Fubini's theorem and as $q_{t+1}^P(\cdot|\w^t)\in\mathcal{Q}$ for all $\w^t\in\Omega^t$, 
$$\mathbb{E}_P |Z|^n = \int_{\Omega^t} \mathbb{E}_{q_{t+1}^P(\cdot|\w^t)}|Z|^n P^t(d\w^t) \leq 
\sup_{p\in\mathcal{Q}}\mathbb{E}_p |Z|^n \leq n!\; \sup_{p\in\mathcal{Q}}\mathbb{E}_p \exp(|Z|) \leq n!\; C,$$
using the exponential series inequality $|x|^n / n! \leq \exp(|x|)$ and the definition of $\mathcal{Q}$.
Thus, 
$$\sup_{P\in\mathcal{Q}^t}\mathbb{E}_P |\Delta S_{t+1}|^n \leq 2^{n-1}(|r|^n + \sigma^n \sup_{P\in\mathcal{Q}^t}\mathbb{E}_P |Z|^n ) \leq 2^{n-1}(|r|^n + \sigma^n n! \; C )<+\infty$$ and as $n$ is arbitrary, $|\Delta S_{t+1}|\in\mathcal{W}^{t+1}$. All the conditions of Theorem \ref{optimality_M_typeA} are satisfied and there exists an 
optimal strategy for all  random utility $U$ of type (A).\\
\end{example}

\begin{remark}
\label{remoptiadmi}
The integrability condition on $1/\alpha_t^P$ is necessary in Theorem \ref{optimality_M_typeA} even in a uni-prior context. We exhibit a nonrandom utility function of type (A) for which one-step optimal strategies exist (see the first part of Theorem \ref{one_step_strategy}) but all of them are not admissible and are therefore not optimal. This means that the dynamic programming procedure fails. Set $T=2$, $\Omega_1 := \mathbb{R}$, $\Omega_2 := \{1,3\}$ and $d=1$. Fix $0.5 <a<1$. Let $U(x):= a x$ if $x\leq 0$ and $U(x):=(1+x)^{a}-1$ if $x> 0$. It is easy to see that $U$ is concave, nondecreasing, non-constant and continuously differentiable. 
Now, $\mbox{AE}_{+\infty}(U)= a<1$ and $\lim_{x\to +\infty}{U}(x)=+\infty $. Moreover,  $U(x) \geq 0$ for all $x\geq 0$ and  $U(x) = a x = - a |x| \geq - a(1+|x|)$ for all $x <0$ and \eqref{ineq_b_inf_det_bb} holds true with $\widetilde{b}=a$ and $p=1$. 
Thus, $U$ satisfies Definition \ref{ineq_b_inf_det_bb} with  $Z=0$ and is a (nonrandom) utility function of type (A), see Proposition \ref{utility_bench_is_typeA}. Let $S_0 := 2$, $S_1(\w^1) := 2$ for all $\w^1\in\Omega_1$ and $S_2(\w^2) := S_2(\w_2) := \w_2$. Then, Assumption \ref{S_borel} holds true. Let $l : \Omega^1 \to (0,1)$ be a Borel-measurable function and $q_1\in \mathfrak{P}(\Omega^1)$ be such that the law of $l$ under $q_1$ is uniform on $(0,1)$. Let $q_2(\cdot|\w^1) := l(\w^1) \delta_{\{3\}}(\cdot) + (1- l(\w^1)) \delta_{\{1\}}(\cdot)$. Set $P:= q_1\otimes q_2$ and $\mathcal{Q}^2= \{P\}$. Then, Assumption \ref{analytic_graph} holds true (see \citep[Corollary 7.14.1, p121]{ref1}). 
As $\Delta S_1=0$, $\Delta S_1 \in \mathcal{W}^1$ and $D_1 = D_1^P = \{0\}$. Thus, there is nothing to check in \eqref{set_qt_na_H} and we can take for $\alpha_0$ any number in $(0,1]$. We choose $\alpha_0=1$. Now, $\Delta S_2(\w^2) = \w_2-2$ and recalling the definition of $q_2(\cdot|\w^1),$ 
 $D_2=D_2^P(\w^1) = \{-1,1\}$ for all $\w^1\in\Omega^1$ and $|\Delta S_2|=1$. Thus, $\Delta S_2 \in \mathcal{W}^2.$ 
 We now determine $\alpha_1(\cdot) \in (0,1]$ that satisfies \eqref{set_qt_na_H}. We need that 
$ q_{2}(\Delta S_{2}(\w^1,\cdot)<-\alpha_1(\w^1) | \w^1 ) \geq \alpha_1(\w^1),$ i.e. $1-l(\w^1)\geq \alpha_1(\w^1)$ and $ q_{2}(\Delta S_{2}(\w^1,\cdot)>\alpha_1(\w^1) | \w^1 )\geq \alpha_1(\w^1),$ i.e. $l(\w^1)\geq \alpha_1(\w^1).$ So, we choose $\alpha_1(\w^1) := \min(l(\w^1),1-l(\w^1))$. Thus, the 
$NA(\mathcal{Q}^2)$ condition holds true. This follows also from  $\mathcal{H}^2 = \mathcal{Q}^2 = \{P\}$ and  Lemma \ref{lemmaH_nonempty}. 
However, it is easy to see that $\mathbb{E}_P 1/\alpha_1 = +\infty$ as the law of $l$ under $q_1$ is uniform on $(0,1)$. \\
We claim now that the dynamic programming approach fails. Let $u_1 : \Omega^1 \times \mathbb{R}\times \mathbb{R} \to \mathbb{R}$  be defined by $u_1(\w^1,x,h) := \mathbb{E}_{q_2(\cdot|\w^1)} U(x+h\Delta S_2(\w^1,\cdot))= l(\w^1) U(x+h) +(1-l(\w^1))U(x-h).$ Recall that $\mathcal{U}_1(\w^1,x):=\sup_{h\in\mathbb{Q}} u_1(\w^1,x,h)$.  As $U$ is continuous and concave, $u_1(\w^1,\cdot,\cdot)$ is continuous and concave. Thus,  for all $\w^1\in \Omega^1, $  $\mathcal{U}_1(\w^1,\cdot)$ is concave and 
$\mathcal{U}_1(\w^1,x)= \sup_{h\in\mathbb{R}} u_1(\w^1,x,h).$ Simple computations show that $\mathcal{U}_1(\w^1,x)<\infty$ and as $\mathcal{U}_1(\w^1,\cdot)$ is concave, it is also continuous and  $\mathcal{U}_1={U}_1.$ 
Set $\phi_1^{*,0} := h^*$ for any $h^* \in \mathbb{R}$. As $\Delta S_1 = 0$, the candidate for optimal cash at time $1$ (starting with initial cash $0$) is then $V_1^{0,\phi^*}(\w^1):= h^* \Delta S_1(\w^1) = 0$ for all $\w^1\in\Omega^1$. Now, for all $(\w^1,h)\in \Omega^1\times \mathbb{R}$, we have that
\begin{equation*}
    u_1\left(\w^1,V_1^{0,\phi^{*,0}}(\w^1),h\right) = u_1(\w^1,0,h) =
    \begin{cases}
     ((1+h)^a - 1)l(\w^1) - a h (1-l(\w^1)) & \mbox{if}\;  h\geq 0 \\
     ((1-h)^a - 1)(1-l(\w^1)) + ah l(\w^1) & \mbox{if}\;  h < 0
    \end{cases}.
\end{equation*}
Moreover, $U_1(\w^1,V_1^{0,\phi^{*,0}}(\w^1))=\mathcal{U}_1(\w^1,0)= u_1(\w^1,0,H^*(\w^1))$ where
\begin{equation*}
    H^*(\w^1) := \left(\left(\frac{l(\w^1)}{1-l(\w^1)}\right)^{\frac{1}{1-a}} - 1\right) \;\; \mbox{if}\;\;  l(\w^1)\geq \frac12  \mbox{ and } H^*(\w^1) := \left(1-\left(\frac{1-l(\w^1)}{l(\w^1)}\right)^{\frac{1}{1-a}}\right) \;\;\mbox{if}\;\;  l(\w^1)< \frac12.
\end{equation*}
Let $\phi^{*,0}_2(\w^1):= H^*(\w^1)$ for all $\w^1\in\Omega^1$. Then, the candidate for optimal cash is for all $\w^2\in\Omega^2$, $$V_2^{0,\phi^{*,0}}(\w^2):= h^* \Delta S_1(\w^1) + H^*(\w^1)\Delta S_2(\w^2) = H^*(\w^1)\Delta S_2(\w^2).$$ 

\noindent We show now that the strategy $\phi^{*,0}$ is not admissible. Indeed,
\begin{multline*}
\mathbb{E}_P \; U^-\left(V_2^{0,\phi^{*,0}}\right) = a \mathbb{E}_P \left(V_2^{0,\phi^{*,0}}\right)^-\geq a\mathbb{E}_{P} ((H^*)^+ (\Delta S_2)^-)=  a \mathbb{E}_{q_1} \left((H^*)^+ (1-l(\cdot))\right) \\ 
= a \mathbb{E}_{q_1} \left(\left(\left(\frac{l(\cdot)}{1-l(\cdot)}\right)^{\frac{1}{1-\alpha}} - 1\right) (1-l(\cdot)) 1_{l(\cdot)\geq \frac12}\right) =\infty,
\end{multline*} 
where we have used Fubini's theorem for the second equality and  $0.5<a<1$ (and thus $\frac{1}{1-\alpha}\geq 2$) for the last one. 
It follows that $\mathbb{E}_P U^-(V_2^{0,\phi^{*,0}}) = +\infty,$ $\phi^{*,0}$ is not admissible and thus not optimal (see convention \eqref{cvt_inf}). 
One can also show that $\mathcal{U}_0(0) = \mathbb{E}_{q_1} U_1(\cdot,0) =+ \infty.$ As $U_0 (0)\geq \mathcal{U}_0(0),$  
${U}_0(0) =+ \infty$ and Assumption \ref{U0} is not satisfied. 


\end{remark}

\section{One-period case}
\label{op}
As mentioned after Theorem \ref{one_step_strategy}, we construct the optimal solutions of \eqref{RUMP} by gluing together one-step optimal strategies. This is the reason why we start with a one-period model. Let $(\overline{\Omega},\mathcal{G})$ be a measurable space,  $\mathfrak{P}(\overline{\Omega})$ be the set of all probability measures on $(\overline{\Omega},\mathcal{G})$ and $\mathcal{Q}$ be a nonempty convex subset of $\mathfrak{P}(\overline{\Omega})$. Let $Y(\cdot):=(Y_1(\cdot),...,Y_d(\cdot))$ be a $\mathcal{G}$-measurable $\mathbb{R}^d$-valued random variable, which could represent the price process change of value  during the period. Like previously, $D\subset \mathbb{R}^d$ is the support of the distribution of $Y(\cdot)$ under $\mathcal{Q}$ and $D_p\subset \mathbb{R}^d$ is the one of $Y(\cdot)$ under $p\in\mathcal{Q}$. We first assume the $\mbox{NA}(\mathcal{Q})$ condition, see \citep[Proposition 5.8]{ref4} and \citep[Lemma 2.7]{refBaZh}.

\begin{assumption}
There exists $p^*\in \mathcal{Q}$ such that $0\in \textup{ri}(\textup{conv}(D_{p^*}))$ and $\textup{Aff}(D)= \textup{Aff}(D_{p^*})$.
\label{P^*}
\end{assumption}
In the rest of this Section, we fix some $p^*$ as in Assumption \ref{P^*}. 
As $NA(p^*)$ holds true, the ``quantitative" no-arbitrage condition is satisfied and \citep[Proposition 3.3]{ref2} shows that there exists some $0<\alpha^*\leq 1$ such that for all $h\in \textup{Aff}(D_{p^*})=\textup{Aff}(D)$ (see Assumption \ref{P^*}), $h\neq 0$,  
\begin{eqnarray}
p^*(h Y<-\alpha^* |h|)\geq \alpha^*.
\label{qNA}
\end{eqnarray} 
\begin{assumption}
A random utility $V$ : $\overline{\Omega}\times \mathbb{R} \to \mathbb{R}\cup\{-\infty,+\infty\}$ is a function such that for all $x\in \mathbb{R}$, $V(\cdot,x)$ is $\mathcal{G}$-measurable and $V(\w,\cdot)$ is nondecreasing, concave and usc for all $\w\in\overline{\Omega}$.
\label{V}
\end{assumption}
Under Assumption \ref{V}, $V$ is $\mathcal{G}\otimes \mathcal{B}(\mathbb{R})$ measurable, see Remark \ref{joint_mes_U}.\\ 

We aim to solve the one-period multiple-priors utility maximization problem: 
\begin{eqnarray}
v(x) &:=& \sup_{h\in\mathbb{R}^d} \inf_{p\in \mathcal{Q}} \mathbb{E}_p V(\cdot, x+h Y(\cdot)).
\label{v}
\end{eqnarray}
This problem have already been solved in the PhD thesis of R. Blanchard under a different set of assumptions, see \citep{ref11}. We will comment as we go along on the difference between our assumptions and his assumptions. We start with the integrability conditions. 
\begin{assumption}
For all $x \in \mathbb{Q}$, $h\in \mathbb{Q}^d$, $\sup_{p\in\mathcal{Q}} \mathbb{E}_p V^-(\cdot,x+hY(\cdot))<+\infty$.
\label{integ_V-}
\end{assumption}

\begin{assumption}
$\mathbb{E}_{p^*} V^+(\cdot,1)<+\infty$.
\label{integ_V+}
\end{assumption}
Assumption \ref{integ_V+} is of course satisfied when $V$ is deterministic. We have seen in Remark \ref{remark_CE1} that assuming only that $v(x)<+\infty$ is not enough to get the existence of an optimal solution (see also \citep[Example 2.3]{ref9} in the case of a bounded from above utility function defined on the positive real axis). Assumption \ref{integ_V+} is similar to \citep[Assumption 3.5.6 (3.19)]{ref11} and \citep[Assumption 3.16]{ref3} and provides some upper bound for the value function. Nevertheless, these two assumptions were postulated for all $p\in\mathcal{Q}$ and not only for $p^*$. Using Assumption \ref{integ_V+}, we can not apply Fatou's lemma in a straightforward manner for any other prior than $p^*$ and regularity results are more difficult to obtain. Assumption \ref{integ_V-} acts as a counterbalance to Assumption \ref{integ_V+} and is stronger than \citep[Assumption 3.5.6 (3.20)]{ref11} or \citep[Assumption 3.13]{ref3} which are stated only for $h=0$. One may wonder if it is possible to postulate Assumption \ref{integ_V-} only for $p^*$. In this case, $\Psi : (x,h) \mapsto \inf_{p\in\mathcal{Q}} \mathbb{E}_p V(\cdot,x+hY(\cdot))$ may fail to be continuous (or even usc) which raises non trivial mathematical issues for finding an optimal strategy, see Remark \ref{remark_assumpV-}. A notable consequence of Assumption \ref{integ_V-} is that $\Psi$ will have full domain, see Proposition \ref{well_def_one}. We now give the assumption related to RAE in discrete time.
\begin{assumption}
There exist some constants $0<\underline{\gamma}\leq 1 \leq \overline{\gamma}$ such that $\underline{\gamma}\neq \overline{\gamma}$ and a $\mathcal{G}$-measurable random variable $C : \overline{\Omega}\to \mathbb{R}^+ \cup \{+\infty\}$  such that $c^*:=\mathbb{E}_{p^*}(C)<+\infty$ and such that for all $\w\in \overline{\Omega}$ satisfying $C(\w)<+\infty$, for all $\lambda \geq 1$, $x\in \mathbb{R}$,
\begin{eqnarray}
V(\w,\lambda x)&\leq& \lambda^{\underline{\gamma}}(V(\w,x)+C(\w))\quad \mbox{and} \quad V(\w,\lambda x)\leq \lambda^{\overline{\gamma}}(V(\w,x)+C(\w)). \label{V_pos} \label{V_minus}
\end{eqnarray}
\label{AE_one}
From now, we choose some $0<\eta<1$ such that $\underline{\gamma}<\eta \overline{\gamma}$.
\end{assumption}
The right-hand side of \eqref{V_minus} will give a control on the negative part of $V$. Indeed, let $\w\in\overline{\Omega}$ such that $C(\w)<+\infty$, $x\in\mathbb{R}$ and $\lambda\geq 1$. Then, as $C$ is non-negative, we easily see that
\begin{eqnarray*}
V^-(\w,\lambda x)+\lambda^{\overline{\gamma}} C(\w) \geq \max(-V(\w,\lambda x)+\lambda^{\overline{\gamma}} C(\w),0)&\geq  &\max(-\lambda^{\overline{\gamma}} V(\w,x),0) =\lambda^{\overline{\gamma}} V^-(\w,x). 
\end{eqnarray*}
A somewhat similar inequality for $V^+$ follows from the left-hand side of \eqref{V_pos} as $x\mapsto x^+$ is nondecreasing and sub-additive,
\begin{eqnarray}
V^+(\w,\lambda x)\leq \lambda^{\underline{\gamma}} (V^+(\w,x)+C(\w)). \label{ineq_pos_part}
\end{eqnarray}
The coefficient $\eta$ will play an important role to establish bounds for the value function and the optimal strategy: it is crucial that the control $\overline{\gamma}$ on $V^-$ is strictly larger than the control $\underline{\gamma}$ on $V^+$.
\begin{assumption}
There exists some $n_0^*\in \mathbb{N}\setminus \{0\}$ such that, 
\begin{eqnarray*}
p^*\left(V(\cdot,-n_0^*)\leq -\left(1+2\frac{c^*}{\alpha^*}\right)\right)\geq 1-\frac{\alpha^*}{2},
\end{eqnarray*}
where $\alpha^*$ is defined in (\ref{qNA}) and $c^*$ in Assumption \ref{AE_one}.
\label{pb_inequality}
\end{assumption}
Assumption \ref{pb_inequality}, already introduced in \citep[Assumption 8]{ref7}, is the one-period counterpart of Assumption \ref{nncst} and ensures that the value function $v$ (see \eqref{v}) can take arbitrary negative values.
Note that if $\lim_{x\to -\infty}V(\cdot,x)=-\infty$ $p^*-$almost surely, then Assumption \ref{pb_inequality} is verified. This assumption is similar to \citep[Assumption 3.5.8]{ref11} but which  was again postulated for all $p\in\mathcal{Q}$.
\begin{remark}
If Assumption \ref{pb_inequality} fails, then there may be no solution to the one-step utility maximization problem even in the uni-prior case. Fix $p_0\in\mathfrak{P}(\overline{\Omega})$ and set $\mathcal{Q}:=\{p_0\}$. We choose a bounded $\mathcal{G}$-measurable random variable $Y$ with a continuous cumulative distribution function under $p_0$ such that $\textup{NA}(p_0)$ holds true. So, Assumption \ref{P^*} is satisfied and recall $\alpha^*\in (0,1]$ from \eqref{qNA}. As $x \mapsto p_0(Y\leq x)$ is continuous, one can find some $M>0$ such that $0<p_0(Y> M)<1-\alpha^*/2.$ Let $\widetilde{V} : \mathbb{R} \to \mathbb{R}$ be some continuously differentiable, nondecreasing, non-constant and concave function satisfying $\lim_{x\to +\infty} \widetilde{V}(x)=+\infty$ and such that $\mbox{AE}_{+\infty}(\widetilde{V})< 1$ or $\mbox{AE}_{-\infty}(\widetilde{V})> 1$. Set for all $\w\in\overline{\Omega}$ and $x\in\mathbb{R}$, $V(\w,x):=\widetilde{V}(x)1_{\{Y>M\}}(\w)$. Then, Assumption \ref{V} holds true. 
Applying Lemma \ref{lemmaAE1} and Proposition \ref{annex_AE2} in the appendix to $\widetilde{V}$, we get that \eqref{AE_gamma1} and \eqref{det_AE} hold true for some $\gamma\neq 1$. So, Assumption \ref{AE_one} holds true with $\overline{\gamma}:=\max(\gamma,1)$ and $\underline{\gamma}:=\min(\gamma,1)$. Moreover, Assumptions \ref{integ_V-} and \ref{integ_V+} hold true as $Y$ is bounded and $\mathcal{Q}=\{p_0\}$. However, Assumption \ref{pb_inequality} fails as for any $n$ and $\mu>0$, $p_0(V(\cdot,-n)\leq -\mu)=1_{\{\widetilde{V}(-n)\leq -\mu\}}p_0(Y> M)<1-\alpha^*/2$. Moreover, we have for all $\w\in\overline{\Omega}$, $x\in\mathbb{R}$ and $h>0$ that $V(\w,x+hY) \geq \widetilde{V}(x+hM)1_{\{Y>M\}}(\w)$, so that, $\mathbb{E}_{p_0} V(\cdot,x+hY) \geq \widetilde{V}(x+hM) p_0(Y> M)$. Recalling that $p_0(Y> M)>0$, $\lim_{x\to +\infty} \widetilde{V}(x)=+\infty$ and taking the limit when $h\to +\infty$ show that there is no solution to the utility maximization problem. 
Note that Assumption \ref{nncst} fails also. Indeed, for every random variables $\underline{X}$ and $C(\cdot)\geq 0$
$$p_0(V(\cdot,\underline{X}(\cdot))<-C(\cdot))=p_0(\{\widetilde{V}(\underline{X}(\cdot))<-C(\cdot)\} \cap \{Y(\cdot)> M\}) \leq p_0( Y(\cdot)> M)<1-\alpha^*/2<1. $$
\label{rm_mino}
\end{remark}


Let $p\in\mathcal{Q}$. We introduce the functions $\Psi_p$ and $\Psi$ $:$ $\mathbb{R}\times \mathbb{R}^d \to \mathbb{R}\cup\{-\infty,+\infty\}$ which are defined for all $x\in \mathbb{R}$ and  $h\in \mathbb{R}^d$ by
\begin{eqnarray}
\Psi_p(x,h):=\mathbb{E}_p V(\cdot,x+h Y(\cdot)) \quad \mbox{and} \quad \Psi(x,h):=\inf_{p\in \mathcal{Q}} \Psi_p(x,h).\label{Phi}
\end{eqnarray}
Then, $v(x)=\sup_{h\in\mathbb{R}^d}\Psi(x,h)$, see \eqref{v}. Recall that the expectations in \eqref{Phi} are defined in the generalized sense using the convention $+\infty-\infty=-\infty+\infty=-\infty$ and that $\Psi_p$ and $\Psi$ are well-defined although being potentially infinite. The function $\Psi_p$ is introduced as a tool to prove properties on $\Psi$ and $v$ but also to pass Assumption \ref{U0} through the dynamic programming procedure. The results on $\Psi_p$ can be found in \citep{ref2} or \citep{ref11} but under a different set of assumptions.

\begin{proposition}
Fix $p\in\mathcal{Q}$. If Assumption \ref{V} holds true, then $\Psi_p$ and $\Psi$ are concave. 
If Assumption \ref{integ_V-} holds true, then for all $(x,h)\in\mathbb{R} \times \mathbb{R}^d$, 
$\sup_{p\in\mathcal{Q}} \mathbb{E}_p V^-(\cdot,x+h Y(\cdot)) <+\infty$ and $\Psi_p\geq \Psi > -\infty$. 
Assume that Assumptions \ref{P^*}, \ref{V},  \ref{integ_V-}, \ref{integ_V+} and \ref{AE_one} hold true.  Then, $\Psi$ is finite and continuous and for all $(x,h)\in\mathbb{R} \times \mathbb{R}^d$,
\begin{eqnarray}
\mathbb{E}_{p^*} V^+(\cdot,x+h Y(\cdot)) \leq (|h|\vee x^+ \vee 1)^{\underline{\gamma}} (l^*+c^*),
\label{inequ_imp_positive}
\end{eqnarray}
where $c^*= \mathbb{E}_{p^*} C<+\infty$, $l^*:=\sum_{\theta\in\{-1,1\}^d} \mathbb{E}_{p^*}V^+(\cdot,1 + \theta Y(\cdot))<+\infty$ and $a\vee b= \max(a,b)$.
\label{well_def_one}
\end{proposition}
\proof{Proof.}
Fix some $p\in\mathcal{Q}$. First, assume that  Assumption \ref{V} holds true. Recall that a (extended) function $f : \mathbb{R} \to \mathbb{R}\cup\{-\infty,+\infty\}$ is said to be concave if for all $0<t<1$ and $x,y\in\mathbb{R}$,
$$f(tx+(1-t)y)\geq tf(x)+(1-t)f(y),$$ using convention \eqref{cvt_inf}. Let $(x_1,h_1)$ and $(x_2,h_2)$ in $\mathbb{R}\times \mathbb{R}^d$ and $0<t<1$. Using the concavity of $V$ and in \eqref{eq_ccv_psi}, \citep[Lemma 7.11, (a), p140]{ref1} adapted to convention \eqref{cvt_inf}, we get that
\begin{eqnarray}
\Psi_p\left(t(x_1,h_1)+(1-t)(x_2,h_2)\right)&=&\mathbb{E}_p V\left(\cdot,t(x_1+h_1 Y(\cdot))+(1-t)(x_2+h_2 Y(\cdot))\right)\nonumber\\
&\geq & \mathbb{E}_p\; \left(t\; V(\cdot,x_1+h_1 Y(\cdot))+(1-t)V(\cdot,x_2+h_2 Y(\cdot))\right)\nonumber\\
&\geq & t\; \mathbb{E}_p V(\cdot,x_1+h_1 Y(\cdot)) + (1-t) \mathbb{E}_p V(\cdot,x_2+h_2 Y(\cdot))\label{eq_ccv_psi}\\
&=& t\; \Psi_p(x_1,h_1) + (1-t)\Psi_p(x_2,h_2)\nonumber,
\end{eqnarray}
which shows the concavity of $\Psi_p$. 
Then, as $\Psi$ is the pointwise infimum of concave functions, $\Psi$ is also concave, see \citep[Proposition 2.9]{ref5}.

We now show that Assumption \ref{integ_V-} can be extended to all $(x,h)\in \mathbb{R}\times \mathbb{R}^d$. Let $(x,h)\in \mathbb{R}\times \mathbb{R}^d$. For $i\in\{1,\cdot\cdot\cdot,d\}$, let $\widehat{\theta}_i(\cdot):= \mbox{sgn}(Y_i(\cdot))$ where for all $y\in\mathbb{R}$, $\mbox{sgn}(y):=1$ if $y\geq 0$ and $\mbox{sgn}(y):=-1$ otherwise. Then, $\widehat{\theta}$ is a $\{-1,1\}^d$-valued process and 
\begin{eqnarray}
|Y(\cdot)|\leq |Y(\cdot)|_1=\sum_{i=1}^d \textup{sgn}(Y_i(\cdot))Y_i(\cdot)= \widehat{\theta}(\cdot)Y(\cdot).
\label{Y_ineq}
\end{eqnarray}
Let $\w\in \overline{\Omega}$, as $V^-(\w,\cdot)$ is nonincreasing, using Cauchy-Schwarz inequality and (\ref{Y_ineq}), we get that
\begin{eqnarray}
V^-(\w,x+hY(\w))\leq V^-(\w,-\lceil x^- \rceil - \lceil |h| \rceil |Y(\w)|)&\leq&  V^-(\w,-\lceil x^- \rceil - \lceil |h| \rceil\widehat{\theta}(\w)Y(\w))\nonumber\\
&\leq & \sum_{\theta\in \{-1,1\}^d}V^-(\w,-\lceil x^- \rceil - \lceil |h| \rceil\theta Y(\w)),\label{temp_Q_R}
\end{eqnarray}
where $ \lceil x \rceil$ is the smallest natural number greater than $x$. Then, Assumption \ref{integ_V-} shows that
\begin{eqnarray}
\sup_{p\in\mathcal{Q}} \mathbb{E}_p V^-(\cdot,x+hY(\cdot)) \leq \sum_{\theta\in \{-1,1\}^d} \sup_{p\in\mathcal{Q}} \mathbb{E}_p V^-(\cdot,-\lceil x^- \rceil - \lceil |h| \rceil \theta Y(\cdot))<+\infty.
\label{supV-_inequ}
\label{ineq_sup_one_minus}
\end{eqnarray}
This shows that Assumption \ref{integ_V-} holds true for all $(x,h)\in\mathbb{R}\times\mathbb{R}^d$. Moreover,
\begin{eqnarray}
\Psi(x,h)&\geq &\inf_{p\in\mathcal{Q}} \left(- \mathbb{E}_p V^-(\cdot,x+h Y(\cdot))\right)= -\sup_{p\in\mathcal{Q}} \mathbb{E}_p V^-(\cdot,x+h Y(\cdot))>-\infty
\label{temp_domain_psi}
\end{eqnarray}
and $\Psi_p\geq \Psi > -\infty.$ \\
Assume now that Assumptions \ref{P^*}, \ref{V},  \ref{integ_V-}, \ref{integ_V+} and \ref{AE_one} hold true. 
Let $\w\in \overline{\Omega}$ such that $C(\w)<+\infty$ and $(x,h)\in \mathbb{R}\times \mathbb{R}^d$. As $V^+(\w,\cdot)$ is nondecreasing, using Cauchy-Schwarz inequality, (\ref{ineq_pos_part}) and (\ref{Y_ineq}), we obtain that
\begin{eqnarray}
V^+(\w,x+h Y(\w))&\leq&  V^+(\w,(|h|\vee x^+ \vee 1) (1 + |Y(\w)|))\leq  (|h|\vee x^+ \vee 1)^{\underline{\gamma}} (V^+(\w,1 + |Y(\w)|)+C(\w))\nonumber\\
& \leq & (|h|\vee x^+ \vee 1)^{\underline{\gamma}} \left(\sum_{\theta\in\{-1,1\}^d}V^+(\w,1 + \theta Y(\w))+C(\w)\right).\label{temp_well_def_psi}
\end{eqnarray}
As $c^*<+\infty$ by Assumption \ref{AE_one}, $p^*(C<+\infty)=1$ and taking the expectation under $p^*$ in \eqref{temp_well_def_psi} shows \eqref{inequ_imp_positive}. If $\Psi_{p^*}(x_0,h_0)=+\infty$ for some $(x_0,h_0)\in\mathbb{R}\times \mathbb{R}^d$, as $\Psi_{p^*}$ is concave and $\Psi_{p^*}>-\infty$, \eqref{eq_ccv_psi} will imply that $\Psi_{p^*}(x,h)=+\infty$ for all $(x,h)\in\mathbb{R}\times \mathbb{R}^d$ which contradicts $\Psi_{p^*}(1,0)<+\infty$, see Assumption \ref{integ_V+}. So, for all $\theta\in \{-1,1\}^d$, $\Psi_{p^*}(1,\theta)<+\infty$ and again, as $\Psi_{p^*}>-\infty$, we must have that $\mathbb{E}_{p^*} V^+(\cdot,1+\theta Y(\cdot))<+\infty$. It follows that $l^*<+\infty$ and $\Psi_{p^*}$ and $\Psi$ are finite. As $\Psi$ and $\Psi_{p^*}$ are concave and finite, they are continuous (see for example \citep[Proposition 2.35]{ref5}). 
\Halmos \endproof

\begin{remark}
The regularity of $\Psi$ is the core argument to solve the one-period optimization problem and Assumptions \ref{integ_V-}, \ref{integ_V+} and the concavity of $V$ are crucial to get the continuity of $\Psi$. Otherwise, $\Psi$ would only be continuous on the interior of its domain\footnote{Let $f : \mathbb{R}^n \to \mathbb{R}\cup\{-\infty,+\infty\}$, $\mbox{Dom}(f):= \{x\in\mathbb{R}^n, f(x)>-\infty\}$.} (see \citep[Theorem 2.35, p59]{ref5}) which may be empty. 
It is also important to note that convention \eqref{cvt_inf} is crucial to show the concavity of $\Psi$ (and $\Psi_p$), see \eqref{eq_ccv_psi}.
\label{remark_assumpV-}
\end{remark}

The next proposition provides a polynomial control on $\Psi$ as well as a bound on the optimal strategies. 
It is a crucial step in order to solve the maximization problem \eqref{v}. This proposition provides similar results as \citep[Lemma 3.5.12]{ref11} in the multiple-priors case and \citep[Lemma 3]{ref7} in the uni-prior case. However, recall that \citep[Assumptions 3.5.6 and 3.5.8]{ref11}) are stated for all $p\in\mathcal{Q}$.
\begin{proposition}
Suppose that Assumptions \ref{P^*}, \ref{V}, \ref{integ_V-}, \ref{integ_V+}, \ref{AE_one}  and \ref{pb_inequality} hold true.
For $x\in\mathbb{R}$, let
\begin{eqnarray*}
K_0(x)&:=& \max\left(1,x^+,\frac{x^++n_0^*}{\alpha^*},\left(\frac{x^++n_0^*}{\alpha^*}\right)^{\frac{1}{1-\eta}}\right),\\
K_1(x)&:=&\max\Bigg(K_0(x),\left(\frac{6 l^*}{\alpha^*}\right)^{\frac{1}{\eta \overline{\gamma}-\underline{\gamma}}},\left(\frac{6c^*}{\alpha^*}\right)^{\frac{1}{\eta \overline{\gamma}-\underline{\gamma}}},\left(\frac{6}{\alpha^*}\sup_{p\in\mathcal{Q}}\mathbb{E}_p V^-(\cdot,-x^-)\right)^{\frac{1}{\eta\overline{\gamma}}}\Bigg),
\end{eqnarray*} 
where $\alpha^*$ is defined in \eqref{qNA}, $c^*$, $\eta$, $\overline{\gamma}$ and $\underline{\gamma}$ in Assumption \ref{AE_one}, $l^*$ in Proposition \ref{well_def_one} and $n_0^*$ in Assumption \ref{pb_inequality}.
Then, $K_0(x)<+\infty$, $K_1(x)<+\infty$ for all $x\in\mathbb{R}$. Let $x\in\mathbb{R}$ and $h\in \textup{Aff}(D)$. We get 
\begin{eqnarray}
|h|\geq K_0(x) &\implies& \Psi(x,h)\leq |h|^{\underline{\gamma}} (l^*+c^*)-|h|^{\eta\overline{\gamma}}\frac{\alpha^*}{2}\label{coercive_ineq}\\
|h|\geq K_1(x) &\implies &
\Psi(x,h) \leq \Psi(x,0).
\label{sub_opt_ineq}
\end{eqnarray}
Moreover, for all $x\in\mathbb{R}$,
\begin{eqnarray}
v(x)&=&\sup_{h\in\mathbb{R}^d}\Psi(x,h)=\sup_{h\in  \textup{Aff}(D)}\Psi(x,h) = \sup_{\substack{|h|\leq K_1(x) \\ h\in \textup{Aff}(D)}}\inf_{p\in\mathcal{Q}} \mathbb{E}_p V(\cdot,x+h Y).\label{eq_value_valboun}
\end{eqnarray}
\label{sub_optimal}
\label{inequ_borne}
\end{proposition}
\proof{Proof.}
Let $(x,h)\in \mathbb{R}\times\mathbb{R}^d$.
Using (\ref{inequ_imp_positive}), if $|h|\geq \max(1,x^+)$, we get that 
\begin{eqnarray}
\mathbb{E}_{p^*} V^+(\cdot,x+h Y(\cdot)) \leq |h|^{\underline{\gamma}} (l^*+c^*).
\label{ineq_psi+}
\end{eqnarray}
We introduce the set 
\begin{eqnarray}
B_{n_0,h}^*:=\left\{V(\cdot,-n_0^*)\leq -\left(1+\frac{2 c^*}{\alpha^*}\right),\; h Y(\cdot)< -\alpha^* |h|\right\}. \label{setB}
\end{eqnarray}
Assume that $x^+ - \alpha^*|h|\leq -n_0^*$. Let $\w\in B_{n_0,h}^*$. As $V(\w,\cdot)$ is nondecreasing, we have that
$$V(\w,x+h Y(\w))\leq V(\w,x^+ - \alpha^* |h|)\leq V(\w,-n_0^*)\leq 0.$$
So, $B_{n_0,h}^*\subset \{V(\cdot,x+h Y(\cdot))\leq 0\}.$ Assume now that $|h|\geq \max\left(1,x^+,\frac{x^++n_0^*}{\alpha^*}\right)$. We get that
\begin{eqnarray*}
V^-(\cdot,x+h Y(\cdot))&=&-V(\cdot,x+h Y(\cdot))1_{\{V(\cdot,x+h Y(\cdot))\leq 0\}}(\cdot)\geq  -V(\cdot,x+h Y(\cdot))1_{B_{n_0,h}^*}(\cdot).
\end{eqnarray*}
Let $\w\in\overline{\Omega}$ such that $C(\w)<+\infty$. Using (\ref{V_minus}) with $\lambda=|h|^\eta \geq 1$ as $|h|\geq 1$, we obtain that
\begin{eqnarray*}
V(\w,x+h Y(\w))&\leq & |h|^{\eta\overline{\gamma}}V\left(\w\;,\frac{x+hY(\w)}{|h|^{\eta}} \right)+ |h|^{\eta\overline{\gamma}} C(\w).
\end{eqnarray*}
Thus, as $C\geq 0$ and $V$ is nondecreasing, we deduce that
\begin{eqnarray*}
V^-(\w,x+h Y(\w))&\geq & -V(\w,x+h Y(\w)) 1_{B_{n_0,h}^*}(\w)\geq  -|h|^{\eta\overline{\gamma}} V\left(\w\;,\frac{x+hY(\w)}{|h|^{\eta}} \right)1_{B_{n_0,h}^*}(\w)-|h|^{\eta\overline{\gamma}} C(\w)\\
&\geq& -|h|^{\eta\overline{\gamma}}V\left(\w\;,x^+ -\alpha^* |h|^{1-\eta}\right) 1_{B_{n_0,h}^*}(\w)-|h|^{\eta\overline{\gamma}} C(\w).
\end{eqnarray*}
Suppose furthermore that $x^+ -\alpha^* |h|^{1-\eta}\leq -n_0^*$. Then,
\begin{eqnarray}
V^-(\w,x+h Y(\w))&\geq & |h|^{\eta\overline{\gamma}}\left(1+\frac{2 c^*}{\alpha^*}\right)1_{B_{n_0,h}^*}(\w)-|h|^{\eta\overline{\gamma}} C(\w).\label{temp_ineq_bound_1}
\end{eqnarray}
Using Assumption \ref{pb_inequality} and (\ref{qNA}), which holds true under Assumption \ref{P^*} if $h\in \textup{Aff}(D)$ and $h\neq 0$,
\begin{eqnarray}
p^*(B_{n_0,h}^*)&\geq & p^*\left(V(\cdot,-n_0^*)\leq -\left(1+\frac{2 c^*}{\alpha^*}\right)\right)  +p^*(h Y(\cdot)< -\alpha^* |h|) -1 \geq  1-\frac{\alpha^*}{2}+\alpha^* -1 =\frac{\alpha^*}{2}.
\label{p_B}
\end{eqnarray}
So, as $p^*(C<+\infty)=1$ since $c^*<+\infty$, \eqref{temp_ineq_bound_1} shows that if $|h| \geq K_0(x)$ (recall that $0<\eta<1$, see Assumption \ref{AE_one})
\begin{eqnarray}
\mathbb{E}_{p^*}V^-(\cdot,x+h Y(\cdot)) \geq |h|^{\eta\overline{\gamma}} \frac{\alpha^*}{2}.
\label{ineq_psi-}
\end{eqnarray}
Finally, as $\Psi(x,h)\leq \Psi_{p^*}(x,h)$, (\ref{ineq_psi+}) and (\ref{ineq_psi-}) show \eqref{coercive_ineq}. Then, as $0<\eta<1$, $\alpha^*>0$ (see \eqref{qNA}) and $n_0^*\in\mathbb{N}\setminus \{0\}$ (recall Assumption \ref{pb_inequality}), we get that $K_0(x)<+\infty$.\\ 
Now, using (\ref{ineq_sup_one_minus}), we find that $\sup_{p\in\mathcal{Q}}\mathbb{E}_p V^-(\cdot,-x^-)<+\infty$. As $l^*<+\infty$ (see Proposition \ref{well_def_one}), $\eta\overline{\gamma}-\underline{\gamma}>0$ and $c^*<+\infty$ (see Assumption \ref{AE_one}), we obtain that $K_1(x)<+\infty$.\\ 
Assume now that $|h|\geq K_1(x)$. As $\underline{\gamma}<\eta \overline{\gamma}$, we get that $|h|\geq K_0(x)$,  $|h|^{\underline{\gamma}} l^* \leq (|h|^{\eta\overline{\gamma}}\alpha^*)/6$, $|h|^{\underline{\gamma}} c^* \leq (|h|^{\eta\overline{\gamma}}\alpha^*)/6$ and $\sup_{p\in\mathcal{Q}}\mathbb{E}_p V^-(\cdot,-x^-) \leq (|h|^{\eta\overline{\gamma}}\alpha^*)/6.$ 
So, using successively \eqref{coercive_ineq}, that $V$ is nondecreasing and \eqref{temp_domain_psi}, we obtain that
$$\Psi(x,h)\leq -|h|^{\eta\overline{\gamma}}\frac{\alpha^*}{6} \leq -\sup_{p\in\mathcal{Q}}\mathbb{E}_p V^-(\cdot,-x^-)  \leq \Psi(x,0)$$
and \eqref{sub_opt_ineq} is proved.\\ Assumption \ref{P^*} shows that $0\in \textup{ri}(\textup{conv}(D_{p^*}))\subset\textup{Aff}(D_{p^*})=\textup{Aff}(D)$ and $\textup{Aff}(D)$ is a vector space. For $h\in\mathbb{R}^d$, let $h^\perp$ be the orthogonal projection of $h$ on $\textup{Aff}(D)$. Using \citep[Remark 3.10]{ref3}, we get that for all $x\in\mathbb{R}$ and $h\in\mathbb{R}^d$, $hY=h^{\perp} Y$ $\mathcal{Q}-\mbox{q.s.}$ and $\Psi(x,h)=\Psi(x,h^{\perp}).$ So, we have that $\sup_{h\in\mathbb{R}^d}\Psi(x,h)=\sup_{h\in\textup{Aff}(D)}\Psi(x,h)$. Thus, \eqref{eq_value_valboun} follows from \eqref{sub_opt_ineq} as $0\in\textup{Aff}(D)$.
\Halmos \endproof

We now define the auxiliary function $v_\mathbb{Q}$ as follows
\begin{eqnarray}
v_\mathbb{Q}(x) &:=& \sup_{h\in\mathbb{Q}^d} \inf_{p\in \mathcal{Q}} \mathbb{E}_p V(\cdot, x+h Y(\cdot)). \label{vq}
\end{eqnarray}
The function $v_\mathbb{Q}$ will be useful for solving measurability issues arising in the multiple-period setting. We will show in Proposition \ref{existence_uni} that $v_{\mathbb{Q}}$ equals $v$ under all the previous assumptions. First, we prove some fundamental properties of $v$, $v_\mathbb{Q}$ and $\textup{Cl}(v_\mathbb{Q})$ when only Assumption  \ref{V} is postulated. 
\begin{lemma}
\label{minimal_V_pp}
Assume that Assumption \ref{V} holds true. Then, $\textup{Cl}(v_{\mathbb{Q}})(x)= \lim_{\delta\to 0, \delta>0} v_\mathbb{Q}(x+\delta)$ and $v$, $v_\mathbb{Q}$ and $\textup{Cl}(v_{\mathbb{Q}})$ are nondecreasing and concave.
\end{lemma}
\proof{Proof.}
First, $v$, $v_{\mathbb{Q}}$ and $\textup{Cl}(v_\mathbb{Q})$ are clearly nondecreasing. Using for example \citep[p14]{ref5}, we find that for all $x\in\mathbb{R}$,
\begin{eqnarray*}
\textup{Cl}(v_\mathbb{Q})(x)&=&\lim_{\delta\to 0,\delta>0} \sup_{x-\delta<y<x+\delta}v_{\mathbb{Q}} (y) 
=\lim_{\delta\to 0, \delta>0} v_\mathbb{Q}(x+\delta).
\end{eqnarray*}
The concavity of $v$ and $v_\mathbb{Q}$ relies on a midpoint concavity argument and on the Ostrowski theorem, see \citep[p12]{ref6}. The proof is very similar to \citep[Proposition 4.2]{ref2} (see also \citep[Proposition 3.5]{ref9}) and is thus omitted. The concavity of $\textup{Cl}(v_\mathbb{Q})$ follows then from \citep[Proposition 2.32, p57]{ref5}.
\Halmos \endproof

We now show that under our assumptions an optimal strategy exists for \eqref{v}.
\begin{proposition}
Assume that Assumptions \ref{P^*}, \ref{V}, \ref{AE_one}, \ref{integ_V-}, \ref{integ_V+} and \ref{pb_inequality} hold true. Then, $v$ is finite and for all $x\in\mathbb{R}$, there exists an optimal strategy $\widehat{h}_x \in\textup{Aff}(D)$ such that 
\begin{eqnarray}
v(x)=\inf_{p\in \mathcal{Q}}\mathbb{E}_p V(\cdot,x+\widehat{h}_x Y(\cdot)).
\label{maximizer}
\end{eqnarray}
Moreover, for all $x\in \mathbb{R}$, 
\begin{eqnarray}
v(x)=&v_\mathbb{Q}(x)&=\textup{Cl}(v_\mathbb{Q})(x).\label{v_v_q}
\end{eqnarray}  
\label{existence_uni}
\end{proposition}
\proof{Proof.}
Recalling that $\Psi$ is continuous from Proposition \ref{well_def_one}, $$v(x)= \sup_{h\in\mathbb{R}^d} \Psi(x,h) = \sup_{h\in\mathbb{Q}^d }\Psi(x,h)= v_{\mathbb{Q}}(x)$$ and the first equality in (\ref{v_v_q}) holds true. Let $x\in\mathbb{R}$. Noting that $\Psi(x,\cdot)$ is continuous and finite according to Proposition \ref{well_def_one} again, we have that $\Psi(x,\cdot)$ admits a maximizer $\widehat{h}_x$ on any compact set and in particular on the intersection of $\textup{Aff}(D)$ and of the closed ball centered on $0$ and of radius $K_1(x)$. So, \eqref{maximizer} follows from \eqref{eq_value_valboun} and $v(x)= \Psi(x,\widehat{h}_x)$. So, $v$ is finite as $\Psi$ is finite. As $v_\mathbb{Q}$  is concave and $v_\mathbb{Q}=v$ is finite, $v_\mathbb{Q}$ is continuous on $\mathbb{R}$ and the second equality in (\ref{v_v_q}) holds true. \Halmos \endproof

Finally, we show that $v$ can take arbitrary negative values. This result will be essential to prove that Assumption \ref{pb_inequality} is preserved by dynamic programming.
\begin{proposition}
\label{as_little_v}
Suppose that Assumptions \ref{P^*}, \ref{V}, \ref{integ_V-}, \ref{integ_V+}, \ref{AE_one}  and \ref{pb_inequality} hold true and
let 
\begin{eqnarray}
\bar{K}:=\max\left(1,\frac{n_0^*}{\alpha^*},\left(\frac{n_0^*}{\alpha^*}\right)^{\frac{1}{1-\eta}},\left(\frac{8c^*}{\alpha^*}\right)^{\frac{1}{\eta \overline{\gamma}-\underline{\gamma}}},\left(\frac{8l^*}{\alpha^*}\right)^{\frac{1}{\eta \overline{\gamma}-\underline{\gamma}}}\right). 
\end{eqnarray}
Then, $\bar{K}<+\infty$. For all $m\geq 1$, let $n_m:=\lceil N_m  \rceil$ where 
\begin{eqnarray}
N_m:= n_0^* \left(\frac{4}{\alpha^*}\left(m+(\bar{K}^{\underline{\gamma}}+1)(l^*+c^*)\right)\right)^{\frac{1}{{\underline{\gamma}}}}. \label{N_exp}
\end{eqnarray}
Then, $n_m<+\infty$ and 
\begin{eqnarray}
v(-n_m) \leq -m.
\label{v_ineq_m}
\end{eqnarray}
Recall that $\alpha^*$ is defined in \eqref{qNA}, $c^*$, $\eta$, $\overline{\gamma}$ and $\underline{\gamma}$ in Assumption \ref{AE_one}, $l^*$ in Proposition \ref{well_def_one} and $n_0^*$ in Assumption \ref{pb_inequality}.
\end{proposition}
\proof{Proof.}
Recall that $0< \alpha^* \leq 1$ from \eqref{qNA}, $n_0^*\in \mathbb{N}\setminus \{0\}$ from Assumption \ref{pb_inequality} and that $0<\eta<1$, $\eta\overline{\gamma}-\underline{\gamma}>0$ and $c^*<+\infty$ from Assumption \ref{AE_one}. Moreover, $l^*<+\infty$ by Proposition \ref{well_def_one}. So, $\bar{K}<+\infty$, $N_m<+\infty$ and $n_m<+\infty$ for any $m\geq 1$.\\ We are looking for an upper bound for $\Psi_{p^*}(x,h)$ when $x\leq -n_0^*$ and $h\in\textup{Aff}(D)$. We first provide a lower bound for $\mathbb{E}_{p^*} V^-(\cdot,x+h Y(\cdot))$. Let $x\leq -n_0^*$ and $h\in\textup{Aff}(D)$, $h\neq 0$. Using (\ref{V_pos}) and $C\geq 0$, we get as in Proposition \ref{inequ_borne} that for all $\w\in\overline{\Omega}$ such that $C(\w)<+\infty$,
\begin{eqnarray*}
-V^-(\w,x+h Y(\w)) &=& V(\w,x+h Y(\w))1_{\{V(\w,x+h Y(\w))\leq 0\}}\leq V(\w,x+h Y(\w))1_{B_{n_0,h}^*}(\w)\\&\leq& V(\w,x-\alpha^* |h|)1_{B_{n_0,h}^*}(\w)\leq V(\w,x)1_{B_{n_0,h}^*}(\w)\\
&\leq& \left(\frac{x}{-n_0^*}\right)^{\underline{\gamma}} \left(V(\w,-n_0^*)+C(\w)\right)1_{B_{n_0,h}^*}(\w)\\
&\leq& -\left(\frac{x}{-n_0^*}\right)^{\underline{\gamma}} \left(1+\frac{2c^*}{\alpha^*}\right)1_{B_{n_0,h}^*}(\w)+ \left(\frac{x}{-n_0^*}\right)^{\underline{\gamma}} C(\w),
\end{eqnarray*} 
where $B_{n_0,h}^*$ is defined in \eqref{setB}.
Then, using (\ref{p_B}) and recalling that $p^*(C<+\infty)=1$, we get that
\begin{eqnarray}
\mathbb{E}_{p^*} V^-(\cdot,x+h Y(\cdot))\geq \left(\frac{x}{-n_0^*}\right)^{\underline{\gamma}} \frac{\alpha^*}{2}.
\label{V-*N}
\end{eqnarray}
Now, we turn to the majoration of $\mathbb{E}_{p^*}V^+(\cdot,x+h Y(\cdot))$. Fix $x\leq 0$ and $h\in \textup{Aff}(D)$. We distinguish two cases. Assume first that $|h|< \bar{K}$. Using 
\eqref{inequ_imp_positive} in Proposition \ref{well_def_one} (recall that $\bar{K}\geq 1$ and $x^+=0$),
\begin{eqnarray}
\mathbb{E}_{p^*}V^+(\cdot,x+h Y(\cdot))\leq (|h| \vee 1)^{\underline{\gamma}} (l^*+c^*)\leq 
\bar{K}^{\underline{\gamma}}(l^*+c^*).
\label{lemme_V+*}
\end{eqnarray}
Thus, using (\ref{V-*N}) and (\ref{lemme_V+*}), we obtain for $x\leq -n_0^*$, $h\in \textup{Aff}(D)$, $h\neq 0$ and $|h|<\bar{K}$ that
\begin{eqnarray}
\Psi_{p^*}(x,h)=\mathbb{E}_{p^*}V(\cdot,x+h Y(\cdot))\leq \bar{K}^{\underline{\gamma}}(l^*+c^*)-\left(\frac{x}{-n_0^*}\right)^{\underline{\gamma}} \frac{\alpha^*}{2}.
\label{E*Vleq2}
\end{eqnarray}
We now study the case of $|h|>\bar{K}$.
First, we show that if $h\in \textup{Aff}(D)$, $|h|\geq \bar{K}$ and $x\leq 0$, 
\begin{eqnarray}
\mathbb{E}_{p^*}V(\cdot,x+h Y(\cdot))&=& \mathbb{E}_{p^*}V^+(\cdot,x+h Y(\cdot))-\mathbb{E}_{p^*}V^-(\cdot,x+h Y(\cdot))\leq  -\frac{1}{2} \mathbb{E}_{p^*} V^-(\cdot,x+hY(\cdot)).
\label{lemme_E*Vleq}
\end{eqnarray}
Indeed, using \eqref{inequ_imp_positive} and \eqref{ineq_psi-}, as $\bar{K}\geq K_0(x)$ (recall that $x^+=0$ and $|h|\geq \bar{K} \geq 1$),
\begin{eqnarray*}
\mathbb{E}_{p^*}V^+(\cdot,x+h Y(\cdot)) -\frac{1}{2} \mathbb{E}_{p^*} V^-(\cdot,x+hY(\cdot))\leq |h|^{\underline{\gamma}} (l^*+c^*)-|h|^{\eta\overline{\gamma}}\frac{\alpha^*}{4}\leq 0,
\end{eqnarray*}
as $|h|^{\underline{\gamma}}l^* \leq (|h|^{\eta \overline{\gamma}}\alpha^*)/8$ and $|h|^{\underline{\gamma}} c^* \leq (|h|^{\eta \overline{\gamma}}\alpha^*)/8$.
So, \eqref{lemme_E*Vleq} is proved. Now, combining (\ref{V-*N}) and (\ref{lemme_E*Vleq}), we get that when  $x\leq -n_0^*$, $h\in\textup{Aff}(D)$ and $|h|\geq \bar{K}$,
\begin{eqnarray}
\Psi_{p^*}(x,h)=\mathbb{E}_{p^*} V(\cdot,x+h Y(\cdot))\leq -\left(\frac{x}{-n_0^*}\right)^{\underline{\gamma}} \frac{\alpha^*}{4}.
\label{E*Vleq1}
\end{eqnarray}
Comparing (\ref{E*Vleq2}) and (\ref{E*Vleq1}), we finally find that for all $x\leq -n_0^*$ and $h\in\textup{Aff}(D)$, $h\neq 0$
\begin{eqnarray*}
\Psi(x,h)\leq \Psi_{p^*}(x,h)\leq \bar{K}^{\underline{\gamma}}(l^*+c^*)-\left(\frac{x}{-n_0^*}\right)^{\underline{\gamma}} \frac{\alpha^*}{4}.
\end{eqnarray*}
Recall from \eqref{inequ_imp_positive} that $\Psi(x,0)\leq l^*+c^*$. So, the second equality in \eqref{eq_value_valboun} implies that for $x\leq -n_0^*$, 
\begin{eqnarray*}
v(x)\leq (\bar{K}^{\underline{\gamma}}+1)(l^*+c^*)-\left(\frac{x}{-n_0^*}\right)^{\underline{\gamma}} \frac{\alpha^*}{4}.
\end{eqnarray*}
Consequently, for $m\geq 1$, recalling the definition of $n_m$ and $N_m$, we have that $n_m\geq N_m\geq n_0^*$ and 
$$v(-n_m)\leq v(-N_m)\leq (\bar{K}^{\underline{\gamma}}+1)(l^*+c^*)-\left(\frac{N_m}{n_0^*}\right)^{\underline{\gamma}} \frac{\alpha^*}{4}=-m.$$ This concludes the proof.
\Halmos \endproof

\section{Dynamic Programming} 
\label{muti_per}
In this section, we prepare the proofs of Theorems \ref{one_step_strategy} and \ref{optimality_M_typeA}. For that we will apply the one-period results in two contexts. The first one, called the robust context, assume that $\mathcal{Q}:=\mathcal{Q}_{t+1}(\w^t)$ and $V:= U_{t+1}(\w^t,\cdot,\cdot)$ and is used to prove Theorem \ref{one_step_strategy}. For $P\in\mathcal{H}^T$, the second one, called the $P$-prior context, suppose that $\mathcal{Q}:=\{q_{t+1}^P(\cdot|\w^t)\}$ 
 and $V:= U_{t+1}^P(\w^t,\cdot,\cdot)$ and is used to prove Theorem \ref{optimality_M_typeA}. 
 Note that in the $P$-prior context, $\textup{Graph}(\mathcal{Q})=\textup{Graph}(q_{t+1}^P)$ may not be an analytic set. This will not be an issue as 
 in the one-period case, we did not assume that $\textup{Graph}(\mathcal{Q})$ is analytic. 
 We will construct a $\mathcal{Q}^t$-full-measure set $\widetilde{\Omega}^t$ (resp. a $P^t$-full-measure set $\widetilde{\Omega}^{t,P}$) where Assumptions  \ref{P^*} to 
\ref{pb_inequality}  hold true in the robust (resp. $P$-prior) context (see Lemma \ref{lemma_Assumption_true} and Proposition \ref{U_t^-&C}). To do that, we first introduce and prove properties for a lower bound of $U_t$ and for the dynamic version of $C$ that appears in Assumption \ref{AE} (see Proposition \ref{CJ_t}). Then, Proposition \ref{U_t_well} gives fundamental properties of the values functions $U_t$ and $U_t^P$. The proof of Proposition \ref{U_t_well} will be given in Appendix \ref{proof_U_t_well}. To prove measurability results, we will also use Proposition \ref{lsa_cvt} and Lemma \ref{lemma_mes_u} stated and proved in Appendix \ref{inf_cvt_sub}. 
For the rest of this section, we fix a random utility function $U$ in the sense of Definition \ref{U_hp}. For $0\leq t\leq T$, we define by induction $J_t : \Omega^t \times \mathbb{R} \to \mathbb{R}\cup\{-\infty,+\infty\}$ and $C_t : \Omega^t \to \mathbb{R}\cup\{-\infty,+\infty\}$ as follows: for all $\w^t\in\Omega^t$ and $x\in\mathbb{R}$, 
\begin{eqnarray}
&&\left\{
    \begin{array}{ll}
    J_T(\w^T,x)&:=U^-(\w^T,x)\\
	J_t(\w^t,x)&:=
	\sup_{p\in\mathcal{Q}_{t+1}(\w^t)}\mathbb{E}_p J_{t+1}(\w^t,\cdot,x). \label{J_t_eq}
    \end{array}
\right.\\
&&\left\{
    \begin{array}{ll}
    C_T(\w^T)&:=C(\w^T)\\
	C_t(\w^t)&:=
	\sup_{p\in\mathcal{Q}_{t+1}(\w^t)}\mathbb{E}_p C_{t+1}(\w^t,\cdot). \label{C_t_eq}
    \end{array}
\right.
\end{eqnarray}
The function $-J_t$ will serve as a lower bound for $U_t$, while $C_t$ will appear in \eqref{elas_gammaf} in Assumption \ref{AE} stated for $U_t$. We first show  crucial properties for $J_t$ and $C_t$. The definition of lower-semianalytic (lsa) and upper-semianalytic (usa) functions is recalled in Definition \ref{lsausadef} in the appendix. 
\begin{proposition}
\label{CJ_t}
Assume that Assumptions \ref{S_borel}, \ref{analytic_graph}, \ref{AE} and \ref{simple_adm} hold true. For all $0\leq t\leq T$, $J_t$ and $C_t$ are non-negative, usa and satisfy for all $x\in\mathbb{R}$ that
\begin{eqnarray}
J_t(\cdot,x)<+\infty\;\; \mathcal{Q}^t -q.s \;\; \mbox{and} \;\; C_t(\cdot)<+\infty \;\; \mathcal{Q}^t -q.s.
\label{finite_CJ}
\end{eqnarray}
Moreover, for all $0\leq t \leq T-1$, $x\in\mathbb{R}$ and $h\in\mathbb{R}^d$, the set 
\begin{eqnarray}
\Omega^t_{J,x,h}&:=&\left\{\w^t\in\Omega^t,\;\; \sup_{p\in\mathcal{Q}_{t+1}(\w^t)} \mathbb{E}_p J_{t+1}(\w^t,\cdot, x+h\Delta S_{t+1}(\w^t,\cdot))<+\infty  \right\}
\label{set_J_t}
\end{eqnarray}
belongs to $\mathcal{B}_c(\Omega^t)$ and is of $\mathcal{Q}^t$-full-measure.
\end{proposition}
\proof{Proof.}
\noindent We first show the following claim:\\
\textit{$J_t$ and $C_t$ are non-negative, usa and satisfy for all $1\leq k\leq t$, $x\in\mathbb{R}$, $h\in\mathbb{R}^d$ that \begin{eqnarray}
\sup_{P\in \mathcal{Q}^t} \mathbb{E}_P J_t(\cdot,x+h\Delta S_k(\cdot))<+\infty \quad \mbox{and} \quad \sup_{P\in\mathcal{Q}^t}\mathbb{E}_P C_t<+\infty.
\label{property_temp}
\end{eqnarray}}\\ 
We show the claim by backward induction. At time $T$, $U$ is $\mathcal{B}(\Omega^T) \otimes \mathcal{B}(\mathbb{R})$-measurable (see Remark \ref{joint_mes_U}) and $C$ is non-negative and $\mathcal{B}(\Omega^T)$-measurable (see Assumption \ref{AE}). Thus, $J_T=U^-$ and $C_T=C$ are non-negative and usa. The second inequality in \eqref{property_temp} at time $T$ is given in Assumption \ref{AE}. For the first one, fix $1\leq k\leq T$, $x\in\mathbb{R}$ and $h\in\mathbb{R}^d$. As $U^-(\w^T,\cdot)$ is nonincreasing for all $\w^T\in\Omega^T$, \eqref{temp_Q_R} shows that,
\begin{eqnarray*}
U^-(\cdot,x+h\Delta S_t(\cdot))&\leq &\sum_{\theta\in\{-1,1\}^d} U^-(\cdot,-\lceil x \rceil-\lceil |h| \rceil\theta \Delta S_t(\cdot)).
\end{eqnarray*}
So, the first inequality in \eqref{property_temp} follows from Assumption \ref{simple_adm}.\\
Let $0\leq t\leq T-1$. Assume that $J_{t+1}$ and $C_{t+1}$ are non-negative, usa and that \eqref{property_temp} holds at time $t+1$. Let ${\jmath}_t : \Omega^t\times \mathbb{R}\times  \mathfrak{P}(\Omega_{t+1})\times \mathbb{R}^d \to \mathbb{R}\cup\{-\infty,+\infty\}$ and $\widetilde{\jmath}_t : \Omega^t\times \mathbb{R}\times  \mathbb{R}^d \to \mathbb{R}\cup\{-\infty,+\infty\}$ be  defined for all $(\w^t,x,p,\overline{h})\in \Omega^t\times \mathbb{R}\times \mathfrak{P}(\Omega_{t+1}) \times \mathbb{R}^d$ by
\begin{eqnarray}
{\jmath}_t(\w^t,\overline{x},p,\overline{h}) &:=& \mathbb{E}_p J_{t+1}(\w^t,\cdot,\overline{x}+\overline{h}\Delta S_{t+1}(\w^t,\cdot)) \quad \mbox{and} \quad 
\widetilde{\jmath}_t(\w^t,\overline{x},\overline{h}) := \sup_{p\in\mathcal{Q}_{t+1}(\w^t)}{\jmath}_t(\w^t,\overline{x},p,\overline{h}).\label{tilde_j_t} 
\end{eqnarray}
As $J_{t+1}$ is usa, Lemma \ref{lemma_mes_u} (iii) shows that ${\jmath}_t$ and $\widetilde{\jmath}_t$ are usa. Now, using \citep[Lemma 7.30 (3), p177]{ref1}, we get that $J_t(\cdot,\cdot)=\widetilde{\jmath}_t(\cdot,\cdot,0)$ is usa. The non-negativity of $J_t$ (resp. of $\widetilde{\jmath}_t$) follows from the one of $J_{t+1}$ and of \eqref{J_t_eq} (resp. \eqref{tilde_j_t}). Let $0\leq k\leq t$, $x\in\mathbb{R}$ and $h\in\mathbb{R}^d$. Recalling Assumption \ref{S_borel} and using again \citep[Lemma 7.30 (3), p177]{ref1}, $(\w^t,p,\overline{h}) \mapsto {\jmath}_t(\w^t,x+h\Delta S_k(\w^t),p,\overline{h})$ is usa. Then, Assumption \ref{analytic_graph} and \citep[Proposition 7.50, p184]{ref1} show that given any $\epsilon>0$, there exists $q^\epsilon : \Omega^t\times \mathbb{R}^d \to \mathfrak{P}(\Omega_{t+1})$, which is $\mathcal{B}_c(\Omega^t\times \mathbb{R}^d)$-measurable, such that for all $\w^t\in\Omega^t$ and $\overline{h}\in\mathbb{R}^d$, $q^\epsilon(\cdot|\w^t,\overline{h})\in \mathcal{Q}_{t+1}(\w^t)$ and 
\begin{eqnarray}
{\jmath}_t(\w^t,x+h\Delta S_{k}(\w^t),q^{\epsilon}(\cdot|\w^t,\overline{h}),\overline{h}) \geq \left\{
    \begin{array}{ll}
    \frac{1}{\epsilon} \;\; \mbox{if} \; \widetilde{\jmath}_t(\w^t,x+h\Delta S_{k}(\w^t),\overline{h})=+\infty,\\
	 \widetilde{\jmath}_t(\w^t,x+h\Delta S_{k}(\w^t),\overline{h})-\epsilon \;\; \mbox{otherwise.}
    \end{array}
\right.
\label{temp_J_qeps}
\end{eqnarray} 
For all $\overline{P}\in\mathcal{Q}^t$, taking the expectation under $\overline{P}$ and using Fubini's theorem as $J_{t+1}\geq 0$ we get 
\begin{multline}
\mathbb{E}_{\overline{P}\otimes q^{\epsilon}} J_{t+1}(\cdot,x+h\Delta S_{k}(\cdot)+\overline{h}\Delta S_{t+1}(\cdot)) \geq \frac{1}{\epsilon} \overline{P}\left(\widetilde{\jmath}_t(\cdot,x+h\Delta S_{k}(\cdot),\overline{h})=+\infty\right) \\+ \mathbb{E}_{\overline{P}}\left(\left(\widetilde{\jmath}_t(\cdot,x+h\Delta S_{k}(\cdot),\overline{h})-\epsilon\right)1_{\{\widetilde{\jmath}_t(\cdot,x+h\Delta S_{k}(\cdot),\overline{h})<+\infty\}}\right).\label{temp_CJt_gen}
\end{multline}
As $\overline{P}\otimes q^{\epsilon}\in\mathcal{Q}^{t+1}$ and $\widetilde{\jmath}_t$ is non-negative, we get that
\begin{eqnarray}
\sup_{P\in\mathcal{Q}^{t+1}}\mathbb{E}_{P} J_{t+1}(\cdot,x+h\Delta S_{k}(\cdot)+\overline{h}\Delta S_{t+1}(\cdot)) \geq \frac{1}{\epsilon} \overline{P}\left(\widetilde{\jmath}_t(\cdot,x+h\Delta S_{k}(\cdot),\overline{h})=+\infty\right)-\epsilon.
\label{ineq_temp_J_t_2}
\end{eqnarray}
As $J_t(\cdot,\cdot)=\widetilde{\jmath}_t(\cdot,\cdot,0)$, if $\overline{P}\left(J_t(\cdot,x+h\Delta S_{k}(\cdot))=+\infty\right)>0$, taking the limit when $\epsilon$ goes to $0$ in \eqref{ineq_temp_J_t_2} applied to $\overline{h}=0$, we find that $\sup_{P\in\mathcal{Q}^{t+1}}\mathbb{E}_{P} J_{t+1}(\cdot,x+h\Delta S_{k}(\cdot))=+\infty$, which contradicts \eqref{property_temp} at time $t+1$. Thus, $\overline{P}\left(J_t(\cdot,x+h\Delta S_{k}(\cdot))=+\infty\right)=0$. 
So, \eqref{temp_CJt_gen} for $\overline{h}=0$ implies that
\begin{eqnarray*}
\sup_{P\in\mathcal{Q}^{t+1}}\mathbb{E}_{P} J_{t+1}(\cdot,x+h\Delta S_{k}(\cdot)) \geq  \mathbb{E}_{\overline{P}} J_t(\cdot,x+h\Delta S_{k}(\cdot))-\epsilon.
\end{eqnarray*}
So, letting $\epsilon$ go to $0$, taking the supremum over all $\overline{P} \in\mathcal{Q}^t$ and using \eqref{property_temp} for $t+1$, we get that $\sup_{P\in\mathcal{Q}^{t}}\mathbb{E}_{P} J_t(\cdot,x+h\Delta S_{k}(\cdot)) <+\infty$. Similar arguments show that $C_t$ is non-negative, usa and that $\sup_{P\in\mathcal{Q}^t}\mathbb{E}_P C_t<+\infty$, which gives \eqref{property_temp} at time $t$ and concludes the backward induction.\\

\noindent\textit{Proof of \eqref{finite_CJ}.}\\
Let $0\leq t\leq T-1$. Assume that there exists some $P\in\mathcal{Q}^t$ such that $P(J_t(\cdot,x)=+\infty)>0$. Then, as $J_t\geq 0$, $\mathbb{E}_P J_t(\cdot,x) =+\infty$ and also $\sup_{P\in\mathcal{Q}^t}\mathbb{E}_P J_t(\cdot,x) =+\infty$, a contradiction to \eqref{property_temp} with $h=0$. So, $P(J_t(\cdot,x)<+\infty)=1$ for all $P\in\mathcal{Q}^t$. The proof for $C_t$ is similar and thus omitted.\\

\noindent\textit{The set $\Omega^t_{J,\overline{x},\overline{h}}$ belongs to $\mathcal{B}_c(\Omega^t)$ and is of $\mathcal{Q}^t$-full-measure.}\\ Fix some $0\leq t \leq T-1$, $\overline{x}\in\mathbb{R}$ and $\overline{h}\in\mathbb{R}^d$. As $\widetilde{\jmath}_t$ is usa and $\Omega^t_{J,\overline{x},\overline{h}}=\{\w^t\in\Omega^t,\; \widetilde{\jmath}_t(\w^t,\overline{x},\overline{h})<+\infty\},$ we get that $\Omega^t_{J,\overline{x},\overline{h}}\in \mathcal{B}_c(\Omega^t).$ Assume that $\Omega^t_{J,\overline{x},\overline{h}}$ is not a $\mathcal{Q}^t$-full-measure set. Then, there exists $\overline{P}\in\mathcal{Q}^t$ such that $\overline{P}(\Omega^t_{J,\overline{x},\overline{h}})<1$. 
 Using \eqref{ineq_temp_J_t_2} with $x=\overline{x}$ and $h=0$, we get that
\begin{eqnarray}
\sup_{P\in\mathcal{Q}^{t+1}}\mathbb{E}_{P} J_{t+1}(\cdot,\overline{x}+\overline{h}\Delta S_{t+1}(\cdot)) \geq \frac{1}{\epsilon} (1-\overline{P}(\Omega^t_{J,\overline{x},\overline{h}}))-\epsilon.
\label{ineq_temp_J_t_22}
\end{eqnarray}
Taking the limit in \eqref{ineq_temp_J_t_22} when $\epsilon$ goes to $0$, we find that $\sup_{P\in\mathcal{Q}^{t+1}}\mathbb{E}_{P} J_{t+1}(\cdot,\overline{x}+\overline{h}\Delta S_{t+1}(\cdot)) =+\infty$, which contradicts \eqref{property_temp} at time $t+1$. Thus, for all $\overline{P}\in\mathcal{Q}^t$, $\overline{P}(\Omega^t_{J,\overline{x},\overline{h}})=1$ and $\Omega^t_{J,\overline{x},\overline{h}}$ is a $\mathcal{Q}^t$-full-measure set.
\Halmos \endproof



The next proposition gives some fundamental properties of the value functions $U_t$ and $U_t^P$.

\begin{proposition}
\label{U_t_well}
Assume that Assumptions \ref{S_borel}, \ref{analytic_graph}, \ref{AE} and \ref{simple_adm} hold true. Let $P\in\mathcal{Q}^T$ and $0\leq t \leq T$. We have that (i) $U_t$ is lsa, (ii) $U_t^P$ is $\mathcal{B}_c(\Omega^t\times \mathbb{R})$-measurable, (iii) for all $\w^t\in\Omega^t$, $U_t(\w^t,\cdot)$, $U_t^P(\w^t,\cdot)~:$ $\mathbb{R}\to \mathbb{R}\cup\{-\infty,+\infty\}$ are nondecreasing, usc and concave, (iv) $U_t^- \leq J_t$, (v) $U_t\leq U_t^P$.
Moreover, for all $\w^t\in\Omega^t$ such that $C_t(\w^t)<+\infty$, $\lambda\geq 1$, $x\in\mathbb{R}$, we get that
\begin{eqnarray}
U_t(\w^t,\lambda x)&\leq& \lambda^{\gamma}(U_t(\w^t,x)+C_t(\w^t))\label{elas_gammaft}\\
U_t^P(\w^t,\lambda x)&\leq& \lambda^{\gamma}(U_t^P(\w^t,x)+C_t(\w^t))\label{elas_gammafPt}.
\end{eqnarray}
\end{proposition}
\proof{Proof.}
See Appendix \ref{proof_U_t_well}.
\Halmos \endproof
Assume that $NA(\mathcal{Q}^T)$ and Assumptions \ref{S_borel},  \ref{analytic_graph} and \ref{AE} hold true.
Then, Lemma \ref{lemmaH_nonempty} shows that $\mathcal{H}^T\neq \emptyset$ and we fix for the rest of the paper some $P^*\in\mathcal{H}^T$ with the following fix disintegration 
\begin{eqnarray}
P^*:=q^{P^*}_1\otimes \cdot\cdot\cdot \otimes q^{P^*}_T.
\label{P^*_exp}
\end{eqnarray}
Lemma \ref{simi_qt_na} shows the existence and the $\mathcal{B}_c(\Omega^t)$-measurability of the functions $\alpha_t^{P^*}$ : $\Omega^t\to (0,1]$ and also that $\Omega^{t,P^*}_{qNA}$ defined in \eqref{set_qt_na_H} is a $\mathcal{Q}^t$-full-measure set. The stochastic kernels $(q^{P^*}_t)_{1\leq t\leq T}$ will be of special interest for the statements of the one-period Assumptions \ref{P^*}, \ref{integ_V+} and  \ref{AE_one} in the multiple-period contexts. On the other hand, $(\alpha_t^{P^*})_{0\leq t\leq T-1}$ will serve for the one of Assumption \ref{pb_inequality}.\\

We now present the two different contexts where we will apply the one-period results. The robust context will be used to prove Theorem \ref{one_step_strategy} while the $P$-prior one will be used to prove Theorem \ref{optimality_M_typeA}.
\begin{definition}
\label{def_one_prior}
Let $0\leq t\leq T-1$. For any $\w^t\in\Omega^t$, we call context $(t+1)$, the following one-period market: $\overline{\Omega}:=\Omega_{t+1}$, $\mathcal{G}:=\mathcal{B}_c(\Omega_{t+1})$, $Y(\cdot):=\Delta S_{t+1}(\w^t,\cdot)$, $C(\cdot):= C_{t+1}(\w^t,\cdot)+J_{t+1}(\w^t,\cdot,0)$, $\overline{\gamma}:=\max(1,\gamma)$ and $\underline{\gamma}:=\min(1,\gamma)$, where $\gamma$ is introduced in Assumption \ref{AE}.\\ Then, we are in the robust $(t+1)$ context if in addition $\mathcal{Q}:=\mathcal{Q}_{t+1}(\w^t)$, $p^*:=q^{P^*}_{t+1}(\cdot|\w^t)$, $\alpha^*:=\alpha_t^{P^*}(\w^t)$ and 
$V(\cdot,\cdot):=U_{t+1}(\w^t,\cdot,\cdot)$. As a consequence, $v_{\mathbb{Q}}(x)= \mathcal{U}_t(\w^t,x)$ and $\mbox{cl}(v_{\mathbb{Q}})(x)= U_t(\w^t,x)$, see \eqref{state_val_t_rob} and \eqref{vq}.\\
Now, let $P:=q_1^P\otimes \cdots \otimes q_T^P$. We are in the $P$-prior $(t+1)$ context if $\mathcal{Q}:=\{q^{P}_{t+1}(\cdot|\w^t)\}$, $p^*:=q^{P}_{t+1}(\cdot|\w^t)$, $\alpha^*:=\alpha_t^{P}(\w^t)$ and 
$V(\cdot,\cdot):=U_{t+1}^P(\w^t,\cdot,\cdot)$. As a consequence, $v_{\mathbb{Q}}(x)= \mathcal{U}_t^P(\w^t,x)$ and $\mbox{cl}(v_{\mathbb{Q}})(x)= U_t^P(\w^t,x)$, see \eqref{state_val_t_p} and \eqref{vq}.\\
\end{definition} 
Let $0\leq t\leq T-1$ and $P:=q_1^P \otimes \cdots \otimes q_T^P\in \mathcal{H}^T$. For all $\w^t\in\Omega^t$, we define
\begin{eqnarray}
c_t^P(\w^t)&:=&\mathbb{E}_{q_{t+1}^P(\cdot|\w^t)}\; C_{t+1}(\w^t,\cdot)+\mathbb{E}_{q_{t+1}^P(\cdot|\w^t)} J_{t+1}(\w^t,\cdot,0)\label{ct^P} \label{ct^*} \\
i_t^P(\w^t)&:=&1+2\frac{c_t^P(\w^t)}{\alpha_t^P(\w^t)}.\label{It_P}
\end{eqnarray}
Fix $\w^t\in\Omega^t$. The multiple-period counterpart of $c^*$, $l^*$ and $n_0^*$ (see Assumption \ref{AE_one}, Proposition \ref{well_def_one} and Assumption \ref{pb_inequality}) in the robust $(t+1)$ context are respectively $c_t^{P^*}(\w^t)$, 
\begin{eqnarray}
l_t^*(\w^t)&:=&\sum_{\theta\in\{-1,1\}^d}\mathbb{E}_{q^{P^*}_{t+1}(\cdot|\w^t)}\; U_{t+1}^+(\w^t,\cdot,1+\theta \Delta S_{t+1}(\w^t,\cdot)),\nonumber \\ 
N_t^*(\w^t) &:=&\inf \left\{k\geq 1,\;  q^{P^*}_{t+1}\left (U_{t+1}(\w^t,\cdot, -k)\leq - i_t^{P^*}(\w^t) | \w^t \right)\geq 1-\frac{\alpha_t^{P^*}(\w^t)}{2}\right\},\label{eq_Nt}
\end{eqnarray}
with the convention (which will be used until the end of the paper) that $\inf \emptyset = +\infty$.\\
Now, the counterpart of $c^*$, $l^*$ and $n_0^*$ in the $P$-prior $(t+1)$ context are respectively  $c_t^P(\w^t)$, 
\begin{eqnarray}
l_t^{P}(\w^t)&:=&\sum_{\theta\in\{-1,1\}^d}\mathbb{E}_{q_{t+1}^P(\cdot|\w^t)}\; (U_{t+1}^P)^+(\w^t,\cdot,1+\theta \Delta S_{t+1}(\w^t,\cdot)), \label{l_tP}\\
N_t^P(\w^t) &:=&\inf \left\{k\geq 1,\;  q_{t+1}^P\left (U_{t+1}^P(\w^t,\cdot, -k)\leq - i_t^P(\w^t) | \w^t \right)\geq 1-\frac{\alpha_t^P(\w^t)}{2}\right\}.\label{eq_NtP}
\end{eqnarray}

Note that for all $\w^t\in\Omega^t$, using \eqref{ct^P}, \eqref{J_t_eq}, \eqref{C_t_eq} and Proposition \ref{CJ_t},
\begin{eqnarray}
0\leq c_t^P(\w^t) &\leq& \sup_{p\in\mathcal{Q}_{t+1}(\w^t)}\mathbb{E}_{p}\; C_{t+1}(\w^t,\cdot)+\sup_{p\in\mathcal{Q}_{t+1}(\w^t)}\mathbb{E}_{p} J_{t+1}(\w^t,\cdot,0)= C_t(\w^t)+ J_t(\w^t,0). \label{ineq_ctp2}
\end{eqnarray} 

We first show that all the previous random variables are measurable.
\begin{lemma}
Assume that the $NA(\mathcal{Q}^T)$ condition as well as Assumptions \ref{S_borel}, \ref{analytic_graph}, \ref{AE} and \ref{simple_adm} hold true. Let $P\in\mathcal{H}^T$ and $0\leq t \leq T-1$. Then, $l_t^*$, $N_t^*$, $i_t^P$, $c_t^P$, $l_t^P$ and $N_t^P$ are $\mathcal{B}_c(\Omega^t)$-measurable.
\label{lemma_mes_N}
\end{lemma}
\proof{Proof.}
Recalling that $\alpha^P_t$ is $\mathcal{B}_c(\Omega^t)$-measurable (see Lemma \ref{simi_qt_na}) and that $C_{t+1}$ and $J_{t+1}(\cdot,0)$ are usa (see Proposition \ref{CJ_t}), Proposition \ref{univ_cvt} (iii) shows that $c_t^P$ and $i_t^P$ are $\mathcal{B}_c(\Omega^t)$-measurable. Now, as $U_{t+1}^P$ and $U_{t+1}$ are $\mathcal{B}_c(\Omega^{t+1}\times \mathbb{R})$-measurable (see Proposition \ref{U_t_well}) and Assumption \ref{S_borel} holds true, Lemma \ref{lemma_mes_u} (i) and \citep[Lemma 7.29, p174]{ref1} prove that $l_t^P$ and $l_t^*$ are $\mathcal{B}_c(\Omega^t)$-measurable. We now show that $N_t^*$ is $\mathcal{B}_c(\Omega^t)$-measurable. The proof for $N_t^P$ is completely similar and thus omitted. Let $n\geq 1$. By definition of $N_t^*$ in \eqref{eq_Nt},
$$\{N_t^*\leq n\} = \bigcup_{k=1}^n \left\{ \w^t\in\Omega^t,\; \int_{\Omega^t} 1_{A(k)}(\w^t,\w_{t+1})q_{t+1}^{P^*}(d\w_{t+1}|\w^t) -1 + \frac{\alpha^{P^*}_t(\w^{t})}{2} \geq 0\right\},$$
where $A(k):= \{(\w^{t},\w_{t+1})\in\Omega^{t}\times \Omega_{t+1},\; U_{t+1}(\w^{t},\w_{t+1},-k) +i_t^{P^*}(\w^t)\leq 0\}\in \mathcal{B}_c(\Omega^t\times \Omega_{t+1})$. So, Proposition \ref{univ_cvt} (iii) shows that $\{N_t^* \leq n\}\in\mathcal{B}_c(\Omega^t)$ and this concludes the proof.
\Halmos \endproof
The following sets describe the paths $\w^t\in\Omega^t$ for which the one-period assumptions are satisfied in the robust $(t+1)$ context and/or in the $P$-prior $(t+1)$ context for a prior $P := q_1^P \otimes \cdots \otimes q_T^P\in\mathcal{H}^T$.
\begin{definition}
Let $0\leq t\leq T-1$. For $i\in\{\ref{P^*}, \ref{V}, \ref{integ_V-}, \ref{integ_V+}, \ref{AE_one}\}$, 
let
\begin{eqnarray*}
\Omega^t_{i}&:=&\{\w^t\in\Omega^t,\; \mbox{Assumption i holds true in the robust } (t+1) \mbox{ context}\}\\
\Omega^t_{\ref{pb_inequality}}&:=& \Omega^{t,P^*}_{qNA} \cap \{ N_t^*<+\infty \}\\
\Omega^{t,P}_{i}&:=&\{\w^t\in\Omega^t,\; \mbox{Assumption i holds true in the } P\mbox{-prior } (t+1) \mbox{ context}\}\\
\Omega^{t,P}_{\ref{pb_inequality}}&:=& \Omega^{t,P}_{qNA} \cap \{N_t^P<+\infty\},
\end{eqnarray*}
recall \eqref{set_qt_na_H} for the definition of $\Omega^{t,P^*}_{qNA}$ and $\Omega^{t,P}_{qNA}$. Moreover, we set
\begin{eqnarray}
\widetilde{\Omega}^t := \bigcap_{i=7}^{12} \Omega^t_{i} \quad \mbox{and} \quad \widetilde{\Omega}^{t,P}:=\bigcap_{i=7}^{12} \Omega^{t,P}_{i}. \label{tildeOmega}\label{tildeOmegaP}
\end{eqnarray}
\label{def_ctxt}
\end{definition}

The next lemma shows that if we choose $\w^t$ in $\widetilde{\Omega}^t$ or $\widetilde{\Omega}^{t,P}$, the one-period assumptions are true in the associated $(t+1)$ context.
\begin{lemma}
Assume that the $NA(\mathcal{Q}^T)$ condition as well as Assumptions \ref{S_borel}, \ref{analytic_graph}, \ref{AE} and \ref{simple_adm} hold true. Let $0\leq t\leq T-1$ and $P := q_1^P \otimes \cdots \otimes q_T^P\in\mathcal{H}^T$. If $\w^t\in \widetilde{\Omega}^t$ (resp. $\w^t\in \widetilde{\Omega}^{t,P}$), then Assumptions \ref{P^*}, \ref{V}, \ref{integ_V-}, \ref{integ_V+}, \ref{AE_one} and \ref{pb_inequality} hold true in the robust $(t+1)$ context (resp. $P$-prior $(t+1)$ context).
\label{lemma_Assumption_true}
\end{lemma}
\proof{Proof.}
We make the proof for $\w^t\in\widetilde{\Omega}^t$. The proof for $\w^t\in \widetilde{\Omega}^{t,P}$ is completely similar and thus omitted. For $7\leq i\leq 11$, we trivially have that if $\w^t\in\Omega^t_{i}$, Assumption $i$ holds true in the robust $(t+1)$ context. 
Now, for $\w^t\in\Omega^t_{\ref{pb_inequality}}$, $N_t^*(\w^t)<+\infty$, so that $$q^{P^*}_{t+1}\left (U_{t+1}(\w^t,\cdot, -N_t^*(\w^t))\leq - \left(1+2 \frac{c_t^{P^*}(\w^t)}{\alpha_t^{P^*}(\w^t)}\right) \bigg| \w^t \right)\geq 1-\frac{\alpha_t^{P^*}(\w^t)}{2}.$$ The fact that $\w^t\in\Omega_{qNA}^{t,P^*}$ shows \eqref{qNA} and Assumption \ref{pb_inequality} holds true in robust $(t+1)$ context.
\Halmos \endproof

We now prove that the $\Omega^t_{i}$ are $\mathcal{Q}^t$-full-measure sets while the $\Omega^{t,P}_{i}$ are $P^t$-full-measure sets. The proof needs the technical Lemmata \ref{lemma1+} and \ref{lemma_N_finite} which are relegated to Appendix \ref{miss_proof}.

\begin{proposition}
Assume that the $NA(\mathcal{Q}^T)$ condition as well as Assumptions \ref{S_borel}, \ref{analytic_graph}, \ref{AE} and \ref{simple_adm} hold true. Let $P\in\mathcal{H}^T$ and $0\leq t \leq T-1$. Then, for all $i \in \{\ref{P^*}, \ref{V}, \ref{integ_V-},  \ref{AE_one}\}$, 
$\Omega_{i}^t$ is a $\mathcal{Q}^t$-full-measure set and $\Omega_{i}^{t,P}$ is a $P^t$-full-measure set. Assume furthermore that Assumptions \ref{nncst} and \ref{U0} hold true. Then, $\Omega_{\ref{integ_V+}}^t$ and $\Omega_{\ref{pb_inequality}}^t$ are $\mathcal{Q}^t$-full-measure sets while $\Omega_{\ref{integ_V+}}^{t,P}$ and $\Omega_{\ref{pb_inequality}}^{t,P}$ are $P^t$-full-measure sets. So, there exists a $\mathcal{Q}^t$-full-measure set $\widehat{\Omega}^t \in\mathcal{B}_c(\Omega^t)$ that satisfies $\widehat{\Omega}^t\subset\widetilde{\Omega}^t$.
\label{U_t^-&C}
\end{proposition}
\proof{Proof.}
Fix $P:=q_1^P\otimes \cdots \otimes q_T^P \in\mathcal{H}^T$.\\
\noindent\textit{Let $0\leq t\leq T-1$ and $i \in \{\ref{P^*}, \ref{V}, \ref{integ_V-},  \ref{AE_one}\}$. The sets $\Omega_{i}^t$ and $\Omega_{i}^{t,P}$ are of $\mathcal{Q}^t$-full-measure and also of $P^t$-full-measure under $NA(\mathcal{Q}^T)$ and Assumptions \ref{S_borel}, \ref{analytic_graph}, \ref{AE} and \ref{simple_adm}.}\\ 
By Definitions \ref{def_one_prior} and \ref{def_ctxt}, we see that
\begin{eqnarray*}
\Omega^{t,P}_{\ref{P^*}}&:=& \{\w^t\in\Omega^t,\; 0 \in \textup{ri}(\textup{conv}(D_{P}^{t+1})(\w^t))\}\\
\Omega^{t}_{\ref{P^*}}&:=& \{\w^t\in\Omega^t,\; 0 \in \textup{ri}(\textup{conv}(D_{P^*}^{t+1})(\w^t)), \textup{Aff}(D_{P^*}^{t+1})(\w^t)=\textup{Aff}(D^{t+1})(\w^t)\},
\end{eqnarray*}
and $\Omega^{t}_{\ref{P^*}} \subset \Omega^{t,P^*}_{\ref{P^*}}.$ As $P^*$ and $P$ belong to $\mathcal{H}^T$ 
$\Omega_{\ref{P^*}}^t$, $\Omega^{t,P^*}_{\ref{P^*}}$ and $\Omega_{\ref{P^*}}^{t,P}$ are $\mathcal{Q}^t$-full-measure sets. 
Proposition \ref{U_t_well} at $t+1$ is now in force. Assertions (i) and (ii) together with \citep[Lemma 7.29, p174]{ref1} and Assertion (iii) show that $\Omega^{t}_{\ref{V}}=\Omega_{\ref{V}}^{t,P}=\Omega^t$, which is of course of $\mathcal{Q}^t$-full-measure. Using now (iv) and (v), we have that $\bigcap_{(x,h)\in\mathbb{Q}\times\mathbb{Q}^{d}}\Omega^t_{J,x,h} \subset \Omega_{\ref{integ_V-}}^t\subset \Omega_{\ref{integ_V-}}^{t,P}$ and Proposition \ref{CJ_t} shows that $\Omega_{\ref{integ_V-}}^t$ and $\Omega_{\ref{integ_V-}}^{t,P}$ are $\mathcal{Q}^t$-full-measure sets. Fix $\w^t\in\Omega^t$ such that $J_t(\w^t,0)<+\infty$ and $C_t(\w^t)<+\infty$. Let $\w_{t+1}\in\Omega_{t+1}$ such that $J_{t+1}(\w^t,\w_{t+1},0)<+\infty$ and $C_{t+1}(\w^t,\w_{t+1})<+\infty$. Let $\lambda \geq 1$ and $x\in\mathbb{R}$. Lemma \ref{lemmaAE1} together with Assertions (iii) and (iv) and the fact that $C_{t+1}\geq 0$ show that
\begin{eqnarray*}
U_{t+1}(\w^t,\w_{t+1},\lambda x)&\leq& \lambda (U_{t+1}(\w^t,\w_{t+1},x) + U_{t+1}^-(\w^t,\w_{t+1},0))\\
&\leq &\lambda (U_{t+1}(\w^t,\w_{t+1},x) + J_{t+1}(\w^t,\w_{t+1},0)+C_{t+1}(\w^t,\w_{t+1})).
\end{eqnarray*}
Now, \eqref{elas_gammaft} and $J_{t+1}\geq 0$ implies that
\begin{eqnarray*}
U_{t+1}(\w^t,\w_{t+1},\lambda x)&\leq& \lambda^{\gamma} (U_{t+1}(\w^t,\w_{t+1},x) + J_{t+1}(\w^t,\w_{t+1},0)+C_{t+1}(\w^t,\w_{t+1})).
\end{eqnarray*}
Let $C(\w_{t+1}):= J_{t+1}(\w^t,\w_{t+1},0)+C_{t+1}(\w^t,\w_{t+1}),$ then  $c_t^{P^*}(\w^t)= \mathbb{E}_{q_{t+1}^*(\cdot|\w^t)} C(\cdot) \leq C_{t}(\w^t)+J_{t}(\w^t,0)<+\infty$  using \eqref{ineq_ctp2}. So, the inequalities in \eqref{V_pos} are satisfied in the robust $(t+1)$ context with $\underline{\gamma}:=\min(1,\gamma)$ and $\overline{\gamma}:=\max(1,\gamma)$. Thus, $\{\w^t\in\Omega^t,\; J_{t}(\w^t,0)<+\infty,\; C_t(\w^t)<+\infty\} \subset \Omega^t_{\ref{AE_one}}$. So, \eqref{finite_CJ} in Proposition \ref{CJ_t} shows that $\Omega^t_{\ref{AE_one}}$ is a $\mathcal{Q}^t$-full-measure set. The same arguments apply for $U_{t+1}^P$ (using \eqref{elas_gammafPt} instead of \eqref{elas_gammaft} and $c_t^P(\w^t)$ instead of $c_t^{P^*}(\w^t)$) and $\Omega^{t,P}_{\ref{AE_one}}$ is also a $\mathcal{Q}^t$-full-measure set.\\ Note that, as $\mathcal{H}^T \subset \mathcal{Q}^T$, $\Omega^t_{i}$ and $\Omega^{t,P}_{i}$ are also $P^t$-full-measure sets for all $i \in \{\ref{P^*}, \ref{V}, \ref{integ_V-},  \ref{AE_one}\}$.\\

\noindent\textit{Let $0\leq t\leq T-1$ and $i \in \{\ref{integ_V+}, \ref{pb_inequality}\}$. 
The set $\Omega_{i}^t$ is of $\mathcal{Q}^t$-full-measure and $\Omega^{t,P}_{i}$ is of $P^t$-full-measure if we also assume Assumptions \ref{nncst} and \ref{U0}.}\\
The assertions for $\Omega^t_{\ref{integ_V+}}$ and $\Omega^{t,P}_{\ref{integ_V+}}$ are proved in Lemma \ref{lemma1+} in the Appendix. Recall $\Omega^t_{\ref{pb_inequality}}$ and $\Omega^{t,P}_{\ref{pb_inequality}}$ from Definition \ref{def_ctxt}.
We prove by backward induction that $\Omega^t_{\ref{pb_inequality}}$ is a $\mathcal{Q}^t$-full-measure set and that $\Omega^{t,P}_{\ref{pb_inequality}}$ is a $P^t$-full-measure set for $P\in\mathcal{H}^T$. The initialization step at $T-1$ is a direct consequence of \eqref{NtoPfiniteT} in Lemma \ref{lemma_N_finite} and of the fact that $\Omega^{T-1,P^*}_{qNA}$ and $\Omega^{T-1,P}_{qNA}$ are $\mathcal{Q}^{T-1}$-full-measure sets under $NA(\mathcal{Q}^T)$ (see Lemma \ref{simi_qt_na}). Assume now that the induction hypothesis holds true for some $1\leq t\leq T-1$. We have already proved that for all $7\leq i\leq 11$, the sets $\Omega^{t,P}_{i}$ are of $P^t$-full-measure. So, the induction hypothesis implies that $\widetilde{\Omega}^{t,P}$ (see \eqref{tildeOmegaP}) is also a $P^t$-full-measure set for all $P\in\mathcal{H}^T$ and Lemma \ref{lemma_N_finite} can be applied for $t$. Thus, \eqref{NtoPfinite} and the fact that $\Omega^{t-1,P^*}_{qNA}$ and $\Omega^{t-1,P}_{qNA}$ are $\mathcal{Q}^{t-1}$-full-measure sets show the heredity step. This concludes the backward induction.\\ 

Finally, under all the assumptions, $\widetilde{\Omega}^t$ is a $\mathcal{Q}^t$-full-measure set and we choose $\widehat{\Omega}^t\in\mathcal{B}_c(\Omega^t)$ such that $\widehat{\Omega}^t$ is a $\mathcal{Q}^t$-full-measure set and $\widehat{\Omega}^t\subset \widetilde{\Omega}^t$.
\Halmos \endproof


\section{Proof of Theorem \ref{one_step_strategy}}
\label{proof_th1}
We now turn to the proof of Theorem \ref{one_step_strategy}. First, we show that for all $0\leq t\leq T-1$, there exists a jointly measurable optimal investment strategy at time $t$ when starting with a cash position $x$. For that, we use the results obtained on $\widehat{\Omega}^t$ and $U_t$ in the preceding section.

\begin{proposition}
\label{opt_strat_exist}
Assume that the $NA(\mathcal{Q}^T)$ condition as well as Assumptions \ref{S_borel}, \ref{analytic_graph}, \ref{AE}, \ref{nncst}, \ref{simple_adm} and \ref{U0} hold true. Let $0\leq t\leq T-1$. There exists a $\mathcal{B}_c(\Omega^t)\otimes \mathcal{B}(\mathbb{R})$-measurable function $H_{t+1}^* : \Omega^t\times \mathbb{R} \to \mathbb{R}^d$ such that $H_{t+1}^*(\w^t,\cdot)\in \mbox{Aff}(D^{t+1})(\w^t)$ for all $\w^t\in\widehat{\Omega}^t$, where $\widehat{\Omega}^t$ has been defined in Proposition \ref{U_t^-&C} and does not depend from $x$. Moreover, for all $\w^t\in \widehat{\Omega}^t$ and $x\in\mathbb{R}$, 
\begin{eqnarray}
U_t(\w^t,x)&=& \sup_{h\in\mathbb{R}^d} \inf_{p\in\mathcal{Q}_{t+1}(\w^t)} \mathbb{E}_p U_{t+1}\big(\w^t,\cdot, x+h \Delta S_{t+1}(\w^t,\cdot)\big)\label{exist_opt_t_eq0}\\ 
&=&\inf_{p\in\mathcal{Q}_{t+1}(\w^t)} \mathbb{E}_p U_{t+1}\big(\w^t,\cdot, x+H_{t+1}^*(\w^t,x) \Delta S_{t+1}(\w^t,\cdot)\big) .
\label{exist_opt_t_eq}
\end{eqnarray}
\label{exist_opt_t_pp}
\end{proposition}
\proof{Proof.}
Fix $x\in\mathbb{R}$. First note that the set $\widehat{\Omega}^t$, introduced in Proposition \ref{U_t^-&C}, does not depend from $x$. In this proof, we will apply several results of the one-period section in the robust $(t+1)$ context (see Definition \ref{def_one_prior}). This is possible if we choose $\w^t\in\widehat{\Omega}^t \subset \widetilde{\Omega}^t$. Indeed, Lemma \ref{lemma_Assumption_true} shows that  Assumptions  \ref{P^*} to 
\ref{pb_inequality}  hold in the robust $(t+1)$ context. Now, recalling Definition \ref{def_one_prior}, $v_{\mathbb{Q}}(x)= \mathcal{U}_t(\w^t,x)$, $\mbox{cl}(v_{\mathbb{Q}})(x)= U_t(\w^t,x)$ and \eqref{exist_opt_t_eq0} is an immediate consequence of \eqref{state_val_t_rob}, \eqref{v} and from (\ref{v_v_q}) in Proposition \ref{existence_uni}. Now, \eqref{Phi} implies that
\begin{eqnarray}
\Psi(x,h)=\inf_{p\in\mathcal{Q}_{t+1}(\w^t)} \mathbb{E}_p U_{t+1}(\w^t, \cdot, x+h\Delta S_{t+1}(\w^t,\cdot))=: \widetilde{u}_t(\w^t,x,h).\label{tilde_u_t}
\end{eqnarray}
Proposition \ref{U_t_well} shows that $U_{t+1}$ is lsa and Lemma \ref{lemma_mes_u} (ii) for $f=U_{t+1}$ proves that $\widetilde{u}_t$ is lsa. Let $\widehat{u}_t : \Omega^t \times \mathbb{R} \times \mathbb{R}^d \to \mathbb{R}\cup\{-\infty,+\infty\}$ be defined by $\widehat{u}_t(\w^t,x,h):=1_{\widehat{\Omega}^t}(\w^t)\widetilde{u}_t(\w^t,x,h).$ As $\widehat{\Omega}^t\in\mathcal{B}_c(\Omega^t)$, $\widehat{u}_t$ is $\mathcal{B}_c(\Omega^t\times \Omega_{t+1}\times \mathbb{R}^d)$-measurable. Fix $\w^t\in\widehat{\Omega}^t \subset \widetilde{\Omega}^t$, Proposition \ref{well_def_one} shows that $\widehat{u}_t(\w^t,\cdot,\cdot)$ is finite-valued and continuous. Fix some $h\in\mathbb{R}^d$. We get that for all $x\in\mathbb{R}$, $\widehat{u}_t(\cdot,x,h)$ is $\mathcal{B}_c(\Omega^t)$-measurable, see \citep[Lemma 7.29, p174]{ref1}. So, $\widehat{u}_t(\cdot,\cdot,h)$ is a Caratheodory integrand and thus a normal integrand (with respect to $\mathcal{B}_c(\Omega^t)$), see \citep[Example 14.29, p662]{ref5}. Note that here $f$ is a normal integrand if $-f$ satisfies \citep[Definition 14.27, p661]{ref5}. Now, \citep[Corollary 14.34, p664]{ref5} shows that $\widehat{u}_t(\cdot,\cdot,h)$ is $\mathcal{B}_c(\Omega^t)\otimes \mathcal{B}(\mathbb{R})$-measurable for all $h\in\mathbb{R}^d$. Then, as $\widehat{u}_t(\w^t,x,\cdot)$ is continuous for every $(\w^t,x)\in \Omega^t\times \mathbb{R}$, we get that $\widehat{u}_t$ is also a Caratheodory integrand and thus a normal integrand (with respect to $\mathcal{B}_c(\Omega^t)\otimes \mathcal{B}(\mathbb{R})$). Set for all $(\w^t,x)\in\Omega^t\times \mathbb{R}$, $Z(\w^t,x):=\mbox{argmax}\; \widehat{u}_t(\w^t,x,\cdot)$.
Then, \citep[Theorem 14.37, p665]{ref5} shows that  $Z : \Omega^t\times \mathbb{R} \twoheadrightarrow \mathbb{R}^d$ is closed-valued and $\mathcal{B}_c(\Omega^t)\otimes \mathcal{B}(\mathbb{R})$-measurable in the sense of \citep[Definition 14.1, p643]{ref5}. Moreover, $\textup{Aff}(D^{t+1}) : \Omega^t \twoheadrightarrow \mathbb{R}^d$ is non-empty, closed-valued and $\mathcal{B}_c(\Omega^t)$-measurable using \citep[Lemma 2.6]{ref4}. So, $Z\cap \textup{Aff}(D^{t+1})$ is closed-valued and $\mathcal{B}_c(\Omega^t) \otimes \mathcal{B}(\mathbb{R})$-measurable, see \citep[Proposition 14.11 (a), p651]{ref5}. Remark that for all $(\w^t,x)\in\widehat{\Omega}^t \times \mathbb{R}$, $Z(\w^t,x)=\mbox{argmax}\; \widetilde{u}_t(\w^t,x,\cdot).$ Thus, using \eqref{maximizer} in Proposition \ref{existence_uni}, we get that
\begin{eqnarray*}
\widehat{\Omega}^t \times \mathbb{R}\subset \{(\w^t,x)\in\Omega^t\times \mathbb{R},\; Z(\w^t,x)\cap \mbox{Aff}(D^{t+1})(\w^t) \neq \emptyset \}. 
\end{eqnarray*}
So, \citep[Corollary 14.6, p647]{ref5} applies to $Z\cap \mbox{Aff}(D^{t+1})$ and shows the existence of a $\mathcal{B}_c(\Omega^t)\otimes \mathcal{B}(\mathbb{R})$-measurable function $H_{t+1} : \widehat{\Omega}^t \times \mathbb{R} \to \mathbb{R}^d$ such that for all $(\w^t,x)\in\widehat{\Omega}^t \times \mathbb{R}$, $H_{t+1}(\w^t,x)\in \mbox{Aff}(D^{t+1})(\w^t)$ and (\ref{exist_opt_t_eq}) is true. We extend $H_{t+1}$ on $\Omega^t\times \mathbb{R}$ as follows. Let $H^*_{t+1}$ be defined by $H^*_{t+1}(\w^t,x) := H_{t+1}(\w^t,x)1_{\widehat{\Omega}^t}(\w^t)$ for all $(\w^t,x)\in \Omega^t \times \mathbb{R}$. As $\widehat{\Omega}^t\in\mathcal{B}_c(\Omega^t)$, $H^*_{t+1}$ is $\mathcal{B}_c(\Omega^t)\otimes\mathcal{B}(\mathbb{R})$-measurable. Moreover, for all $(\w^t,x)\in\widehat{\Omega}^t \times \mathbb{R}$, $H_{t+1}^*(\w^t,x)=H_{t+1}(\w^t,x)\in\textup{Aff}(D^{t+1})(\w^t)$ and (\ref{exist_opt_t_eq}) remains true (on $\widehat{\Omega}^t$). 
\Halmos \\ \endproof 


We are now in position to prove Theorem \ref{one_step_strategy}. 

\proof{Proof of Theorem \ref{one_step_strategy}}
First, we show that $U_0(x)\geq u(x)$ and that the admissibility of a strategy is stable in time. Then, we prove that there exists an appropriate strategy satisfying \eqref{one_step_strategy_temp2}. Finally, we show that when this strategy is admissible, it achieves the maximum in \eqref{RUMP}.\\

\noindent\textit{Upper bound for $u(x)$ and stability of admissibility.}\\ 
We show that for any $x\in\mathbb{R}$, \begin{eqnarray}
U_0(x)\geq \sup_{\phi\in \Phi(x,U,\mathcal{Q}^T)}\inf_{P\in \mathcal{Q}^T}\mathbb{E}_P U(\cdot,V_T^{x,\phi}(\cdot))=u(x),
\label{ineg_fdmt}
\end{eqnarray}
and if $\phi\in \Phi(x,U,\mathcal{Q}^T)$, then $\phi\in \Phi_{|t}(x,U_t,\mathcal{Q}^t)$ for all $1\leq t\leq T$, where $\Phi_{|t}(x,U_t,\mathcal{Q}^t)$ is the set of admissible strategies for the random utility $U_t$ with time horizon $t$, see Definition \ref{admissibility_def}. Note that $\Phi_{|T}(x,U_T,\mathcal{Q}^T)=\Phi(x,U,\mathcal{Q}^T)$. Fix $x\in\mathbb{R}$ and $\phi\in \Phi(x,U,\mathcal{Q}^T)$. We proceed by backward induction with the following induction hypothesis: 
\begin{eqnarray}
\phi\in \Phi_{|t}(x,U_t,\mathcal{Q}^t) \;\; \mbox{and} \;\; \inf_{P\in\mathcal{Q}^T} \mathbb{E}_P U(\cdot,V_T^{x,\phi}(\cdot)) \leq \inf_{P\in\mathcal{Q}^t} \mathbb{E}_P U_t(\cdot,V_t^{x,\phi}(\cdot)).
\label{temp_ind_eq}
\end{eqnarray}
 The initialization step is trivial as $U_T=U$. Let $0\leq t\leq T-1$ and assume that the induction hypothesis holds true at time $t+1$. Proposition \ref{U_t_well} shows that $U_{t+1}$ is lsa and Lemma \ref{lemma_mes_u} (ii) for $f=U_{t+1}$ that 
\begin{eqnarray}
(\w^t,x,h,p)\mapsto u_t(\w^t,x,h,p):= \mathbb{E}_p U_{t+1}(\w^t,\cdot,x+h\Delta S_{t+1}(\w^t,\cdot))\label{u_t}
\end{eqnarray}
is lsa. Note that $u_t(\w^t,\cdot,\cdot,p)$ is equal to $\Psi_p$ (see \eqref{Phi}) in the robust $(t+1)$ context. Let $\epsilon>0$. Assumption \ref{analytic_graph} and \citep[Proposition 7.50, p184]{ref1} show that there exists a $\mathcal{B}_c(\Omega^t\times \mathbb{R} \times \mathbb{R}^d)$-measurable $q^\epsilon_{t+1} : \Omega^t\times \mathbb{R} \times \mathbb{R}^d \to \mathfrak{P}(\Omega_{t+1})$ such that  $ \forall (\w^t,x,h)\in\Omega^t \times \mathbb{R} \times \mathbb{R}^d$, $q^\epsilon_{t+1}(\cdot|\w^t,x,h)\in \mathcal{Q}_{t+1}(\w^t)$ and  
\begin{eqnarray}
u_t(\w^t,x,h,q^{\epsilon}_{t+1}(\cdot|\w^t,x,h))\leq -\frac{1}{\epsilon}1_{\{\widetilde{u}_t(\w^t,x,h)=-\infty\}} +(\widetilde{u}_t(\w^t,x,h)+\epsilon)1_{\{\widetilde{u}_t(\w^t,x,h)>-\infty\}},
\label{temp_ind_eq2}
\end{eqnarray}
where $\widetilde{u}_t$ is defined in \eqref{tilde_u_t}. Recall $\widehat{\Omega}^t$ from Proposition \ref{U_t^-&C}. Fix $\w^t\in\widehat{\Omega}^t\subset\widetilde{\Omega}^t$. Proposition \ref{well_def_one} applies in the robust $(t+1)$ context thanks to Lemma \ref{lemma_Assumption_true} and $\widetilde{u}_t(\w^t,x,h)>-\infty$ for all $(x,h)\in\mathbb{R}\times \mathbb{R}^d$. Moreover, recalling \eqref{state_val_t_rob} and \eqref{tilde_u_t}, $U_t(\w^t,x)\geq \mathcal{U}_t(\w^t,x) \geq \widetilde{u}_t(\w^t,x,h)$ for all $(x,h)\in\mathbb{R}\times \mathbb{R}^d$. So, using \eqref{temp_ind_eq2} with $x=V_t^{x,\phi}(\w^t)$ and $h=\phi_{t+1}(\w^t)$ and setting $\overline{q}^{\epsilon}_{t+1}(\cdot|\w^t):=q^{\epsilon}_{t+1}(\cdot |\w^t,V_t^{x,\phi}(\w^t),\phi_{t+1}(\w^t))$, we have that for all $\w^t\in \widehat{\Omega}^t$, 
\begin{eqnarray*}
u_t\left(\w^t,V_t^{x,\phi}(\w^t),\phi_{t+1}(\w^t),\overline{q}^{\epsilon}_{t+1}(\cdot|\w^t)\right) & \leq &  U_t(\w^t,V_t^{x,\phi}(\w^t))+\epsilon.
\end{eqnarray*}
Then, taking the negative part and using Jensen inequality, we obtain that,
\begin{eqnarray*}
\int_{\Omega_{t+1}} U^-_{t+1}(\w^t,\w_{t+1},V_{t+1}^{x,\phi}(\w^t,\w_{t+1})) \overline{q}^{\epsilon}_{t+1}(d\w_{t+1}|\w^t)  &\geq & U_t^-(\w^t,V_{t}^{x,\phi}(\w^t))-\epsilon.
\end{eqnarray*}
Let $P\in\mathcal{Q}^t$. As $\widehat{\Omega}^t$ is a $\mathcal{Q}^t$-full-measure set (see Proposition \ref{U_t^-&C}), we obtain from the two preceding inequalities that
\begin{eqnarray}
\int_{\Omega^t}\int_{\Omega_{t+1}} U_{t+1}(\w^t,\w_{t+1},V_{t+1}^{x,\phi}(\w^t,\w_{t+1})) \overline{q}^{\epsilon}_{t+1}(d\w_{t+1}|\w^t)  P(d\w^t)&\leq&  \mathbb{E}_{P}U_{t}(\cdot,V_{t}^{x,\phi}(\cdot)) +\epsilon
\label{temp_ineq_decreasing_1}\\
\int_{\Omega^t}\int_{\Omega_{t+1}} U_{t+1}^-(\w^t,\w_{t+1},V_{t+1}^{x,\phi}(\w^t,\w_{t+1})) \overline{q}^{\epsilon}_{t+1}(d\w_{t+1}|\w^t)  P(d\w^t)&\geq&  \mathbb{E}_{P}U_{t}^-(\cdot,V_{t}^{x,\phi}(\cdot)) -\epsilon.
\label{temp_ineq_decreasing_3}
\end{eqnarray}
As $\phi\in\Phi$ and Assumption \ref{S_borel} holds true, $\phi_{t+1}$ and $V_t^{x,\phi}$ are $\mathcal{B}_c(\Omega^t)$-measurable.
Recalling that $q_{t+1}^\epsilon$ is $\mathcal{B}_c(\Omega^t\times \mathbb{R} \times \mathbb{R}^d)$-measurable, \citep[Proposition 7.44, p172]{ref1} shows that $\overline{q}^{\epsilon}_{t+1} \in SK_{t+1}$. So, $P\otimes \overline{q}^{\epsilon}_{t+1} \in\mathcal{Q}^{t+1}$ and Fubini's theorem applied to \eqref{temp_ineq_decreasing_3}, together with the induction hypothesis $\phi\in \Phi_{|t+1}(x,U_{t+1},\mathcal{Q}^{t+1})$ show that
$$+\infty> \mathbb{E}_{P\otimes \overline{q}^{\epsilon}_{t+1}}U_{t+1}^-(\cdot,V_{t+1}^{x,\phi}(\cdot)) \geq \mathbb{E}_{P}U_{t}^-(\cdot,V_{t}^{x,\phi}(\cdot)) -\epsilon.$$ As this holds true for all $P\in\mathcal{Q}^t$, $\phi\in \Phi_{|t}(x,U_t,\mathcal{Q}^t)$ and the first part of the induction is proved. Similarly, using \eqref{temp_ineq_decreasing_1}, we get that
$$\inf_{P\in\mathcal{Q}^{t+1}}\mathbb{E}_{P}U_{t+1}(\cdot,V_{t+1}^{x,\phi}(\cdot)) \leq \mathbb{E}_{P\otimes \overline{q}^{\epsilon}_{t+1}}U_{t+1}(\cdot,V_{t+1}^{x,\phi}(\cdot))\leq \mathbb{E}_{P}U_{t}(\cdot,V_{t}^{x,\phi}(\cdot)) +\epsilon.$$
Taking the infimum over all $P\in\mathcal{Q}^t$ on the right-hand side, letting $\epsilon$ go to $0$ and using the second part of the induction hypothesis in \eqref{temp_ind_eq}, we get that
$$\inf_{P\in\mathcal{Q}^T} \mathbb{E}_P U(\cdot,V_T^{x,\phi}(\cdot)) \leq \inf_{P\in\mathcal{Q}^{t+1}}\mathbb{E}_{P}U_{t+1}(\cdot,V_{t+1}^{x,\phi}(\cdot)) \leq \inf_{P\in\mathcal{Q}^t}\mathbb{E}_{P}U_{t}(\cdot,V_{t}^{x,\phi}(\cdot)).$$
This concludes the induction.\\
Now, \eqref{temp_ind_eq} shows that for $\phi\in\Phi(x,U,\mathcal{Q}^T)$,
$\inf_{P\in\mathcal{Q}^T} \mathbb{E}_P U(\cdot,V_T^{x,\phi}(\cdot))  \leq U_0(x)$
and \eqref{ineg_fdmt} follows taking the supremum over all such $\phi$.\\

\noindent\textit{Existence of a one-step optimal strategy.}\\
Let $x\in\mathbb{R}$. We define recursively the strategy $\phi^{*,x}$ as follows. Let $\phi^{*,x}_1 := x$ and for all $1\leq t\leq T-1$ and $\w^t\in\Omega^t$, $\phi^{*,x}_{t+1}(\w^t):=H_{t+1}^*(\w^t,x+ \sum_{s=0}^t \phi_s^{*,x}(\w^t) \Delta S_{s}(\w^t))$ where $H_{t+1}^* : \Omega^t\times \mathbb{R} \to \mathbb{R}^d$ is defined in Proposition \ref{exist_opt_t_pp}. Let $1\leq t\leq T-1$. Proposition \ref{exist_opt_t_pp} shows that for all $(\w^t,x)\in \widehat{\Omega}^t\times \mathbb{R}$, $\phi^{*,x}_{t+1}(\w^t)\in \textup{Aff}(D^{t+1})(\w^t)$ and that \eqref{exist_opt_t_eq0} and \eqref{exist_opt_t_eq} hold true for $x=V_t^{x,\phi^{*,x}}(\w^t)$ (recall that $\widehat{\Omega}^t$ does not depend from $x$). So, \eqref{onevc v c _step_strategy_temp0} and (\ref{one_step_strategy_temp2}) hold also true.\\ 
We show by induction that $\phi^{*,x}_{t+1}$ is $\mathcal{B}_c(\Omega^{t})$-measurable  for all $0\leq t \leq T-1$. At $t=0$, this is trivial (recall that $\Omega^0$ is a singleton). Suppose that this holds true for all $0\leq s \leq t-1$. Then, recalling Assumption \ref{S_borel}, $V_t^{x,\phi^{*,x}}$ is $\mathcal{B}_c(\Omega^t)$-measurable. Thus, as $H_{t+1}^*$ is $\mathcal{B}_c(\Omega^t)\otimes \mathcal{B}(\mathbb{R})$-measurable (see Proposition \ref{exist_opt_t_pp}), we find that $\phi^{*,x}_{t+1}$ is $\mathcal{B}_c(\Omega^{t})$-measurable.
This concludes the induction and $\phi^{*,x}\in\Phi$.\\

\noindent\textit{Optimality of an admissible one-step optimal strategy.}\\
Assume now that $\phi^{*,x} \in \Phi(x,U,\mathcal{Q}^T)$. Let $0\leq t \leq T-1$ and $P:=q_1^P \otimes \cdot\cdot\cdot \otimes q_T^P\in\mathcal{Q}^T$. We have proved in the preceding step that (\ref{one_step_strategy_temp2}) holds true. So, we have for all $\w^t\in\widehat{\Omega}^t$ that
\begin{eqnarray*}
U_{t}(\w^t,V_t^{x,\phi^{*,x}}(\w^t)) \leq \int_{\Omega_{t+1}} U_{t+1}\left(\w^t,\w_{t+1},V_{t+1}^{x,\phi^{*,x}}(\w^t,\w_{t+1})\right) q_{t+1}^P(d\w_{t+1}|\w^t).
\end{eqnarray*}
As $\widehat{\Omega}^t$ is a $\mathcal{Q}^t$-full-measure set (see Proposition \ref{U_t^-&C}), we get that
\begin{eqnarray*}
\mathbb{E}_{P^t}U_{t}(\cdot,V_t^{x,\phi^{*,x}}(\cdot))\leq \int_{\Omega^t}\int_{\Omega_{t+1}} U_{t+1}\left(\w^t,\w_{t+1},V_{t+1}^{x,\phi^{*,x}}(\w^t,\w_{t+1})\right) q_{t+1}^P(d\w_{t+1}|\w^t) P^t(d\w^{t}).
\end{eqnarray*}
As $\phi^{*,x} \in \Phi(x,U,\mathcal{Q}^T)$, we have proved in the first step that $\phi^{*,x}\in\Phi_{|t+1}(x,U_{t+1},\mathcal{Q}^{t+1})$. So, $\mathbb{E}_{P^{t+1}}U_{t+1}^-(\cdot,V_{t+1}^{x,\phi^{*,x}}(\cdot))<+\infty$ and using Fubini's theorem, we get that
$$\mathbb{E}_{P^t} U_{t}\left(\cdot,V_t^{x,\phi^{*,x}}(\cdot)\right) \leq \mathbb{E}_{P^{t+1}}U_{t+1}\left(\cdot,V_{t+1}^{x,\phi^{*,x}}(\cdot)\right). $$
Iterating the process, we get that $U_0(x)\leq \mathbb{E}_{P}U(\cdot,V_T^{x,\phi^{*,x}}(\cdot))$ for all $P\in \mathcal{Q}^T$ and thus
$$U_0(x)\leq \inf_{P\in\mathcal{Q}^T}\mathbb{E}_{P}U(\cdot,V_T^{x,\phi^{*,x}}(\cdot))\leq u(x),$$ as $\phi^{*,x} \in \Phi(x,U,\mathcal{Q}^T)$. Using \eqref{ineg_fdmt}, we get that \eqref{eg_fdmt} holds true which concludes the proof.
\Halmos \endproof

\section{Proof of Theorem \ref{optimality_M_typeA}}
\label{proof_optimality_M}
In this part, we prove Theorem \ref{optimality_M_typeA}. We will apply Theorem \ref{one_step_strategy} and verify the different conditions needed for that. We will prove that if Assumptions \ref{S_borel} and \ref{analytic_graph} as well as $NA(\mathcal{Q}^T)$ hold true and if $U$ is a random utility of type $(A)$, then Assumptions \ref{simple_adm} and \ref{U0} hold true. We will also show that the optimal strategy is admissible. Recall that Assumptions \ref{AE} and \ref{nncst} hold true by definition of a random utility of type $(A)$. For that, we use the result of Section \ref{muti_per} for $\Omega^{t,P}$ and $U_t^P$.\\

 
\noindent\textit{Assumption \ref{simple_adm} holds true.}\\ 
Let $x\in\mathbb{R}$, $h\in\mathbb{R}^d$ and $1\leq t\leq T$. Let $r\geq 1$ and $P\in\mathcal{Q}^T$. Thanks to Remark \ref{joint_mes_U} and Assumption \ref{S_borel}, $U^-(\cdot,x+h\Delta S_t(\cdot))$ is $\mathcal{B}_c(\Omega^t)$-measurable. Using (\ref{ineq_b_inf_det_A}) and Cauchy-Schwarz inequality, we get that
\begin{small}
\begin{eqnarray*}
\mathbb{E}_P \left(U^-\Big(\cdot,x+h\Delta S_t(\cdot)\Big)\right)^r &\leq & \mathbb{E}_P(C_1^r(\cdot)(1+|x+h\Delta S_t(\cdot)|^p)^r) 
\leq 2^{r-1} \mathbb{E}_P\left(C_1^r(\cdot)(1+|x+h\Delta S_t(\cdot)|^{pr})\right)\\
&\leq & 2^{r-1}(1+2^{pr-1}|x|^{pr})\mathbb{E}_P C_1^r(\cdot) 
+2^{r-1}2^{pr-1}|h|^{pr} \sqrt{\mathbb{E}_P C_1^{2r}(\cdot)}\sqrt{\mathbb{E}_P |\Delta S_{t}(\cdot)|^{2pr}}.
\end{eqnarray*}
\end{small}
As $C_1 \in \mathcal{W}^T$ and $|\Delta S_t| \in\mathcal{W}^t$, we get that
\begin{eqnarray}
U^-(\cdot,x+h\Delta S_t(\cdot))\in\mathcal{W}^T \label{temp_U^-W}
\end{eqnarray} 
and, in particular, that Assumption \ref{simple_adm} holds true.\\ 

\noindent\textit{Assumption \ref{U0} holds true.}\\
Let $P\in\mathcal{H}^T$. Recall that $c_t^P$, $i_t^P$, $l_t^P$, $N_t^P$ and $\widetilde{\Omega}^{t,P}$ are defined respectively in \eqref{ct^P}, \eqref{It_P}, \eqref{l_tP}, \eqref{eq_NtP} and \eqref{tildeOmegaP} and the sets $\mathcal{M}^t(P)$, $\mathcal{M}^t$ and $\mathcal{W}^t$ are defined in Definition \ref{M_tP} for all $0\leq t\leq T-1$. We need to define some of them also for $t=-1$ and $t=T$ and we set $N_{-1}^P := 0$, $l_T^P :=0$, $\widetilde{\Omega}^{T,P}:= \Omega^T$ and $\mathcal{M}^{-1}(P) :=\{0\}$. For all $0\leq t\leq T$, we prove by backward induction the following induction hypothesis: $(U_{t}^{P})^+(\cdot,1)$ and $l_{t}^P$ belong to $\mathcal{M}^{t}(P)$, $C_t$ and $J_t(\cdot,x)$ belong to $\mathcal{M}^{t}$ for all $x\in\mathbb{R}$, $N_{t-1}^P\in\mathcal{M}^{t-1}(P)$ and $\widetilde{\Omega}^{t,P}$ is a $P^t$-full-measure set. Then, we will obtain from the induction hypothesis at $t=0$ that Assumption \ref{U0} holds true.\\
\noindent\textit{Initialization step.}\\ 
We trivially have that $l_T^P=0 \in \mathcal{M}^T(P)$ and that $\widetilde{\Omega}^{T,P}=\Omega^T$ is a $P$-full-measure set. Moreover, $J_T(\cdot,x)=U^{-}(\cdot,x)\in\mathcal{W}^T \subset \mathcal{M}^T$ for all $x\in\mathbb{R}$ by \eqref{temp_U^-W}.  Additionally, $C_T=C\in\mathcal{M}^T$ and $U^+(\cdot,1)\in \mathcal{M}^T$ as $U$ is of type (A). As $\mathcal{M}^T \subset \mathcal{M}^T(P)$, $(U_T^P)^+(\cdot,1)=U^+(\cdot,1)\in\mathcal{M}^T(P)$. By assumption of Theorem \ref{optimality_M_typeA}, we have that $1/\alpha_{T-1}^P\in\mathcal{M}^{T-1}\subset\mathcal{M}^{T-1}(P)$. Again, as U is of type (A), we know that $\underline{X}$ and $1/|U(\cdot,\underline{X}(\cdot))+C(\cdot)|$ belong to $\mathcal{M}^T$ and also to $\mathcal{M}^T(P)$. So, we can use assertion (A1) in Lemma \ref{lemma_N_finite} (recall from \eqref{temp_U^-W} that $U^-(\cdot,0)\in\mathcal{M}^T(P)$) and we get that $N_{T-1}^P \in\mathcal{M}^{T-1}(P)$. \\

\noindent Assume that the induction hypothesis holds true at time $t+1$ for some $0\leq t\leq T-1$.\\

\noindent\textit{Heredity step 1: $C_t\in\mathcal{M}^t$ and $J_t(\cdot,x)\in\mathcal{M}^t$ for all $x\in\mathbb{R}$.} \\
Using \eqref{finite_CJ},  we have that $J_t(\cdot,x)<+\infty$ $\mathcal{Q}^t$-q.s. Recalling the definitions of $J_t$ and $\widetilde{\jmath}_t$ (see \eqref{J_t_eq} and \eqref{tilde_j_t}), $\widetilde{\jmath}_t(\w^t,x,0)=J_t(\w^t,x)$. So, using \eqref{temp_J_qeps} with $h=\overline{h}=0$ and $\epsilon=1$, we see that there exists some $
q_{t+1}\in SK_{t+1}$ such that for all $\w^t\in \Omega^t$, $q_{t+1}(\cdot|\w^t)\in\mathcal{Q}_{t+1}(\w^t)$ and for all $\w^t$ in the $\mathcal{Q}^t$-full-measure set where $J_t(\cdot,x)<+\infty$,
\begin{eqnarray}
\mathbb{E}_{q_{t+1}(\cdot|\w^t)} J_{t+1}(\w^t,\cdot,x) \geq J_t(\w^t,x)-1.
\label{jt_temp_w}
\end{eqnarray}
By the induction hypothesis $J_{t+1}(\cdot,x) \in \mathcal{M}^{t+1}$ and Lemma \ref{M_hered} shows that $\w^t\mapsto \mathbb{E}_{q_{t+1}(\cdot|\w^t)} J_{t+1}(\w^t,\cdot,x)$ belongs to $\mathcal{M}^t$. Proposition \ref{CJ_t} shows that $J_t(\cdot,x)$ is non-negative and  $\mathcal{B}_c(\Omega^t)$-measurable. So, \eqref{jt_temp_w} and Lemma \ref{lemme_fdmt_W} ensure that $J_t(\cdot,x)\in\mathcal{M}^t$. A very similar reasoning shows that $C_t\in\mathcal{M}^t$.\\

\noindent\textit{Heredity step 2: $l_{t}^P\in\mathcal{M}^{t}(P)$, $\widetilde{\Omega}^{t,P}$ is a $P^t$-full-measure set and $(U_t^P)^+(\cdot,1)\in\mathcal{M}^t(P)$.}\\
We first show that $l_t^P\in\mathcal{M}^t(P)$. Let $\theta\in\{-1,1\}^d$. Using \eqref{finite_CJ}, $C_{t+1}<+\infty$ $\mathcal{Q}^{t+1}$-q.s. So, Cauchy-Schwarz inequality and \eqref{elas_gammafPt} at time $t+1$ in Proposition \ref{U_t_well} show  that $\mathcal{Q}^{t+1}$-q.s. 
\begin{eqnarray}
(U_{t+1}^P)^+(\cdot,1+\theta \Delta S_{t+1}(\cdot))&\leq&  (U_{t+1}^P)^+(\cdot,1+|\theta ||\Delta S_{t+1}(\cdot)|)\nonumber\\
&\leq & \left(1+\sqrt{d}|\Delta S_{t+1}(\cdot)|\right)^\gamma((U_{t+1}^P)^+(\cdot,1)+C_{t+1}(\cdot)).\label{temp_iCP}
\end{eqnarray}
As $U_{t+1}^P$ is $\mathcal{B}_c(\Omega^t\times \mathbb{R})$-measurable (see Proposition \ref{U_t_well}) and Assumption \ref{S_borel} holds true, $(U_{t+1}^P)^+(\cdot,1+\theta \Delta S_{t+1}(\cdot))$ is also $\mathcal{B}_c(\Omega^t)$-measurable. Now, recalling that $(U_{t+1}^P)^+(\cdot,1)\in\mathcal{M}^{t+1}(P)$ and $C_{t+1}\in\mathcal{M}^{t+1}$ by the induction hypothesis and that $|\Delta S_{t+1}|\in\mathcal{W}^{t+1}$ by assumption of Theorem \ref{optimality_M_typeA}, 
we deduce from \eqref{temp_iCP} and Lemma \ref{lemme_fdmt_W} that $(U_{t+1}^P)^+(\cdot,1+\theta \Delta S_{t+1}(\cdot))\in\mathcal{M}^{t+1}(P)$. Thus, 
Lemma \ref{M_hered} (recall Assumption \ref{analytic_graph}) shows that 
$\w^t\mapsto \mathbb{E}_{q_{t+1}^P(\cdot|\w^t)} (U_{t+1}^P)^+(\w^t,\cdot,1+\theta \Delta S_{t+1}(\w^t,\cdot))$ belongs to $\mathcal{M}^t(P)$ for all $\theta\in\{-1,1\}^d$. So,  $l_t^P\in\mathcal{M}^t(P)$, see \eqref{l_tP} and Lemma \ref{lemme_fdmt_W}.\\
We now prove that the set $\widetilde{\Omega}^{t,P}$ is a $P^t$-full-measure set. By the induction hypothesis, $N_t^P\in\mathcal{M}^t(P)$, which implies that $N_t^P<+\infty\; P^t-\mbox{a.s.}$ Lemma \ref{simi_qt_na} shows that $\Omega^{t,P}_{qNA}$ is a $\mathcal{Q}^{t}$-full-measure set and also a $P^t$-full-measure set as $\mathcal{H}^T \subset \mathcal{Q}^T$. Thus, $\Omega^{t,P}_{\ref{pb_inequality}}$ is a $P^t$-full-measure set. 
Moreover, using the first part of Proposition \ref{U_t^-&C}, which do not require Assumption \ref{U0}, we also have that $\Omega_{i}^{t,P}$ is a $P^t$-full-measure set for all $i \in \{\ref{P^*}, \ref{V}, \ref{integ_V-},  \ref{AE_one}\}$. As $(U_{t+1}^{P})^+(\cdot,1)\in \mathcal{M}^{t+1}(P)$, Assumption \ref{analytic_graph} and Lemma \ref{M_hered} show that $\w^t\mapsto \mathbb{E}_{q_{t+1}^P(\cdot|\w^t)} (U_{t+1}^P)^+(\w^t,\cdot,1)$ belongs to $\mathcal{M}^t(P)$. As a result, $\Omega^{t,P}_{\ref{integ_V+}}$ and also $\widetilde{\Omega}^{t,P}$ are $P^t$-full-measure sets. Thus, we can find a $P^t$-full-measure set $\widehat{\Omega}^{t,P} \in \mathcal{B}_c(\Omega^t)$ such that 
$\widehat{\Omega}^{t,P}\subset \widetilde{\Omega}^{t,P}$.\\
We now turn to the proof of $(U_t^P)^+(\cdot,1)\in \mathcal{M}^t(P)$. First, we introduce a bound for the strategies. We define $K_t^P(\w^t)$ as follows. If $\w^t\in\widehat{\Omega}^{t,P}$, let $K_t^P(\w^t):= K_1(1)$ where $K_1$ is defined in Proposition \ref{sub_optimal} applied in the $P$-prior $(t+1)$ context (see Definition \ref{def_one_prior}). This is possible as $\w^t\in \widetilde{\Omega}^{t,P}$ and Assumptions  \ref{P^*} to 
\ref{pb_inequality}  are satisfied in this context. Recall that in the $P$-prior $(t+1)$ context, $\alpha^* = \alpha_{t}^P(\w^t)$, $n_0^* = N_t^P(\w^t)$, $c^*=c_t^P(\w^t)$ and $l^*=l_t^P(\w^t)$. When $\w^t\notin\widehat{\Omega}^{t,P}$, we set $K_t^P(\w^t):=1$. We now prove that $K_t^P\in\mathcal{M}^t(P)$. As $C_{t}$ and $J_{t}(\cdot,0)$ belong to $\mathcal{M}^{t}(P)$ (see Heredity step 1), we have using \eqref{ineq_ctp2}, Lemma \ref{lemma_mes_N} and Lemma \ref{lemme_fdmt_W} that $c_t^P \in \mathcal{M}^t(P)$. Recall then that $N_t^P\in\mathcal{M}^t(P)$ by the induction hypothesis and that $1/\alpha_t^P\in\mathcal{M}^t(P)$ by assumption. We have proved in the beginning of Heredity step 2 that $l_t^P\in\mathcal{M}^t(P)$. So, it remains to prove that for all $x\in\mathbb{R}$, $\w^t\mapsto \mathbb{E}_{q_{t+1}^P(\cdot|\w^t)} (U_{t+1}^P)^- (\w^t,\cdot,x)$ belongs to $\mathcal{M}^t$. This function is $\mathcal{B}_c(\Omega^t)$-measurable (see Lemma \ref{lemma_mes_u}). As $(U_{t+1}^P)^- \leq U_{t+1}^-  \leq J_{t+1} $ (see Proposition \ref{U_t_well}), we have that for all $\w^t\in\Omega^t$ and $x\in\mathbb{R}$,  
\begin{eqnarray}
\mathbb{E}_{q_{t+1}^P(\cdot|\w^t)} (U_{t+1}^P)^-(\w^t,\cdot,x) &\leq & \sup_{p\in\mathcal{Q}_{t+1}(\w^t)}\mathbb{E}_p J_{t+1}(\w^t,\cdot,x) = J_t(\w^t,x),\label{tempjtututp}
\end{eqnarray}
see \eqref{J_t_eq} for the last equality. As $J_{t}(\cdot,x)\in\mathcal{M}^t$ for all $x\in\mathbb{R}$ (see Heredity step 1), we get that $\w^t\mapsto \mathbb{E}_{q_{t+1}^P(\cdot|\w^t)} (U_{t+1}^P)^-(\w^t,\cdot,x)$ belongs to $\mathcal{M}^t$ by Lemma \ref{lemme_fdmt_W}. So, we deduce (again from Lemma  \ref{lemme_fdmt_W}) that $K_t^P \in \mathcal{M}^t(P)$. We now prove that $(U_t^P)^+(\cdot,1)\in \mathcal{M}^t(P)$. As already mentioned, for all $\w^t\in\widehat{\Omega}^{t,P}$,  Assumptions  \ref{P^*} to 
\ref{pb_inequality}  are satisfied in the $P$-prior $(t+1)$ context. We can apply \eqref{v_v_q} in Proposition \ref{existence_uni} and \eqref{eq_value_valboun} in Proposition \ref{sub_optimal} both in the $P$-prior $(t+1)$ context and
\begin{eqnarray*}
U_t^{P}(\w^t,1)&=& \sup_{h\in\mathbb{R}^d} \mathbb{E}_{q_{t+1}^P(\cdot|\w^t)}U_{t+1}^{P}(\cdot,1+h \Delta S_{t+1}(\w^t,\cdot)) =\sup_{|h|\leq K_t^P(\w^t)} \mathbb{E}_{q_{t+1}^P(\cdot|\w^t)}U_{t+1}^{P}(\cdot,1+h \Delta S_{t+1}(\w^t,\cdot)). 
\end{eqnarray*}
So, as $U_{t+1}^{P}(\w^t,\cdot)$ is nondecreasing, we get for all $\w^t\in\widehat{\Omega}^{t,P}$ that
\begin{eqnarray}
(U_t^{P})^+(\w^t,1)& \leq & \mathbb{E}_{q_{t+1}^P(\cdot|\w^t)}(U_{t+1}^{P})^+(\w^t,\cdot,1+K_t^P(\w^t) |\Delta S_{t+1}(\w^t,\cdot)|).\label{temp_Ut+1Ut}
\end{eqnarray}
Using again (\ref{elas_gammafPt}), we get that, 
\begin{eqnarray*}
(U_{t+1}^{P})^+(\cdot,1+K_t^P(\cdot) |\Delta S_{t+1}(\cdot)|) \leq \left(1+K_t^P(\cdot) |\Delta S_{t+1}(\cdot)|\right)^\gamma((U_{t+1}^{P})^+(\cdot,1)+C_{t+1}(\cdot)).
\end{eqnarray*}
Recalling that $K_t^P\in\mathcal{M}^t(P)$, that $|\Delta S_{t+1}(\cdot)|\in\mathcal{W}^{t+1}$ by assumption of Theorem \ref{optimality_M_typeA} and that $C_{t+1}\in\mathcal{M}^{t+1}$ and $(U_{t+1}^{P})^+(\cdot,1)\in\mathcal{M}^{t+1}(P)$ from the induction hypothesis, we get that $(U_{t+1}^{P})^+(\cdot,1+K_t^P(\cdot) |\Delta S_{t+1}(\cdot)|)\in\mathcal{M}^{t+1}(P)$, see Lemma \ref{lemme_fdmt_W}. Indeed, as $U_{t+1}^P$ is $\mathcal{B}_c(\Omega^{t+1}\times \mathbb{R})$-measurable (see Proposition \ref{U_t_well}) and $K_t^P$ is $\mathcal{B}_c(\Omega^t)$-measurable, \citep[Proposition 7.44, p172]{ref1} shows that $(U_{t+1}^{P})^+(\cdot,1+K_t^P(\cdot) |\Delta S_{t+1}(\cdot)|)$ is $\mathcal{B}_c(\Omega^t)$-measurable. Thus, as $\widehat{\Omega}^{t,P}$ is a $P^t$-full-measure set, Lemmata \ref{lemme_fdmt_W} and  \ref{M_hered}  and \eqref{temp_Ut+1Ut} imply that $(U_t^{P})^+(\cdot,1)\in\mathcal{M}^t(P)$.\\

\noindent\textit{Heredity step 3: $N_{t-1}^P\in\mathcal{M}^{t-1}(P)$}.\\ 
If $t=0$, we trivially have that $N_{-1}^P=0\in\mathcal{M}^{-1}(P) = \{0\}$. So, assume that $t> 0$. Recall from Heredity steps 1 and 2 that $l_t^P\in\mathcal{M}^t(P)$, $C_t$, $J_t(\cdot,0)\in\mathcal{M}^t\subset \mathcal{M}^t(P)$, from the induction hypothesis that $N_t^P\in\mathcal{M}^t(P)$ and from assumption of Theorem \ref{optimality_M_typeA} that $1/\alpha_{t-1}^P\in \mathcal{M}^{t-1}\subset \mathcal{M}^{t-1}(P)$ and $1/\alpha_{t}^P\in \mathcal{M}^{t}\subset \mathcal{M}^{t}(P)$. Thus, assertion (A2) in Lemma \ref{lemma_N_finite} for $1\leq t\leq T-1$ (recall from Heredity step 2 that $\widetilde{\Omega}^{t,P}$ is a $P^t$-full-measure set for all $P\in\mathcal{H}^T$) shows that $N_{t-1}^P\in\mathcal{M}^{t-1}(P).$\\ 

\noindent This concludes the heredity step and we can now apply Theorem \ref{one_step_strategy}.\\
\noindent\textit{Application of Theorem \ref{one_step_strategy} and proof of $\phi^{*,x}\in \Phi(x,U,\mathcal{Q}^T)$.}\\
Let $x\in\mathbb{R}$. As the $NA(\mathcal{Q}^T)$ condition as well as Assumptions \ref{S_borel}, \ref{analytic_graph}, \ref{AE}, \ref{nncst}, \ref{simple_adm} and \ref{U0} hold true, we can apply Theorem \ref{one_step_strategy} and there exists $\phi^{*,x}\in \Phi$ such that \eqref{one_step_strategy_temp2} is true. We prove now that $\phi^{*,x}\in \Phi(x,U,\mathcal{Q}^T)$ i.e. that for all $P\in\mathcal{Q}^T$, $\mathbb{E}_P\; U^-(\cdot, V_T^{x,\phi^{*,x}}(\cdot))<+\infty$. As \eqref{ineq_b_inf_det_A} holds true and $C_1\in\mathcal{W}^T$, we only need to check that $V_T^{x,\phi^{*,x}}\in\mathcal{M}^{T}(P)$ for all $P\in\mathcal{Q}^T$. 
Fix $0\leq t \leq T-1$ and recall $\widehat{\Omega}^t$ from Proposition \ref{U_t^-&C}. As for all $\w^t\in\widehat{\Omega}^{t}\subset \widetilde{\Omega}^t$, Assumptions  \ref{P^*} to 
\ref{pb_inequality}  are satisfied in the robust $(t+1)$ context (see Lemma \ref{lemma_Assumption_true} and Definition \ref{def_one_prior}), 
we can apply Proposition \ref{sub_optimal} in the robust $(t+1)$ context and we have for all $\w^t\in\widehat{\Omega}^{t}$ that
$|\phi^{*,x}_{t+1}(\w^t)|\leq K_t(\w^t),$
where for $\w^t\in\widehat{\Omega}^t$, $K_t(\w^t):= K_1(x)$ with $K_1(x)$ defined in Proposition \ref{sub_optimal} in the robust $(t+1)$ context and $K_t(\w^t):=1$ when $\w^t\notin\widehat{\Omega}^{t}$. So, as $\widehat{\Omega}^t$ is a $\mathcal{Q}^t$-full-measure set,
\begin{eqnarray}
|V_T^{x,\phi^{*,x}}|\leq |x|+\sum_{t=0}^{T-1} K_{t}|\Delta S_{t+1}|\; \mathcal{Q}^T-\mbox{q.s.} \label{temp_ineq_V}
\end{eqnarray}
Assume for a moment that for all $P:=q_1^P\otimes\cdots \otimes q_T^P \in\mathcal{Q}^T$ and $0\leq t \leq T-1$, $K_{t}\in \mathcal{M}^{t}(P)$. As for all $0\leq t\leq T-1$, $|\Delta S_{t+1}|\in\mathcal{W}^{t+1}$, we get using \eqref{temp_ineq_V} and Lemma \ref{lemme_fdmt_W} that $V_T^{x,\phi^{*,x}}\in\mathcal{M}^{T}(P)$ for all $P\in\mathcal{Q}^T$. Thus, $\phi^{*,x}\in\Phi(x,U,\mathcal{Q}^T)$ and Theorem \ref{optimality_pp} ensures that \eqref{eg_fdmt} holds true, which concludes the proof of of Theorem \ref{optimality_M_typeA}.\\
Now, we prove that $K_t\in\mathcal{M}^t(P)$. 
For that, we use $\widehat{P}_{t+1} \in\mathcal{H}^T$ defined in \eqref{hatP_lemmaN} in the Appendix. Note that Lemma \ref{lemma_inc_M} shows that $\mathcal{M}^t(\widehat{P}_{t+1})\subset \mathcal{M}^t(P)$. Recall that in the robust $(t+1)$ context, $\alpha^* = \alpha_{t}^{P^*}(\w^{t})$, $n_0^* = N_{t}^*(\w^{t})$, $c^* = c_{t}^{P^*}(\w^{t})$ and $l^* = l_{t}^{*}(\w^{t})$. By assumption of Theorem \ref{optimality_M_typeA}, $1/\alpha_t^{P^*}\in\mathcal{M}^t \subset \mathcal{M}^t(P)$ and also  $1/\alpha_t^{P^*}\in\mathcal{M}^t(\widehat{P}_{t+1})$.  We first prove that $l_{t}^*\in \mathcal{M}^{t}(\widehat{P}_{t+1})\subset \mathcal{M}^{t}(P)$. Using that $U_{t}\leq U_{t}^{\widehat{P}_{t+1}}$ (see Proposition \ref{U_t_well}), we have that $l_{t}^*\leq l_{t}^{\widehat{P}_{t+1}}$. 
So, as $l_{t}^{\widehat{P}_{t+1}}\in\mathcal{M}^{t}(\widehat{P}_{t+1})$ (see Heredity step 2) and $l_{t}^*$ is $\mathcal{B}_c(\Omega^t)$-measurable (see Lemma \ref{lemma_mes_N}), we  get that $l_{t}^*\in \mathcal{M}^{t}(\widehat{P}_{t+1})$ by Lemma \ref{lemme_fdmt_W}. 
We now prove that $N_{t}^*\in\mathcal{M}^{t}(P)$. 
Recall assertions (B1) and (B2) from Lemma \ref{lemma_N_finite}. Assertion (B1) applies if $t=T-1$ and thus shows that $N_{T-1}^*\in\mathcal{M}^{T-1}(P)$ as $1/\alpha_{T-1}^{P^*} \in \mathcal{M}^{T-1}\subset \mathcal{M}^{T-1}(\widehat{P}_{T})$ and $\underline{X}$, $1/|U(\cdot,\underline{X}(\cdot)) + C(\cdot)|$, $C$, $U^-(\cdot,0)$ belong to $\mathcal{M}^T(\widehat{P}_{T})$, see initialization step for $P=\widehat{P}_T$. 
Assertion (B2) applies and shows that $N_t^* \in \mathcal{M}^t(P)$ as $\widetilde{\Omega}^{t+1,P}$ is a $P^{t+1}$-full-measure set for all $P\in\mathcal{H}^T$, $1/\alpha_{t}^{P^*}\in\mathcal{M}^{t}\subset \mathcal{M}^{t}(\widehat{P}_{t+1})$, $1/\alpha_{t+1}^{\widehat{P}_{t+1}}$, $C_{t+1}$, $J_{t+1}(\cdot,0) \in\mathcal{M}^{t+1} \subset \mathcal{M}^{t+1}(\widehat{P}_{t+1})$ and  $l_{t+1}^{\widehat{P}_{t+1}}$, $N_{t+1}^{\widehat{P}_{t+1}}\in \mathcal{M}^{t+1}(\widehat{P}_{t+1})$, see Heredity steps $2$ and $3$ for $P=\widehat{P}_{t+1}$. Now, as $U_{t+1}$ is lsa (see Proposition \ref{U_t_well}), Lemma \ref{lemma_mes_u} (iii) with $f=U_{t+1}^-$ alongside \citep[Lemma 7.30 (3), p177]{ref1} show that $\w^{t}\mapsto \sup_{p\in\mathcal{Q}_{t+1}(\w^{t})}\mathbb{E}_p U_{t+1}^- (\w^{t},\cdot,-x^-)$ is usa and so $\mathcal{B}_c(\Omega^{t})$-measurable. We deduce then from \eqref{J_t_eq}, $U_{t+1}^-\leq J_{t+1}$ (see Proposition \ref{U_t_well}) and the fact that $J_{t}(\cdot,-x^-)\in\mathcal{M}^{t}$ (see Heredity step 1) that $\w^{t}\mapsto \sup_{p\in\mathcal{Q}_{t+1}(\w^{t})}\mathbb{E}_p U_{t+1}^- (\w^{t},\cdot,-x^-)$ belongs to $\mathcal{M}^{t}$, see Lemma \ref{lemme_fdmt_W}. 
Finally, as $C_{t+1}$ and $J_{t+1}(\cdot,0)$ belong to $\mathcal{M}^{t+1}$, $c_{t}^{P^*}\in\mathcal{M}^t$ using \eqref{ct^*} and Lemma \ref{M_hered}. Thus, recalling again that $1/\alpha_{t}^{P^*} \in \mathcal{M}^{t}$, we deduce that $K_{t}\in\mathcal{M}^{t}(P)$. \Halmos

\section{Appendix}
\label{appendix}
The first part of the appendix shows some properties of the asymptotic elasticity. We prove that Assumption \ref{AE} holds true for a deterministic utility function under the condition of Reasonable Asymptotic Elasticity in discrete time (see Proposition \ref{annex_AE2}).
In the second part, we introduce the concept of $(-\infty)$ integrals that plays an important role to prove that concavity is preserved through dynamic programming (see Remark \ref{remark_assumpV-}). We show that several measurability properties are still true for $(-\infty)$ integration (see Proposition \ref{lsa_cvt} and Lemma \ref{lemma_mes_u}). In the third part, we give a sufficient condition so that a probability measure with a given disintegration is dominated by another (see Proposition \ref{domin_lcl}). This is used to extrapolate some properties from $\mathcal{H}^T$ to $\mathcal{Q}^T$. The next part provides the missing results of Section \ref{sec ilus}. We give properties on $\mathcal{W}^t$, $\mathcal{M}^t$ and $\mathcal{M}^t(P)$ (see Lemmata \ref{lemma_inc_M}, \ref{lemme_fdmt_W} and \ref{M_hered}). We show that a random utility with benchmark is of type (A) (see Proposition \ref{utility_bench_is_typeA}). We also prove Lemma \ref{lemma_temp_exemple} which is used to provide an example of application of Theorem \ref{optimality_M_typeA}. 
Finally, the last part collects the missing proofs and results of Section \ref{muti_per}. Proposition \ref{U_t_well} is proved and two technical lemmata and their proofs are given. Lemma \ref{lemma1+} ensures that Assumption \ref{integ_V+} is preserved through dynamic programming while Lemma \ref{lemma_N_finite} shows that $N_t^*$ (see \eqref{eq_Nt}) and $N_t^P$ (see \eqref{eq_NtP}) are almost-surely finite and may be integrable. 

\subsection{Asymptotic elasticity}
\label{AE_app}
First, we show that a concave deterministic function $U$ satisfies the growth condition \eqref{elas_gammaf} in Assumption \ref{AE} with $\gamma=1$. Then, we prove that such a function $U$ satisfies $\mbox{AE}_{+\infty}(U)\leq 1$ and $\mbox{AE}_{-\infty}(U)\geq 1$ and finally that under the Reasonable Asymptotic Elasticity condition in discrete time, Assumption \ref{AE} holds true.

\begin{lemma}
\label{lemmaAE1}
Let $U$ : $\mathbb{R}\to \mathbb{R}\cup\{-\infty,+\infty\}$ be a concave function with $U^-(0)<+\infty$. Then, for all $\lambda\geq 1$ and $x\in\mathbb{R}$,
\begin{eqnarray}
U(\lambda x)\leq \lambda (U(x) + U^-(0)).\label{AE_gamma1}
\end{eqnarray}
\end{lemma}
\proof{Proof.}
Let $\lambda\geq 1$ and $x\in\mathbb{R}$. Using the concavity of $U$, we get that  
$\frac{1}{\lambda} U(\lambda x) + \left(1-\frac{1}{\lambda}\right)U(0) \leq U(x).$ 
As $U^-(0)<+\infty$, we have that 
\begin{eqnarray*}
U(\lambda x) &\leq &  \lambda U(x) - (\lambda-1)U(0) \leq \lambda U(x)+ (\lambda -1)U^-(0)\leq  \lambda U(x) + \lambda U^-(0).\Halmos
\end{eqnarray*}
\endproof

\begin{proposition}
Let $U$ : $\mathbb{R}\to \mathbb{R}$ be a concave, nondecreasing, non-constant and continuously differentiable function such that $\lim_{x\rightarrow +\infty}U(x)>0$. Then,  $\mbox{AE}_{+\infty}(U)\leq 1$ and $\mbox{AE}_{-\infty}(U)\geq 1$.
\label{annex_AE1}
\end{proposition}
\proof{Proof.}
The fact that $\mbox{AE}_{+\infty}(U)\leq 1$ is proved in \citep[Lemma 6.1]{refAE1}. We show analogously that $\mbox{AE}_{-\infty}(U)\geq 1$. Let $x< -1$, using the concavity of $U$, we get that
\begin{eqnarray*}
xU'(x) &=& (x+1)U'(x) - U'(x)\leq  U(x)-U(-1)-U'(x)\leq  U(x)-U(-1)-U'(-1).
\end{eqnarray*}
Observe now that as $U$ is concave, nondecreasing and non-constant,  $\lim_{x\to -\infty} U(x)=-\infty$. So, for $x$ small enough, $U(x)< 0$ and  
$\frac{xU'(x)}{U(x)} \geq 1- \frac{U(-1)+U'(-1)}{U(x)}.$ We conclude as $\lim_{x\to -\infty} U(x)=-\infty$. 
\Halmos \endproof

\begin{proposition}
Let $U$ : $\mathbb{R}\to \mathbb{R}$ be a concave, nondecreasing, non-constant and continuously differentiable function. Assume that either $\mbox{AE}_{+\infty}(U)< 1$ and $\lim_{x\to +\infty}U(x)>0$ or $\mbox{AE}_{-\infty}(U)>1$. Then, there exist some $\gamma\neq 1$ and some $C>0$ such that for all $\lambda \geq 1$ and $x\in\mathbb{R}$,
\begin{eqnarray}
U(\lambda x)\leq \lambda^{\gamma}(U(x)+C).
\label{det_AE}
\end{eqnarray}
Moreover, $\gamma$ and $C$ can be specified as follows.\\
\noindent Assume that $\mbox{AE}_{-\infty}(U)>1$. Let $\gamma\in (1, \mbox{AE}_{-\infty}(U))$. Then, there exists $\underline{x}<0$ such that $U(\underline{x})<0$ and $\frac{x U'(x)}{U(x)}>\gamma$ for all $x\leq \underline{x}$. Now, choosing such an $\underline{x}$ and setting $C:= U^+(0)+U^-(0)+U^-(\underline{x})$, $\gamma$ and $C$ satisfy \eqref{det_AE}.\\
\noindent Assume that $\mbox{AE}_{+\infty}(U)<1$ and $\lim_{x\to +\infty}U(x)>0$. Let $\gamma \in (\mbox{AE}_{+\infty}(U),1)$. Then, there exist $x'<0$ and $\overline{x}>0$ such that $U(x')<0$, $U(\overline{x})>0$ and $\frac{x U'(x)}{U(x)}<\gamma$ for all $x\geq \overline{x}$. Now, choosing such an $x'$ and $\overline{x}$ and setting $C:= U^+(\overline{x})+U^-(x')+U^-(0)$, $\gamma$ and $C$ satisfy \eqref{det_AE}.
\label{annex_AE2}
\end{proposition}
\proof{Proof.}
Assume first that $AE_{-\infty}(U)>1$ and choose $\gamma\in(1, \mbox{AE}_{-\infty}(U))$. We prove that there exists a real number $\underline{x}<0$ such that for all $\lambda\geq 1$ and $x\leq \underline{x}$, 
\begin{eqnarray}
U(\lambda x)< \lambda^\gamma U(x).
\label{temp_AE_ASSUMP_gamma}
\end{eqnarray}
The proof is very similar to the proof of \citep[Lemma 6.3]{refAE1}. 
As $U$ is non-constant, nondecreasing and concave, we have that $\lim_{x\to -\infty}U(x)=-\infty$. Moreover, $\underset{x\rightarrow -\infty}{\lim \inf} \frac{xU'(x)}{U(x)}>\gamma$. So, there exists (and we fix) some $\underline{x}<0$ such that $U(\underline{x})<0$ and $\frac{x U'(x)}{U(x)}>\gamma$ for all $x\leq \underline{x}$. Now, for all $x\leq \underline{x}$, as $U(x)\leq U(\underline{x})<0$, we have that $x U'(x)<\gamma U(x).$ \\
Let $x\leq \underline{x}$. Let $F$ and $G$ be defined by $F(\lambda):=U(\lambda x)$ and $G(\lambda):=\lambda^\gamma U(x)$ for $\lambda\geq 1$. Then, $F$ and $G$ are differentiable and $F'(\lambda)=x U'(\lambda x)$ and $G'(\lambda)=\gamma \lambda^{\gamma-1} U(x)$. 
So, we find that 
$$F'(1)=x U'(x)<\gamma U(x) =G'(1).$$ As $F(1)=G(1)$ and $F'(1)<G'(1)$, there exists $\epsilon>0$ such that for all $\lambda \in (1,1+\epsilon)$, $F(\lambda)<G(\lambda)$. Let $\widehat{\lambda}:= \inf \{\lambda\geq 1, F(\lambda)=G(\lambda)\}$. Assume that $\widehat{\lambda}<+\infty$. By continuity of $F$ and $G$, $F(\widehat{\lambda})=G(\widehat{\lambda})$. Now remark that, by definition of $\widehat{\lambda}$, we have that 
\begin{eqnarray}
0\leq \lim_{h\to 0^+}\frac{F(\widehat{\lambda}-h)-G(\widehat{\lambda}-h)}{-h}= F'(\widehat{\lambda})-G'(\widehat{\lambda}).
\label{temp_AE_det}
\end{eqnarray}
However, as $\widehat{\lambda}x\leq \underline{x}$, 
$F(\widehat{\lambda})=G(\widehat{\lambda})$ shows that 
$$F'(\widehat{\lambda})= x U'(\widehat{\lambda} x)  < \frac{\gamma}{\widehat{\lambda}}U(\widehat{\lambda} x) = \frac{\gamma}{\widehat{\lambda}} F(\widehat{\lambda}) = \frac{\gamma}{\widehat{\lambda}}G(\widehat{\lambda}) = G'(\widehat{\lambda}),$$ 
which contradicts \eqref{temp_AE_det}. Thus, $\widehat{\lambda}=+\infty$ and \eqref{temp_AE_ASSUMP_gamma} is true. Now, we show that \eqref{det_AE} holds true. Set $C:=U^+(0)+U^-(0)+U^-(\underline{x})$. Let $\lambda\geq 1$. Assume first that $x> 0$. Then, $U(x)+U^-(0)\geq 0$. Using \eqref{AE_gamma1} and $\gamma>1$, we have that 
$U(\lambda x)\leq \lambda (U(x)+U^-(0)) \leq \lambda^{\gamma}(U(x)+C).$\\ Assume now that $\underline{x}<x\leq 0$. We have that $$\lambda^{\gamma}(U(x)+C)\geq \lambda^\gamma (U(\underline{x})+C)=\lambda^\gamma (U^+(\underline{x})+U^+(0)+U^-(0))\geq U^+(0)\geq U(0)\geq U(\lambda x),$$ as $x\leq 0$. 
Now, for $x\leq \underline{x}$, \eqref{temp_AE_ASSUMP_gamma} implies \eqref{det_AE}.\\ 
Assume now that $AE_{+\infty}(U)<1$ and choose $\gamma\in (AE_{+\infty}(U),1)$. Furthermore, assume that $\lim_{x\to +\infty} U(x)>0$. Then, there exist (and we fix) some $\overline{x}>0$ such that $U(\overline{x})>0$ and $\frac{x U'(x)}{U(x)}<\gamma$ for all $x>\overline{x}$. Then, \citep[Lemma 6.3]{refAE1} shows that for all $\lambda\geq 1$ and $x\geq \overline{x}$,
\begin{eqnarray}
U(\lambda x)<\lambda^{\gamma} U(x).
\label{temp_AE_det2}
\end{eqnarray}
As $\lim_{x\to -\infty} U(x)=-\infty$, there exists (and we fix) some $x'<0$ such that $U(x')<0$. Let $C:=U^+(\overline{x})+U^-(x')+U^-(0).$ We show that \eqref{det_AE}  holds true. Let $\lambda\geq 1$. If $x\geq \overline{x}$, \eqref{det_AE} follows directly from \eqref{temp_AE_det2}. Assume now that $x<\overline{x}$. As $U$ is nondecreasing, $U(\lambda x)\leq U(\lambda \overline{x})\leq \lambda^{\gamma}U(\overline{x}).$ So, if $x'<x< \overline{x}$, 
$$\lambda^{\gamma}(U(x)+C) \geq \lambda^{\gamma}(U(x\rq{})+C) = \lambda^{\gamma}(U^+(x')+U^+(\overline{x})+U^-(0))\geq \lambda^{\gamma}U(\overline{x})
\geq U(\lambda x)$$ and \eqref{det_AE} holds true. Now, if $x\leq x'$, as $U$ is nondecreasing, $U(\lambda x)<0$. We distinguish two cases. If $U(x)+C\geq 0$, then \eqref{det_AE} holds true trivially. If $U(x)+C<0$, recalling \eqref{AE_gamma1}, 
$U(\lambda x)\leq \lambda (U(x)+U^-(0))\leq \lambda (U(x)+C)$ and \eqref{det_AE} follows as $\gamma<1$. \Halmos \endproof

\subsection{The $(-\infty)$-convention}
\label{inf_cvt_sub}
Let $X$ be a Borel space (see \citep[Definition 7.7]{ref1}). Let $f : X \to \mathbb{R}\cup\{-\infty,+\infty\}$ be a $\mathcal{B}_c(X)$-measurable function and let $p\in \mathfrak{P}(X)$. 
We define the $(-\infty)$ integral, denoted by $\int_{-} f dp$, and the $(+\infty)$ integral, denoted by $\int^{-} f dp$, as follows. If $\int f^+ dp<+\infty$ or $\int f^- dp<+\infty$, both integrals are equal and are defined as the extended integral of $f$ i.e.
\begin{eqnarray}
\int_{-} f dp = \int^{-} f dp:=\int f^+ dp - \int f^- dp. \label{base_integral}
\end{eqnarray}
Otherwise, $\int_{-} f dp := -\infty \;\;\mbox{and}\;\; \int^{-} f dp := +\infty.$
We adopt the usual arithmetic rules in calculations involving $+\infty$ and $-\infty$ described in \citep[p26-27]{ref1} except that we assume that $+\infty-\infty = -\infty+\infty =-\infty.$ In particular, we have that 
\begin{eqnarray}
+(-\infty)=-\infty\;\; \mbox{and} \;\; -(+\infty)=-\infty.\label{cvt_prod_inf}
\end{eqnarray}

We state now a first lemma that links $\int^{-}$ and $\int_{-}$.

\begin{lemma}
Let $X$ be a Borel space. Let $f : X\to \mathbb{R}\cup\{-\infty,+\infty\}$ be a $\mathcal{B}_c(X)$-measurable function and $p\in\mathfrak{P}(X)$. Then,
\begin{eqnarray}
\int^{-}(-f) dp=-\int_{-}f dp.
\label{link_integrals}
\end{eqnarray}
\label{lemma_integ_+-}
\end{lemma}
\proof{Proof.}
The function $-f$ is $\mathcal{B}_c(X)$-measurable. When $\int f^+ dp = \int (-f)^- dp<+\infty$ or $\int f^- dp=\int (-f)^+ dp<+\infty$, the usual definition through the extended integral can be used:
\begin{eqnarray}
\int^{-}(-f) dp &=& \int (-f) dp = \int (-f)^+ dp - \int(-f)^- dp = \int f^- dp - \int f^+ dp\nonumber\\ &=& -\left(\int f^+ dp - \int f^- dp \right)=-\int f dp = -\int_{-}f dp,\label{temp_integ_inf}
\end{eqnarray}
where the first equality in \eqref{temp_integ_inf} may follow from \eqref{cvt_prod_inf}.
Otherwise, $\int_{-} f dp = -\infty$ and $\int^{-} (-f) dp = +\infty.$ Thus, \eqref{link_integrals} follows again from  \eqref{cvt_prod_inf}.
\Halmos \endproof


\begin{definition}
A function $f : X \to \mathbb{R}\cup\{-\infty,+\infty\}$ is lower-semianalytic or lsa (resp. upper-semianalytic or usa) if $\{x\in X,\; f(x)\leq a\}$ (resp. $\{x\in X,\; f(x)\geq a\}$) is an analytic set for all $a\in\mathbb{R}$.
\label{lsausadef}
\end{definition}
Any lsa or usa function is $\mathcal{B}_c(X)$-measurable as $\mathcal{A}(X)\subset \mathcal{B}_c(X).$ 
As already mentioned, \citep{ref1} uses the convention $+\infty-\infty = -\infty + \infty = +\infty$ and thus the $(+\infty)$ integral. 
We now adapt \citep[Proposition  7.46, p177]{ref1} and \citep[Proposition  7.48, p180]{ref1} to our convention $+\infty - \infty = -\infty + \infty = -\infty$. 
\begin{proposition}
Let $X$ and $Y$ be Borel spaces. Let $f : X \times Y \to \mathbb{R}\cup\{-\infty,+\infty\}$ and let $q$ be a stochastic kernel on $Y$ given $X$. Let $\lambda : X  \to \mathbb{R}\cup\{-\infty,+\infty\}$ be defined by $$\lambda(x):=\int_{-} f(x,y)q(dy|x).$$

\noindent (i) Assume that $q$ is a Borel measurable stochastic kernel and that $f$ is usa. Then, $\lambda$ is usa.\\
(ii) Assume that $q$ is a Borel measurable stochastic kernel and that $f$ is lsa. Then, $\lambda$ is lsa.\\
(iii) Assume that $q$ is a universally measurable stochastic kernel and that $f$ is $\mathcal{B}_c(X\times Y)$-measurable. Then, $\lambda$ is $\mathcal{B}_c(X)$-measurable.
\label{lsa_cvt}
\label{usc_corol}
\label{univ_cvt}
\end{proposition}
\proof{Proof.}
\noindent\textit{Proof of (i).} Using \eqref{link_integrals}, we get that for all $x\in\mathbb{R}$, 
\begin{eqnarray}
-\lambda(x)= -\int_{-} f(x,y)q(dy|x) = \int^{-} (-f(x,y))q(dy|x).\label{temp_lsa_usa_int_mes}
\end{eqnarray}
As $-f$ is lsa, \citep[Proposition  7.48, p180]{ref1}, which works for the $(+\infty)$ integral,  shows that $-\lambda$ is lsa and thus that $\lambda$ is usa.\\

\noindent\textit{Proof of (ii).}
For all $n\geq 0$, we define $\lambda_{1,n}, \lambda_2 : X \to \mathbb{R}\cup\{-\infty,+\infty\}$ as follows
\begin{eqnarray*}
\lambda_{1,n}(x) &:=& \int \min(f^+(x,y),n)\;q(dy|x) \quad \mbox{and} \quad 
\lambda_{2}(x):= -\int f^-(x,y) q(dy|x) = \int (-f^-(x,y)) q(dy|x).
\end{eqnarray*}
Let $n\geq 0$. As the integrands in the definition of $\lambda_{1,n}$ and $\lambda_2$ are non-negative, the $(+\infty)$ and $(-\infty)$ integrals equal the extended integral and the last equality follows from \eqref{link_integrals}. 
First, we show that $\lambda_{1,n}$ and $\lambda_{2}$ are lsa. As $f$ is lsa, using \citep[Lemma 7.30 (2), p177]{ref1} (which proof does not rely on the convention $+\infty - \infty =+\infty$), $f^+$, $\min(f^+,n)$ and $-f^-=\min(f,0)$ are lsa. So, \citep[Proposition  7.48, p180]{ref1} shows that $\lambda_{1,n}$ and $\lambda_{2}$ are indeed lsa. Now, as $\lambda_{1,n}+\lambda_{2} \in [-\infty,n]$, for any $a\in\mathbb{R}$,
$$\{x\in X, \lambda_{1,n}(x)+\lambda_2(x)<a\}=\bigcup_{r\in\mathbb{Q}}\{x\in X, \lambda_{1,n}(x)<r\} \cap \{x\in X, \lambda_{2}(x)<a-r\}.$$
So, we get that $\lambda_{1,n}+\lambda_{2}$ is also lsa. Assume for a moment that we have proved that for all $x\in E$, 
\begin{eqnarray}
\lambda(x) =\int_{-} f(x,y)q(dy|x)=  \sup_{n\geq 0} \left(\lambda_{1,n} + \lambda_{2}\right)(x).\label{temp_eq_lbd_int}
\end{eqnarray} 
Then, using again \citep[Lemma 7.30 (2), p177]{ref1}, we deduce that $\lambda$ is lsa. Now, we prove \eqref{temp_eq_lbd_int}. Let $x\in\mathbb{R}$. Assume that $\int f^-(x,y) q(dy|x)=+\infty$. Then, $\lambda(x)=-\infty$ and $\lambda_2(x)=-\infty$, see \eqref{cvt_prod_inf}. Recalling that $0\leq \lambda_{1,n}(x)\leq n$, we get that $(\lambda_{1,n}+\lambda_2)(x)=-\infty$ for all $n\geq 0$. Thus, \eqref{temp_eq_lbd_int} holds true. Assume now that $\int f^-(x,y) q(dy|x)<+\infty$. The monotone convergence theorem shows that $\lim_{n\to +\infty} \lambda_{1,n}(x) = \sup_{n\geq 0} \lambda_{1,n}(x) = \int f^+(x,y) q(dy|x) \in [0,+\infty]$. As $\lambda_2(x)>-\infty$, we deduce from \eqref{base_integral} that $\sup_{n\geq 0} (\lambda_{1,n}+\lambda_2)(x) = \int f^+(x,y) q(dy|x) - \int f^-(x,y) q(dy|x) = \lambda(x)$.\\

\noindent\textit{Proof of (iii).} As $-f$ is $\mathcal{B}_c(X)-$measurable, \eqref{temp_lsa_usa_int_mes} and \citep[Proposition  7.46, p177]{ref1} (which again works for the $(+\infty)$ integral) shows that $-\lambda$ is $\mathcal{B}_c(X)$-measurable. So, $\lambda$ is $\mathcal{B}_c(X)$-measurable.\Halmos \endproof

This lemma is a direct application of Proposition \ref{lsa_cvt} and allows to solve measurability issues in Section \ref{muti_per}.

\begin{lemma}
Assume that Assumption \ref{S_borel} holds true. 
Let $0\leq t\leq T-1$. Let $f : \Omega^{t+1}\times \mathbb{R} \to \mathbb{R}\cup\{-\infty,+\infty\}$ and define $\lambda : \Omega^t\times \mathbb{R}\times \mathbb{R}^d \times \mathfrak{P}(\Omega_{t+1}) \to \mathbb{R}\cup\{-\infty,+\infty\}$ and $\widetilde{\lambda}_{\inf}, \widetilde{\lambda}_{\sup} : \Omega^t\times \mathbb{R}\times \mathbb{R}^d \to \mathbb{R}\cup\{-\infty,+\infty\}$ as follows
\begin{eqnarray*}
\lambda(\w^t,x,h,p)&:=& \int_{-} f(\w^t,\w_{t+1},x+h\Delta S_{t+1}(\w^t,\w_{t+1})) p(d\w_{t+1})\\
\widetilde{\lambda}_{\inf}(\w^t,x,h)&:=&\inf_{p\in\mathcal{Q}_{t+1}(\w^t)}\lambda(\w^t,x,h,p) \quad \mbox{and} \quad \widetilde{\lambda}_{\sup}(\w^t,x,h):=\sup_{p\in\mathcal{Q}_{t+1}(\w^t)}\lambda(\w^t,x,h,p).
\end{eqnarray*}
(i) If $f$ is $\mathcal{B}_c(\Omega^{t+1}\times \mathbb{R})$-measurable, then $\lambda$ is $\mathcal{B}_c(\Omega^t\times \mathbb{R}\times \mathbb{R}^d \times \mathfrak{P}(\Omega_{t+1}))$-measurable.\\
Assume furthermore that Assumption \ref{analytic_graph} holds true.\\
(ii) If $f$ is lsa, then $\lambda$ and $\widetilde{\lambda}_{\inf}$ are lsa.\\
(iii) If $f$ is usa, then $\lambda$ and $\widetilde{\lambda}_{\sup}$ are usa.
\label{lemma_mes_u}
\end{lemma}
\proof{Proof.}
Let $g : (\w^t,x,h,p,\w_{t+1})\mapsto f(\w^t,\w_{t+1},x+h \Delta S_{t+1}(\w^t,\w_{t+1}))$ and consider the Borel measurable stochastic kernel $q$ defined by $q(\cdot|\w^t,x,h,p):=p(\cdot)$.
For (i), remark that as $f$ is $\mathcal{B}_c(\Omega^{t+1}\times \mathbb{R})$-measurable and Assumption \ref{S_borel} holds true, $g$ is $\mathcal{B}_c(\Omega^t\times \mathbb{R}\times \mathbb{R}^d \times \mathfrak{P}(\Omega_{t+1})\times \Omega_{t+1})$-measurable and Proposition \ref{univ_cvt} (iii) shows that $\lambda$ is $\mathcal{B}_c(\Omega^t\times \mathbb{R}\times \mathbb{R}^d \times \mathfrak{P}(\Omega_{t+1}))$-measurable. For (ii), assume now that $f$ is lsa. As Assumption \ref{S_borel} holds true, \citep[Lemma 7.30 (3), p177]{ref1} shows that $g$ is lsa. So, Proposition \ref{lsa_cvt} (ii) with the Borel measurable stochastic kernel $q$ shows that $\lambda$ is lsa. Assumption \ref{analytic_graph} provides that $\mbox{proj}_{\Omega^t}(\mbox{graph}(\mathcal{Q}_{t+1}))=\Omega^t$ as $\mathcal{Q}_{t+1}\neq \emptyset$. It also, together with \citep[Proposition 7.47, p179]{ref1}, shows that $\widetilde{\lambda}_{\inf}$ is lsa and (ii) is proved. The proof if $f$ is usa is similar and  omitted. \Halmos \endproof

\subsection{Absolute continuity}
\label{other_prop}

The following proposition is important to prove that Assumption \ref{U0} is preserved by dynamic programming. Let $p,q\in\mathfrak{P}(X),$ we write that $p\ll q$ if for all $A\in \mathcal{B}(X)$, $q(A)=0$ implies that $p(A)=0$.

\begin{proposition}
\label{domin_lcl}
Let $l\geq 1$ and $X_1$, $\cdot \cdot \cdot$, $X_l$ be Borel spaces. Let $Y_k:=X_1\times \cdot\cdot\cdot \times X_k$ for $1\leq k\leq l$. Let $q_1, \widehat{q}_1\in \mathfrak{P}(X_1)$ such that $q_1\ll \widehat{q}_1$ and for all $1\leq k\leq l-1$, let
$q_{k+1}$ and $\widehat{q}_{k+1}$ be universally measurable stochastic kernels on $X_{k+1}$ given $Y_k$ such that $q_{k+1}(\cdot|y_k)\ll \widehat{q}_{k+1}(\cdot|y_k)$  for all $y_k\in Y_k$. Let $P:=q_1\otimes  \cdot \cdot \cdot \otimes q_{l}$ and $\widehat{P}:=\widehat{q}_1\otimes  \cdot \cdot \cdot \otimes \widehat{q}_l$. Then, $P\ll \widehat{P}.$
\end{proposition}
\proof{Proof.}
We prove the claim for $l=2$ as the general case can easily be obtained by induction. Let $A\in\mathcal{B}(Y_2)$ such that $\widehat{P}(A)=\widehat{q}_1\otimes\widehat{q}_{2}(A)=0$.
Fubini's theorem implies that $\int_{X_1}\widehat{q}_2(A_{x_1}|x_1) \widehat{q}_1(dx_1)=0$ where $A_{x_1}:= \{x_2\in X_2,\; (x_1,x_2)\in A\}.$ Let $B:=
\left\{x_1\in X_1,\;\; \widehat{q}_2(A_{x_1}|x_1)=0  \right\}$. Then, $\widehat{q}_1(B)=1$ and also $q_1(B)=1$ as $q_1\ll \widehat{q}_1$. Let $x_1\in B$. Then, as $q_2(\cdot|x_1)\ll \widehat{q}_2(\cdot|x_1)$, we have that $q_2(A_{x_1}|x_1)=0$. This implies that
$\left\{x_1\in X_1,\;\; q_2(A_{x_1}|x_1)=0  \right\} $
is a $q_1\mbox{-full measure set}$. So, using Fubini's theorem again, $P(A)=q_1\otimes q_{2}(A)=\int_{X_1}q_2(A_{x_1}|x_1) q_1(dx_1)=0$ and $P\ll \widehat{P}.$
\Halmos \endproof

\subsection{Missing results of Section \ref{sec ilus}}
\subsubsection{The sets $\mathcal{M}^t(P)$, $\mathcal{M}^t$ and $\mathcal{W}^t$}
We give properties of these sets defined in Definition \ref{def_W}. First, we show that if a random variable has moments of any orders with respect to some convex combination of two priors, then it also have moments of any orders with respect to these priors. 
\begin{lemma}
Let $1\leq l\leq T$. Let $P:=q_1 \otimes \cdots \otimes q_l \in \mathcal{Q}^l$, $\widetilde{P}:= \widetilde{q}_1 \otimes \cdots \otimes \widetilde{q}_l \in \mathcal{Q}^l$ and 
\begin{eqnarray}
\widehat{P} := (\lambda_1 q_1 + (1-\lambda_1)\widetilde{q}_1) \otimes \cdots \otimes  (\lambda_l q_l + (1-\lambda_l)\widetilde{q}_l), \label{widehatP_gen}
\end{eqnarray}
where $\lambda_i\in (0,1]$ for all $1\leq i\leq l$. Then, for all $0 \leq t \leq l$,
$\mathcal{M}^t(\widehat{P}) \subset \mathcal{M}^t(P)$.
\label{lemma_inc_M}
\end{lemma}
\proof{Proof.}
The inclusion is trivial if $t=0$ as $\mathcal{M}^0(\widehat{P})=\mathcal{M}^0(P)=\mathbb{R}$. Let $1\leq t\leq l$. Let $\widehat{P}$ be defined as in \eqref{widehatP_gen} with $l=t$ and let $X\in\mathcal{M}^t(\widehat{P})$.
Noting that for all $1\leq k\leq t$, $\lambda_k q_{k} + (1-\lambda_k)\widetilde{q}_{k}\geq \lambda_k q_{k}$ (recall that $0< \lambda_k \leq 1$) and using Fubini's theorem, we obtain that for all $r\geq 1$,
\begin{eqnarray*}
\mathbb{E}_{\widehat{P}^{t}} |X|^r &=& \int_{\Omega^t} |X(\w^t)|^r\; (\lambda_t q_{t}+(1-\lambda_t)\widetilde{q}_{t})(d\w_{t}| \w^{t-1}) \otimes \cdots \otimes  (\lambda_1 q_{1}+(1-\lambda_1)\widetilde{q}_{1})(d\w_1)\nonumber\\
&\geq& \lambda_1 \cdots \lambda _t \int_{\Omega_1}\cdots \int_{\Omega_{t}} |X(\w^t)|^r\; q_{t}(d\w_{t}| \w^{t-1})\otimes\cdots \otimes q_{1}(d\w_{1})= \lambda_1 \cdots \lambda_t \;\; \mathbb{E}_{{P}^{t}}|X|^r. 
\end{eqnarray*}
As $X\in \mathcal{M}^t(\widehat{P})$, we have that $\mathbb{E}_{\widehat{P}^{t}} |X|^r<+\infty$ and also $\mathbb{E}_{{P}^{t}}|X|^r <+\infty$ (recall that $\lambda_1 \cdots \lambda_t>0$). Thus, $X\in \mathcal{M}^t(P)$ and the inclusion is proved.
\Halmos \endproof

We give without proof some further simple 
 properties of the sets $\mathcal{M}^t$, $\mathcal{W}^t$ and $\mathcal{M}^t(P)$.
\begin{lemma}
Fix $0\leq t\leq T$, $P\in\mathcal{Q}^T$ and $a\geq 0$. If $X,Y \in\mathcal{W}^t$ (resp. $\mathcal{M}^t$, $\mathcal{M}^t(P)$), then $X+Y$, $XY$, $\min(X,Y)$, $\max(X,Y)$ and $X^a$ belong to $\mathcal{W}^t$ (resp. $\mathcal{M}^t$, $\mathcal{M}^t(P)$). Let $Z :\Omega^t\to \mathbb{R}\cup\{-\infty,+\infty\}$ be $\mathcal{B}_c(\Omega^t)\mbox{-measurable}$. If $0\leq Z\leq Y$ $\mathcal{Q}^t$-q.s. with $Y\in\mathcal{W}^t$ (resp. $\mathcal{M}^t$), then $Z\in\mathcal{W}^t$ (resp. $\mathcal{M}^t$). If $0\leq Z\leq Y$ $P^t$-a.s. with $Y\in\mathcal{M}^t(P)$, then $Z\in\mathcal{M}^t(P)$.
\label{lemme_fdmt_W}
\end{lemma}

We now show that the sets $\mathcal{M}^t$ and $\mathcal{M}^t(P)$ are stable by dynamic programming.
\begin{lemma}
Let $0\leq t\leq  T-1$. Let $X : \Omega^{t+1} \to \mathbb{R}\cup\{-\infty,+\infty\}$ be $\mathcal{B}_c(\Omega^{t+1})$-measurable and choose $q_{t+1}\in SK_{t+1}$ such that $q_{t+1}(\cdot|\w^t)\in\mathcal{Q}_{t+1}(\w^t)$ for all $\w^t\in\Omega^t$. Let $\lambda_X : \Omega^t \to \mathbb{R}\cup\{-\infty,+\infty\}$ be defined by $\lambda_X(\w^t):= \mathbb{E}_{q_{t+1}(\cdot|\w^t)} X(\w^t,\cdot)$. Let $P\in\mathcal{Q}^t$. If $X\in \mathcal{M}^{t+1}(P\otimes q_{t+1})$, then $\lambda_X \in \mathcal{M}^t(P)$. If $X\in\mathcal{M}^{t+1}$, then $\lambda_X \in \mathcal{M}^t$.
\label{M_hered}
\end{lemma}
\proof{Proof.}
The function $\lambda_X$ is $\mathcal{B}_c(\Omega^t)$-measurable by Proposition \ref{univ_cvt}. Let $P\in\mathcal{Q}^t$ and assume that $X\in\mathcal{M}^{t+1}(P\otimes q_{t+1})$. Let $r\geq 1$. Jensen\rq{}s inequality applied to the convex function $x\mapsto |x|^{r}$ implies that $|\mathbb{E}_{q_{t+1}(\cdot|\w^t)} X(\w^t,\cdot)|^r  \leq \mathbb{E}_{q_{t+1}(\cdot|\w^t)}\left|X(\w^t,\cdot)\right|^r$ for all $\w^t\in\Omega^t$.
Using Fubini's theorem, we find that 
\begin{eqnarray*}
\mathbb{E}_P |\lambda_X|^r = \int_{\Omega^t} |\mathbb{E}_{q_{t+1}(\cdot|\w^t)}X(\w^t,\cdot))|^r P(d\w^t)\leq \mathbb{E}_{P\otimes q_{t+1}}\left|X\right|^r <+\infty.
\end{eqnarray*}
Thus, $\lambda_X\in\mathcal{M}^{t}(P)$. Assume now that $X\in\mathcal{M}^{t+1}$. Then, $X\in \mathcal{M}^{t+1}(P\otimes q_{t+1})$ for all $P\in\mathcal{Q}^t$. Using the preceding inequality, we deduce that $\lambda_X\in\mathcal{M}^{t}(P)$ for all $P\in\mathcal{Q}^t$. So, $\lambda_X\in\mathcal{M}^{t}$. 
\Halmos \endproof


\subsubsection{Utility function with random benchmark}
We show that it is indeed possible to apply Theorem \ref{optimality_M_typeA} to a random utility with benchmark.
\begin{proposition}
A utility function with random benchmark (see Definition \ref{utility_bench_def}) is a random utility of type $(A)$ (see Definition \ref{typeA}).
\label{utility_bench_is_typeA}
\end{proposition}

\proof{Proof.}
Let $U$ be a utility function with random benchmark as in Definition \ref{utility_bench_def}. Then,  as $\widetilde{U}$ is concave and $\mbox{Dom} \;\widetilde{U} = \mathbb{R}$, $\widetilde{U}$ is continuous and $U$ is trivially a random utility (see Definition \ref{U_hp}).

\noindent\textit{Step 1: for all $A\in\mathcal{W}^T$, $U^-(\cdot,A(\cdot))\in\mathcal{W}^T$ and \eqref{ineq_b_inf_det_A} holds true.}\\ Let $A\in\mathcal{W}^T$. Let $\w^T\in\Omega^T$, recalling \eqref{ineq_b_inf_det_bb} and $p\geq 1$, we obtain that 
\begin{eqnarray}
U(\w^T,A(\w^T))&=&\widetilde{U}(A(\w^T)-Z(\w^T))\geq -\widetilde{b}\left(1+|A(\w^T)-Z(\w^T)|^p\right)\nonumber\\
&\geq &-\widetilde{b}\left(1+2^{p-1}(|Z(\w^T)|^{p}+|A(\w^T)|^{p})\right) 
\geq 
-C_1(\w^T)\left(1+|A(\w^T)|^{p}\right),\label{temp_rand_ben_ineq1}
\end{eqnarray}
where $C_1(\cdot):= \widetilde{b} 2^{p-1}(1+|Z(\w^T)|^{p})$. We obtain that $C_1\in\mathcal{W}^T$ as $Z\in\mathcal{W}^T$. So, for all $x\in\mathbb{R}$, \eqref{ineq_b_inf_det_A} holds true choosing $A(\cdot)=x$. Now, as $A$ is $\mathcal{B}_c(\Omega^T)$-measurable, Remark \ref{joint_mes_U} shows that $U^-(\cdot,A(\cdot))$ is $\mathcal{B}_c(\Omega^T)$-measurable and thus belongs to $\mathcal{W}^T$, see Lemma \ref{lemme_fdmt_W} and \eqref{temp_rand_ben_ineq1}.\\

\noindent\textit{Step 2: for all $A\in\mathcal{W}^T$, $U^+(\cdot,A(\cdot))\in\mathcal{W}^T$.}\\ 
Let $A\in\mathcal{W}^T$. Let $\w^T\in\Omega^T$. 
Recalling that $\widetilde{U}$ is nondecreasing and using \eqref{AE_gamma1} for $\widetilde{U}$, we get that
\begin{eqnarray*}
U^+(\w^T,A(\w^T))&=&\widetilde{U}^+(A(\w^T)-Z(\w^T))\leq\widetilde{U}^+(1+|A(\w^T)|+|Z(\w^T)|)\\&\leq& (1+|A(\w^T)|+|Z(\w^T)|)(\widetilde{U}(1)+\widetilde{U}^-(0))^+
\\ &\leq& (1+|A(\w^T)|+|Z(\w^T)|)(\widetilde{U}^+(1)+\widetilde{U}^-(0)).
\end{eqnarray*}
As $U^+(\cdot,A(\cdot))$ is $\mathcal{B}_c(\Omega^T)$-measurable and $Z\in\mathcal{W}^T$, we deduce from Lemma \ref{lemme_fdmt_W} that $U^+(\cdot,A(\cdot))\in\mathcal{W}^T$. Choosing $A(\cdot)=1$, we get that $U^+(\cdot,1)\in\mathcal{W}^T$.\\

\noindent\textit{Step 3: Assumption \ref{AE} holds true with $C\in\mathcal{W}^T$.}\\ 
We show first that for all $\w^T\in\Omega^T$, $AE_{-\infty}(U(\w^T,\cdot))=AE_{-\infty}(\widetilde{U})$. Fix $\w^T\in\Omega^T$. Then,
\begin{eqnarray*}
AE_{-\infty}(U(\w^T,\cdot))&=& \underset{x\rightarrow -\infty}{\lim \inf} \frac{xU'(\w^T,x)}{U(\w^T,x)} = \underset{x\rightarrow -\infty}{\lim \inf} \left[\ \frac{(x-Z(\w^T))\widetilde{U}'(x-Z(\w^T))}{\widetilde{U}(x-Z(\w^T))} \frac{x}{x-Z(\w^T)}\right]\ \\
&=& AE_{-\infty}(\widetilde{U}).
\end{eqnarray*}
A similar reasoning shows that $AE_{+\infty}(U(\w^T,\cdot))=AE_{+\infty}(\widetilde{U}).$ Now, we distinguish two cases and show that we can use Proposition \ref{annex_AE2} for $U(\w^t,\cdot)$.\\
\noindent\textit{Step 3.1: $AE_{-\infty}(\widetilde{U})>1$}\\
Assume first that $AE_{-\infty}(\widetilde{U})>1$. Let $\gamma$, $\widetilde{\gamma} \in\mathbb{R}$ such that  $AE_{-\infty}(\widetilde{U})>\widetilde{\gamma}>\gamma>1$ and set $\eta:=\gamma/\widetilde{\gamma}\in (0,1)$. Using Proposition \ref{annex_AE2} for $\widetilde{U}$, there exists some $\underline{x}<0$ such that $\widetilde{U}(\underline{x})<0$ and $\frac{x \widetilde{U}'(x)}{\widetilde{U}(x)}>\widetilde{\gamma}$ for all $x\leq \underline{x}$. Let $\underline{X}(\cdot):= \min(\underline{x}, \underline{x}+Z(\cdot), -\frac{\eta}{1-\eta}Z(\cdot))$. Fix $\w^T\in\Omega^T$. Then, $\underline{X}(\w^T)\leq \underline{x}<0$. Moreover, $U(\w^T,\underline{X}(\w^T))= \widetilde{U}(\underline{X}(\w^T)-Z(\w^T))\leq \widetilde{U}(\underline{x})<0$ as $\widetilde{U}$ is nondecreasing. Now, let $x\leq \underline{X}(\w^T)$. First, we show that $\frac{xU'(\w^T,x)}{U(\w^T,x)}>\gamma$. As $x\leq \underline{X}(\w^T)\leq \underline{x}+Z(\w^T)$, we have that \begin{eqnarray}
\frac{(x-Z(\w^T))U'(\w^T,x)}{U(\w^T,x)} = \frac{(x-Z(\w^T))\widetilde{U}'(x-Z(\w^T))}{\widetilde{U}(x-Z(\w^T))} >\widetilde{\gamma}.
\label{temp_tilde_-inf}
\end{eqnarray}
Now, as $x \leq \underline{X}(\w^T)\leq -\frac{\eta}{1-\eta}Z(\w^T)$ and $\eta<1$, we get that $(1-\eta)x \leq -\eta Z(\w^T)$. 
Remarking that $x-Z(\w^T) \leq \underline{X}(\w^T)-Z(\w^T)\leq \underline{x} <0$, we have that $\frac{x}{x-Z(\w^T)}\geq \eta.$ So, using \eqref{temp_tilde_-inf}, we deduce that 
\begin{eqnarray}
\frac{x U'(\w^T,x)}{U(\w^T,x)} = \frac{(x-Z(\w^T))\widetilde{U}'(x-Z(\w^T))}{\widetilde{U}(x-Z(\w^T))} \frac{x}{x-Z(\w^T)}> \widetilde{\gamma}\eta = \gamma. \label{temp_ineq_AE11}
\end{eqnarray}
We can apply Proposition \ref{annex_AE2} to $U(\w^T,\cdot)$. Indeed,  $AE_{-\infty}(U(\w^T,\cdot))=AE_{-\infty}(\widetilde{U})>1$, $\gamma\in (1, AE_{-\infty}(U(\w^T,\cdot))$, $\underline{X}(\w^T)<0$, $U(\w^T,\underline{X}(\w^T))<0$ and \eqref{temp_ineq_AE11} holds true for all $x\leq \underline{X}(\w^T).$
It follows that  \eqref{det_AE} holds true for $U(\w^T,\cdot)$ with $\gamma$ and $C(\w^T) := U^+(\w^T,0) + U^-(\w^T,0)+ U^-(\w^T,\underline{X}(\w^T))$. Now, we show that $C\in\mathcal{W}^T.$  
As $Z\in\mathcal{W}^T$, $\underline{X}\in\mathcal{W}^T$ (see Lemma \ref{lemme_fdmt_W}). Step 1 shows that $U^-(\cdot,\underline{X}(\cdot)), U^-(\cdot,0)\in\mathcal{W}^T$ and Step 2 that $U^+(\cdot,0)\in \mathcal{W}^T$. Thus $C\in\mathcal{W}^T$ and Assumption \ref{AE} holds true with $C\in\mathcal{W}^T$.\\
\noindent\textit{Step 3.2: $AE_{+\infty}(\widetilde{U})<1$ and $\lim_{x\to +\infty}\widetilde{U}(x)>0$}\\
Let $\widetilde{\gamma}'$ and $\gamma'$ be such that $AE_{+\infty}(\widetilde{U})<\widetilde{\gamma}'<\gamma'<1$ with $\widetilde{\gamma}'>0$ and set $\eta':=\gamma'/\widetilde{\gamma}'>1$. Using Proposition \ref{annex_AE2} for $\widetilde{U}$, there exists $x'<0$ and $\overline{x}>0$ such that $\widetilde{U}(x')<0$, $\widetilde{U}(\overline{x})>0$ and $\frac{x \widetilde{U}'(x)}{\widetilde{U}(x)}<\widetilde{\gamma}'$ for all $x\geq \overline{x}$. Let $\overline{X}(\cdot):= \max(\overline{x}, \overline{x}+Z(\cdot), -\frac{\eta'}{1-\eta'}Z(\cdot))$ and $X'(\cdot):= \min\left(x',x'+Z(\cdot)\right)$. Let $\w^T\in \Omega^T$. We trivially have that $\lim_{x\to +\infty} U(\w^T,x)>0$ as $\lim_{x\to +\infty}\widetilde{U}(x)>0$. Then,
$\overline{X}(\w^T)\geq \overline{x} >0$ and  $X'(\w^T)\leq x'<0$. Moreover,
$U(\w^T,\overline{X}(\w^T))=\widetilde{U}(\overline{X}(\w^T)-Z(\w^T))\geq \widetilde{U}(\overline{x})>0$ and $U(\w^T,X'(\w^T))\leq \widetilde{U}(x')<0$ as $\widetilde{U}$ is nondecreasing. Now, let $x\geq \overline{X}(\w^T)$. First, we show that $\frac{xU'(\w^T,x)}{U(\w^T,x)}<\gamma'$. As $x\geq \overline{X}(\w^T)\geq \overline{x}+Z(\w^T)$, we have that \begin{eqnarray}
\frac{(x-Z(\w^T))U'(\w^T,x)}{U(\w^T,x)} = \frac{(x-Z(\w^T))\widetilde{U}'(x-Z(\w^T))}{\widetilde{U}(x-Z(\w^T))} <\widetilde{\gamma}'.
\label{temp_tilde_+inf}
\end{eqnarray}
As $x\geq \overline{X}(\w^T)\geq -\frac{\eta'}{1-\eta'}Z(\w^T)$ and $\eta'>1$, we get that $(1-\eta')x \leq -\eta'Z(\w^T)$. As $x-Z(\w^T) \geq \overline{x} >0$, we get that $0\leq \frac{x}{x-Z(\w^T)}\leq \eta'.$ So, using \eqref{temp_tilde_+inf}, we deduce that 
\begin{eqnarray}
\frac{x U'(\w^T,x)}{U(\w^T,x)} = \frac{(x-Z(\w^T))\widetilde{U}'(x-Z(\w^T))}{\widetilde{U}(x-Z(\w^T))} \frac{x}{x-Z(\w^T)}< \widetilde{\gamma}'\eta' = \gamma'.
\label{temp_ineq_AE21}
\end{eqnarray}
As $AE_{+\infty}(U(\w^T,\cdot))=AE_{+\infty}(\widetilde{U})<1$, $\lim_{x\to +\infty}U(\w^T,x)>0$, $\gamma'\in (AE_{+\infty}(U(\w^T,\cdot)),1)$, $X'(\w^T)<0$ is such that 
$U(\w^T,X'(\w^T))<0$ and $\overline{X}(\w^T)>0$ is such that $U(\w^T,\overline{X}(\w^T))>0$ and as \eqref{temp_ineq_AE21} holds true for all $x\geq \overline{X}(\w^T)$, Proposition \ref{annex_AE2} applies for $U(\w^T,\cdot)$ and \eqref{det_AE} holds true with $\gamma'$ and $C(\w^T) = U^+(\w^T,\overline{X}(\w^T)) + U^-(\w^T,X'(\w^T)) + U^-(\w^T,0)$. As $Z\in\mathcal{W}^T$, $X', \overline{X}\in\mathcal{W}^T$ (see Lemma \ref{lemme_fdmt_W}) and as before Steps 1 and 2 show that $C\in\mathcal{W}^T$. So, Assumption \ref{AE} holds true for $C\in\mathcal{W}^T$.\\ 

\noindent\textit{Step 4: Assumption \ref{nncst} holds true for some $\underline{X}\in\mathcal{W}^T$ such that $1/|U(\cdot,\underline{X}(\cdot))+C(\cdot)|\in\mathcal{W}^T$.}\\ 
As $\widetilde{U}$ is concave, nondecreasing, non-constant and finite, we have that $\widetilde{U}$ is continuous and that $\lim_{x\to -\infty}\widetilde{U}(x)=-\infty$. 
As a result, one can find a real $\underline{x}<0$ such that $\widetilde{U}(\underline{x})=-\widetilde{U}^-(0)-1.$ Set $\underline{Y}(\cdot):=\underline{x}(C(\cdot)+1)$, where $C(\cdot)\in\mathcal{W}^T$ has been defined in Step 3.1 or Step 3.2 and is such that Assumption \ref{AE} holds true. Using \eqref{AE_gamma1} in Proposition \ref{annex_AE1} for $\widetilde{U}$, we see that
\begin{eqnarray*}
\widetilde{U}(\underline{Y}(\cdot)) = \widetilde{U}\left(\frac{\underline{Y}(\cdot)}{\underline{x}}\underline{x}\right) \leq \frac{\underline{Y}(\cdot)}{\underline{x}}\left(\widetilde{U}(\underline{x})+\widetilde{U}^-(0)\right) = -\frac{\underline{Y}(\cdot)}{\underline{x}} = -C(\cdot)-1.
\end{eqnarray*}
Let $\underline{X}(\cdot):= \min(\underline{Y}(\cdot)+Z(\cdot),-1)<0$. Then, $\underline{X}\in\mathcal{W}^T$ as $Z$ and $C$ belong to $\mathcal{W}^T$ (see Lemma \ref{lemme_fdmt_W}). As $\widetilde{U}$ is nondecreasing $U(\cdot,\underline{X}(\cdot))=\widetilde{U}(\underline{X}(\cdot)-Z(\cdot))\leq \widetilde{U}(\underline{Y}(\cdot))\leq -C(\cdot)-1 <-C(\cdot).$ 
So, Assumption \ref{nncst} holds true for $U$. Moreover, $U(\cdot,\underline{X}(\cdot))+C(\cdot)\leq -1$ and so $1/|U(\cdot,X(\cdot))+C(\cdot)|\leq 1$. As $1/|U(\cdot,X(\cdot))+C(\cdot)|$ is $\mathcal{B}_c(\Omega^T)$-measurable (see Remark \ref{joint_mes_U}), it belongs to $\mathcal{W}^T$ (see Lemma \ref{lemme_fdmt_W} again).\\ Putting all the steps together, $U$ is a random utility of type (A).
\Halmos \endproof

\subsubsection{Proof of Lemma \ref{lemma_temp_exemple}}
\label{sec lemma_temp_exemple}
Lemma \ref{lemma_temp_exemple} was used to provide an example of application of Theorem \ref{optimality_M_typeA}. 
We prove it  by contradiction. Assume that there exists $(p_n)_{n\geq 1}\subset \mathcal{Q}$ such that $p_n(Z < -\frac{1}{n})<\frac{1}{n}$ or $p_n(-Z < -\frac{1}{n})>\frac{1}{n}$ for all $n\geq 1$. We only treat the case $p_n(Z < -\frac{1}{n})<\frac{1}{n}$, the other case is completely similar and thus omitted. \\
\textit{The sequence of image laws $(\mu_n)_{n\geq 1} := (p_n \circ Z^{-1})_{n\geq 1}$ is tight.}\\
Let $K>0$ and $n\geq 1$, Markov's inequality and the definition of $p_n \in \mathcal{Q}$ show that 
$$\mu_n(\mathbb{R}\setminus [-K,K]) = p_n (|Z|>K) \leq \exp(-K)\mathbb{E}_{p_n} \exp(|Z|)  \leq C \exp(-K).$$
Thus,  $(\mu_n)_{n\geq 1}$ is tight and using Prokhorov's theorem (see \citep[Theorem 5.1, p59]{refbilli}), we can extract a subsequence (still denoted by $(\mu_n)_{n\geq 1}$)  such that $(\mu_n)_{n\geq 1}$ converges weakly to some $\mu\in \mathfrak{P}(\mathbb{R})$.\\
Using Portmanteau theorem (see \citep[Theorem 2.1, p16]{refbilli}) and $\mu_n((-\infty,-\frac{1}{n}))<\frac{1}{n}$, we get that  
$ \mu((-\infty,-\epsilon)) \leq \lim\inf_{n\geq 1} \mu_n((-\infty,-\epsilon)) \leq 0$ for every $0<\epsilon<1.$\\
\textit{Application of Skorokhod's representation theorem.}\\
Using Skorokhod's representation theorem (see \citep[Theorem 6.7, p70]{refbilli}), there exist a probability space $(\widehat{\Omega}, \widehat{\mathcal{A}},\widehat{p})$ and random variables $X$ and $(X_n)_{n\geq 1}$ such that $\widehat{p} \circ X^{-1} = \mu$,\; $\widehat{p} \circ X_n^{-1} = \mu_n$ for all $n\geq 1$ and $X_n$ converges to $X$ $\widehat{p}$-a.s. Moreover, as 
$\widehat{p}(X<-\epsilon) =\mu((-\infty,-\epsilon)) \leq 0$ for all $0<\epsilon<1,$ we have that  $\widehat{p}(X\geq 0) = 1$.\\
\textit{Contradiction.}\\
Remark that  $\mathbb{E}_{\widehat{p}} X_n =\int x \mu_n (dx) = \mathbb{E}_{p_n} Z = 0 $ by definition of $\mathcal{Q}$ and also that $\mathbb{E}_{\widehat{p}} X_n^2 = 1$.   Assume for a moment that $(X_n)_{n\geq 0}$ and $(X_n^2)_{n\geq 0}$ are uniformly integrable under $\widehat{p}$. Then, as $(X_n)_{n\geq 1}$ converges to $X$ $\widehat{p}$-a.s. we get that  
$0 = \lim_{n\to +\infty}\mathbb{E}_{\widehat{p}} X_n = \mathbb{E}_{\widehat{p}} X$ and  $1 = \lim_{n\to +\infty}\mathbb{E}_{\widehat{p}} X_n^2 = \mathbb{E}_{\widehat{p}}X^2$, see for example \citep[Theorem 3.5, p31]{refbilli}. As $\widehat{p}(X \geq 0)=1$ and $\mathbb{E}_{\widehat{p}} X=0$, we must have that $ X =0$  $\widehat{p}$-a.s. a contradiction to  $\mathbb{E}_{\widehat{p}} X^2=1$.  Finally, to show  that $(X_n)_{n\geq 1}$ and $(X^2_n)_{n\geq 1}$ are uniformly integrable under $\widehat{p},$ it is enough to prove that for all $n\geq 1$, $\mathbb{E}_{\widehat{p}}|X_n|^3$ is uniformly bounded in $n$ (see \citep[(3.18), p31]{refbilli}). This follows from 
$ \mathbb{E}_{\widehat{p}}|X_n|^{3} = \int |x|^3 \mu_n (dx) =\mathbb{E}_{p_n} |Z|^{3} \leq 3!\; \mathbb{E}_{p_n} \exp(|Z|)\leq  3!\; C.$
\Halmos \endproof

\subsection{Proofs of Section \ref{muti_per}}
\label{miss_proof}
We now provide the missing proofs and results of Section \ref{muti_per}.

\subsubsection{Proof of Proposition \ref{U_t_well}}
\label{proof_U_t_well}
We first show the following claim.\\
\noindent\textit{Proof of (i) to (v).}\\ 
Let $P:=q_1^P \otimes \cdot \cdot \cdot \otimes q_T^P\in \mathcal{Q}^T$. The proof is made by backward induction.\\
\noindent\textit{Initialization step.}\\
As $U_T=U_T^P=U$ and $J_T=U^-$,  (iv) and (v) hold true with an equality.
Remark \ref{joint_mes_U} shows that $U$ is $\mathcal{B}(\Omega^T) \otimes \mathcal{B}(\mathbb{R})$-measurable. So, $U_T$ is lsa, $U_T^P$ is $\mathcal{B}_c(\Omega^T\times \mathbb{R})$-measurable and (i) and (ii) are proved. Finally, Definition \ref{U_hp} implies that for all $\w^T\in\Omega^T$, $U_T(\w^T,\cdot)=U_T^P(\w^T,\cdot)=U(\w^T,\cdot)$ is nondecreasing, usc and concave and (iii) holds true.\\
\noindent\textit{Heredity step.}\\ 
Fix $0\leq t\leq T-1$ and assume that the induction hypothesis holds at time $t+1$. Fix $x\in\mathbb{R}$. (i) at $t+1$ implies that $U_{t+1}(\cdot,\cdot,x)$ is $\mathcal{B}_c(\Omega^t\times\Omega_{t+1})$-measurable and so is $U_{t+1}^P(\cdot,\cdot,x)$ thanks to (ii) at $t+1$ and \citep[Lemma 7.29, p174]{ref1}. The same lemma also shows that $U_{t+1}(\w^t,\cdot,x)$ and $U_{t+1}^P(\w^t,\cdot,x)$ are $\mathcal{B}_c(\Omega_{t+1})$-measurable for all $\w^t\in\Omega^t$. So, (iii) at $t+1$ shows that Assumption \ref{V} is satisfied both in the robust and in the $P$-prior $(t+1)$-context (see Definitions \ref{def_one_prior} and \ref{def_ctxt}). Lemma \ref{minimal_V_pp} applies and we get that for all $\w^t\in\Omega^t$, $U_t(\w^t,\cdot)$ and  $U_t^P(\w^t,\cdot)$ are nondecreasing, usc and concave on $\mathbb{R}$ and (iii) at $t$ is proved. Moreover, for all $\w^t\in\Omega^t$, 
\begin{eqnarray}
U_t(\w^t,x)=\lim_{n\to +\infty}\mathcal{U}_t\left(\w^t,x+\frac{1}{n}\right) \quad \mbox{and} \quad U_t^P(\w^t,x)=\lim_{n\to +\infty}\mathcal{U}_t^P\left(\w^t,x+\frac{1}{n}\right).
\label{temp_clos_t}
\end{eqnarray}
Using (i) at time $t+1$, Lemma \ref{lemma_mes_u} (ii) for $f=U_{t+1}$ and \citep[Lemma 7.30, p177]{ref1}, we find that $\mathcal{U}_t$ is lsa. So, \eqref{temp_clos_t} and \citep[Lemma 7.30, p177]{ref1} again imply that $U_t$ is lsa and thus that (i) at $t$ is proved. Similarly, (ii) at $t+1$, Lemma \ref{lemma_mes_u} (i) and \citep[Lemma 7.30, p177]{ref1} show that $\mathcal{U}_t^P$ is $\mathcal{B}_c(\Omega^t\times \mathbb{R})$-measurable and \eqref{temp_clos_t} shows (ii) at $t$.
Now, (v) at $t$ follows from (v) at $t+1$ and \eqref{state_val_t_p} and \eqref{state_val_t_rob}. Similarly, for all $\w^t\in\Omega^t$, starting from (iv) at time $t+1$, we get using \eqref{state_val_t_rob} and \eqref{J_t_eq} that
\begin{eqnarray*}
-U_t(\w^t,x)&\leq & -\mathcal{U}_t(\w^t,x) \leq \sup_{p\in\mathcal{Q}_{t+1}(\w^t)} \mathbb{E}_p U^-_{t+1}(\w^t,\cdot,x)\leq \sup_{p\in\mathcal{Q}_{t+1}(\w^t)} \mathbb{E}_p J_{t+1}(\w^t,\cdot,x) = J_t(\w^t,x). 
\end{eqnarray*}
and (iv) at $t$ follows as $J_t\geq 0$, see Proposition \ref{CJ_t}. \\

\noindent \textit{Proof of \eqref{elas_gammaft} and \eqref{elas_gammafPt}.}\\ 
We proceed again by backward induction. We only show \eqref{elas_gammaft} as the proof of \eqref{elas_gammafPt} is very similar and thus omitted. Assumption \ref{AE} ensures that \eqref{elas_gammaft} holds true at time $T$ as $C_T:=C$. Fix $0\leq t\leq T-1$ and assume that \eqref{elas_gammaft} holds true at $t+1$. Let $\w^t\in \Omega^t$ such that $C_t(\w^t)<+\infty$, $x\in\mathbb{R}$, $h\in\mathbb{Q}^d$ and $\lambda\in\mathbb{Q}\cap [1,+\infty)$.
Take any $p\in\mathcal{Q}_{t+1}(\w^t)$. The set $\left\{\w_{t+1}\in\Omega_{t+1},\;\; C_{t+1}(\w^t,\w_{t+1})<+\infty \right\}$ is a $p$-full measure set. Otherwise, we get a contradiction with \eqref{finite_CJ}. So, (\ref{elas_gammaft}) at time $t+1$ implies that
\begin{eqnarray*}
\mathbb{E}_p U_{t+1}\left(\w^t,\cdot,\lambda x+\lambda h\Delta S_{t+1}(\w^t,\cdot)\right) \leq \lambda^{\gamma} \left(\mathbb{E}_p U_{t+1}\left(\w^t,\cdot,x+h\Delta S_{t+1}(\w^t,\cdot)\right)+\mathbb{E}_p C_{t+1}(\w^t,\cdot)\right).
\end{eqnarray*}
So, taking the infimum over all $p\in\mathcal{Q}_{t+1}(\w^t)$ and using \eqref{C_t_eq}
\small
\begin{eqnarray*}
\inf_{p\in\mathcal{Q}_{t+1}(\w^t)} \mathbb{E}_p U_{t+1}\left(\w^t,\cdot,\lambda x+\lambda h\Delta S_{t+1}(\w^t,\cdot)\right) &\leq &\lambda^\gamma \inf_{p\in\mathcal{Q}_{t+1}(\w^t)} \left(\mathbb{E}_p U_{t+1}\left(\w^t,\cdot,x+h\Delta S_{t+1}(\w^t,\cdot)\right)+\mathbb{E}_p C_{t+1}(\w^t,\cdot)\right)  \\ &\leq &\lambda^\gamma  \inf_{p\in\mathcal{Q}_{t+1}(\w^t)} \mathbb{E}_p U_{t+1}\left(\w^t,\cdot,x+h\Delta S_{t+1}(\w^t,\cdot)\right) + \lambda^\gamma C_t(\w^t).
\end{eqnarray*}
\normalsize
Now, taking the supremum over every $h\in\mathbb{Q}^d$ (recall that $\lambda\in\mathbb{Q}$), we obtain recalling \eqref{state_val_t_rob} that
\begin{eqnarray*}
\mathcal{U}_t(\w^t,\lambda x)\leq \lambda^\gamma \mathcal{U}_t(\w^t,x)+\lambda^\gamma C_t(\w^t)\leq \lambda^\gamma U_t(\w^t,x)+\lambda^\gamma C_t(\w^t).
\end{eqnarray*}
As $x\mapsto \lambda^\gamma U_t(\w^t,x) + \lambda^\gamma C_t(\w^t)$ is usc (see (iii)), by definition of the closure in $U_t$ (see \eqref{state_val_t_rob}), 
 \eqref{elas_gammaft} at $t$ is proved for $\lambda\in\mathbb{Q}$ and $\lambda \geq 1$.
Take now $\lambda\geq 1$ (not necessarily a rational number). Assume first that $x\geq 0$ and take a sequence $(\lambda_n)_{n\geq 0}\subset \mathbb{Q}$ such that $\lambda_n$ is nonincreasing and $\lim_{n\to +\infty}\lambda_n = \lambda$. Let $n\geq 0$. Using that $U_t(\w^t,\cdot)$ is nondecreasing (see (iii)), we obtain that 
\begin{eqnarray*}
U_t(\w^t,\lambda x) \leq U_t(\w^t,\lambda_n x)\leq \lambda_n^\gamma U_t(\w^t,x)+\lambda_n^\gamma C_t(\w^t).
\end{eqnarray*}
Taking the limit gives (\ref{elas_gammaft}) for any $\lambda\geq 1$ and $x\geq 0$. When $x<0$, the same method with a sequence of nondecreasing rational numbers gives the desired result.\Halmos 

\subsubsection{Lemma \ref{lemma1+}}

The following lemma shows that $\Omega^t_{\ref{integ_V+}}$ is a $\mathcal{Q}^t$-full-measure set and $\Omega^{t,P}_{\ref{integ_V+}}$ is a $P^t$-full-measure set. 

\begin{lemma}
Assume that Assumptions \ref{S_borel}, \ref{analytic_graph}, \ref{AE} and \ref{simple_adm} hold true. Let $P:=q^P_1\otimes \cdot \cdot \cdot \otimes q^P_{T}\in\mathcal{Q}^T$ and $0\leq t\leq T-1$. Then, $\Omega^{t}_{\ref{integ_V+}}$ and $\Omega_{\ref{integ_V+}}^{t,P}$ belong to $\mathcal{B}_c(\Omega^t)$. Assume furthermore that $U_0^P(1)<+\infty,$ then $P^t(\Omega_{\ref{integ_V+}}^{t,P})=1$. Moreover, if Assumption \ref{U0} holds true, $\Omega^{t}_{\ref{integ_V+}}$ is a $\mathcal{Q}^t$-full-measure set.
\label{lemma1+}
\label{lemma_2}
\end{lemma}
\proof{Proof.}
Definitions \ref{def_one_prior} and  \ref{def_ctxt} imply that 
\begin{eqnarray*}
\Omega^{t,P}_{\ref{integ_V+}} &=& \left\{\w^t\in\Omega^t,\; \mathbb{E}_{q^{P}_{t+1}(\cdot|\w^t)} (U_{t+1}^P)^+(\w^t,\cdot,1)<+\infty \right\} \quad 
\Omega^t_{\ref{integ_V+}} =\left \{\w^t\in\Omega^t,\; \mathbb{E}_{q^{P^*}_{t+1}(\cdot|\w^t)} U_{t+1}^+(\w^t,\cdot,1)<+\infty \right\}.
\end{eqnarray*}
Recalling that $U_{t+1}^P$ is $\mathcal{B}_c(\Omega^{t+1}\times \mathbb{R})$-measurable (see Proposition \ref{U_t_well} (ii)), \citep[Proposition 7.44, p172]{ref1} shows that $(\w^t,\w_{t+1})\mapsto (U_{t+1}^P)^+(\w^t,\w_{t+1},1)$ is $\mathcal{B}_c(\Omega^t\times \Omega_{t+1})$-measurable. So, Proposition \ref{univ_cvt} (iii) shows that $\w^t\mapsto \mathbb{E}_{q_{t+1}^P(\cdot|\w^t)} (U_{t+1}^P)^+(\w^t,\cdot,1)$ is $\mathcal{B}_c(\Omega^t)$-measurable, which implies that $\Omega^{t,P}_{\ref{integ_V+}}\in\mathcal{B}_c(\Omega^t)$. Similarly, using Proposition \ref{U_t_well} (i), we get that $\Omega^{t}_{\ref{integ_V+}}\in\mathcal{B}_c(\Omega^t)$.\\

\noindent\textit{If $U_0^P(1)<+\infty$, then $P^t(\Omega^{t,P}_{\ref{integ_V+}})=1$.}\\
\noindent Assume now that $U_0^P(1)<+\infty$. We proceed by contraposition. Assume that there exists some $0\leq t\leq T-1$ such that $P^t(\Omega^{t,P}_{\ref{integ_V+}})<1$. Then, for all $\w^t\notin \Omega^{t,P}_{\ref{integ_V+}}$, 
\begin{eqnarray}
\mathbb{E}_{q^P_{t+1}(\cdot|\w^t)}(U^{P}_{t+1})^+(\w^t,\cdot,1) =+\infty. \label{temp_A35P}
\end{eqnarray}
For all $0\leq k\leq t$, we show by backward induction the following property: there exists some $\widetilde{\Omega}^k_+\in\mathcal{B}_c(\Omega^k)$ such that $P^k(\widetilde{\Omega}^k_+)>0$ and $U_k^P(\w^k,1)=+\infty$ for all $\w^k\in\widetilde{\Omega}_+^k$. The property at $k=0$  will show that $U_0^P(1)=+\infty$: a contradiction that proves the claim.\\ 
We start with $k=t$. For all $\w^t\in \Omega^{t}$, \eqref{state_val_t_p} implies that 
\begin{eqnarray}
U_t^P(\w^t,1)\geq\mathcal{U}_t^P(\w^t,1) \geq  \mathbb{E}_{q^P_{t+1}(\cdot|\w^t)} (U^{P}_{t+1})^+(\w^t,\cdot,1)-\mathbb{E}_{q^P_{t+1}(\cdot|\w^t)} (U^{P}_{t+1})^-(\w^t,\cdot,1).\label{temp_A35P2}
\end{eqnarray}
Let $\widetilde{\Omega}^t_+ := (\Omega^{t,P}_{\ref{integ_V+}})^c \cap \Omega^t_{J,1,0}$ (see \eqref{set_J_t} for the definition of $\Omega^t_{J,1,0}$). Then, Proposition \ref{CJ_t} shows that $\widetilde{\Omega}^t_+ \in \mathcal{B}_c(\Omega^t)$ and $P^t(\widetilde{\Omega}^t_+)=P^t((\Omega^{t,P}_{\ref{integ_V+}})^c)>0$ as $P^t(\Omega^t_{J,1,0})=1$. Let $\w^t\in \widetilde{\Omega}^t_+$. Using that $U_{t+1} \leq U_{t+1}^{P}$ and $U_{t+1}^- \leq J_{t+1}$ (see Proposition \ref{U_t_well}), we get that 
$\mathbb{E}_{q^P_{t+1}(\cdot|\w^t)}(U^{P}_{t+1})^-(\w^t,\cdot,1)\leq \mathbb{E}_{q_{t+1}^P(\cdot|\w^t)} J_{t+1}(\w^t,\cdot, 1)<+\infty,$ see \eqref{set_J_t}. Consequently, using \eqref{temp_A35P} and \eqref{temp_A35P2}, $U_t^P(\w^t,1)=+\infty$ and the property is proved for $k=t$. Now, we prove the induction step. Assume that the property holds true for some $1\leq k+1\leq t$. Define $\widehat{\Omega}^{k}_+:=\{\w^k\in\Omega^k,\;\; q_{k+1}^P(\widetilde{\Omega}^{k+1}_{+,\w^k}|\w^k)>0  \}$, where for all $\w^k\in \Omega^k$, the section of $\widetilde{\Omega}^{k+1}_+$ along $\w^k$ is defined by $\widetilde{\Omega}^{k}_{+,\w^k}:= \{\w_{k+1}\in\Omega_{k+1},\; (\w^k,\w_{k+1})\in\widetilde{\Omega}^{k}_+\}.$ As $\widetilde{\Omega}^{k+1}_{+}\in\mathcal{B}_c(\Omega^{k+1})$, \citep[Corollary 7.44.1, p172]{ref1} shows that 
$\widetilde{\Omega}^{k+1}_{+,\w^k} \in \mathcal{B}_c(\Omega_{k+1})$. Moreover, we have that 
\begin{eqnarray*}
P^{k+1}(\widetilde{\Omega}^{k+1}_{+})= P^{k}\otimes q_{k+1}^P(\widetilde{\Omega}^{k+1}_{+})&= &\int_{\widehat{\Omega}^{k}_{+}}q^P_{k+1}(\widetilde{\Omega}^{k+1}_{+,\w^k}|\w^k) P^k(d\w^k)+ \int_{(\widehat{\Omega}^{k}_{+})^c}q^P_{k+1}(\widetilde{\Omega}^{k+1}_{+,\w^k}|\w^k) P^k(d\w^k)\\
&=&\int_{\widehat{\Omega}^{k}_{+}}q^P_{k+1}(\widetilde{\Omega}^{k+1}_{+,\w^k}|\w^k) P^k(d\w^k).
\end{eqnarray*}
As $P^{k+1}(\widetilde{\Omega}^{k+1}_{+})>0$, we get that $P^{k}(\widehat{\Omega}^{k}_{+})>0$. Let $\widetilde{\Omega}^{k}_{+} := \widehat{\Omega}^{k}_{+} \cap \Omega_{J,1,0}^k.$ Then, $\widetilde{\Omega}^{k}_{+}\in\mathcal{B}_c(\Omega^k)$ and $P^k(\widetilde{\Omega}^{k}_{+})=P^k(\widehat{\Omega}^{k}_{+})>0$. Let $\w^k\in \widetilde{\Omega}^{k}_{+}$. As $U_{k+1}\leq U_{k+1}^{P}$ and $U_{k+1}^- \leq J_{k+1}$, we have that $\mathbb{E}_{q_{k+1}^P(\cdot|\w^k)}(U^{P}_{k+1})^-(\w^k,\cdot, 1)<+\infty,$ see \eqref{set_J_t}. Finally, using \eqref{state_val_t_p}, we see that 
\begin{eqnarray*}
U_k^P(\w^k,1)&\geq &\mathcal{U}_k^P(\w^k,1) \geq   \mathbb{E}_{q_{k+1}^P(\cdot|\w^k)} U_{k+1}^P(\w^k,\cdot,1)\\
 &\geq & \mathbb{E}_{q^P_{k+1}(\cdot|\w^k)}\left((+\infty) 1_{(\w^k,\cdot)\in \widetilde{\Omega}^{k+1}_{+}}-   (U^{P}_{k+1})^-(\w^k,\cdot,1)1_{(\w^k,\cdot)\notin \widetilde{\Omega}^{k+1}_{+}}\right)\\
&\geq & (+\infty) q_{k+1}^P(\widetilde{\Omega}^{k+1}_{+,\w^k}|\w^k) - \mathbb{E}_{q^P_{k+1}(\cdot|\w^k)} \left((U^{P}_{k+1})^-(\w^k,\cdot,1)1_{(\w^k,\cdot)\notin \widetilde{\Omega}^{k+1}_{+}}\right),
\end{eqnarray*}
using for the last inequality \citep[Lemma 7.11 (a), p139]{ref1} adapted to convention \eqref{cvt_inf}. As $q_{k+1}^P(\widetilde{\Omega}^{k+1}_{+,\w^k}|\w^k)>0$,
we find that for all $\w^k\in \widetilde{\Omega}^{k}_{+}$, $U_k^P(\w^k,1)=+\infty$.\\

\noindent\textit{If Assumption \ref{U0} holds true, $\Omega^t_{\ref{integ_V+}}$ is a $\mathcal{Q}^t$-full-measure set.}\\
Let $\widetilde{P} := q_{1}^{\widetilde{P}} \otimes \cdots \otimes q_{T}^{\widetilde{P}} \in\mathcal{Q}^T$ and $\widehat{P} := \frac{q_{1}^{P^*}+q_1^{\widetilde{P}}}{2}\otimes \cdot\cdot\cdot \otimes\frac{q_T^{P^*}+q_T^{\widetilde{P}}}{2}.$ As $\widehat{P}\in \mathcal{H}^T$ (see \eqref{setPT}), we have that $U_0^{\widehat{P}}(1)<+\infty$ by Assumption \ref{U0}. So, the preceding step shows that $\Omega^{t,\widehat{P}}_{\ref{integ_V+}}$ is a $\widehat{P}^t$-full-measure set. Using now (v) in Proposition \ref{U_t_well} at $t+1$, $U_{t+1}\leq U_{t+1}^{\widehat{P}}$, $\Omega^{t,\widehat{P}}_{\ref{integ_V+}}\subset \Omega^{t}_{\ref{integ_V+}}$ and 
$\Omega^{t}_{\ref{integ_V+}}$ is also a $\widehat{P}^t$-full-measure set. So, $\Omega^t_{\ref{integ_V+}}$ is a ${\widetilde{P}}^t$-full-measure set using Proposition \ref{domin_lcl} and $\widetilde{P} \ll  \widehat{P}$. As ${\widetilde{P}}$ is arbitrary and $\Omega^{t}_{\ref{integ_V+}}\in\mathcal{B}_c(\Omega^t)$, $\Omega^{t}_{\ref{integ_V+}}$ is a $\mathcal{Q}^t$-full-measure set. \Halmos \endproof

\subsubsection{Lemma \ref{lemma_N_finite}}

The next lemma ensures that Assumption \ref{pb_inequality} is preserved through dynamic programming and provides some useful properties on $N_t^P$ and $N_t^*$.
\begin{lemma}
Assume that the $NA(\mathcal{Q}^T)$ condition as well as Assumptions  \ref{S_borel}, \ref{analytic_graph}, \ref{AE}, \ref{nncst} and \ref{simple_adm} hold true. For all $P:= q_1^P \otimes \cdots \otimes q_T^P\in\mathcal{Q}^T$ and $1\leq t \leq T$, let  
\begin{eqnarray}
\widehat{P}_t:= \frac{q_{1}^{P^*}+q_1^{P}}{2} \otimes \cdots \otimes q_{t}^{P^*} \otimes \cdots \otimes \frac{q_T^{P^*}+q_T^{P}}{2}. \label{hatP_lemmaN}
\end{eqnarray}
Then, $\widehat{P}_t \in\mathcal{H}^T$ and we have that
\begin{eqnarray}
N_{T-1}^{P}<+\infty \;\;P^{T-1}-\mbox{a.s.} \;\; \mbox{and} \;\; N^*_{T-1}<+\infty \;\;\mathcal{Q}^{T-1}-\mbox{q.s.}  \label{NtoPfiniteT} \label{Nt_finiteT}
\end{eqnarray}
Moreover, for all $P\in\mathcal{H}^T$, Assertions (A1) and (B1) below hold true.\\ 
\noindent (A1): If $1/\alpha_{T-1}^P \in \mathcal{M}^{T-1}(P)$ and $\underline{X}$,  $1/|U(\cdot,\underline{X}(\cdot))+C(\cdot)|$, $C$, $U^-(\cdot,0)\in \mathcal{M}^{T}(P)$, then $N_{T-1}^{P}\in\mathcal{M}^{T-1}(P)$.\\
\noindent (B1): If $1/\alpha_{T-1}^{P^*}\in\mathcal{M}^{T-1}(\widehat{P}_T)$ and $\underline{X}$, $1/|U(\cdot,\underline{X}(\cdot))+C(\cdot)|$, $C$, $U^-(\cdot,0) \in \mathcal{M}^T(\widehat{P}_T)$, then $N^*_{T-1} \in \mathcal{M}^{T-1}(P)$.\\
Assume now that there exists some $1\leq t\leq T-1$ such that $\widetilde{\Omega}^{t,P}$ (see \eqref{tildeOmegaP}) is a $P^{t}$-full-measure set for all $P\in\mathcal{H}^T$. Then,
\begin{eqnarray}
N_{t-1}^{P}<+\infty \;\;P^{t-1}-\mbox{a.s.} \;\; \mbox{and} \;\; N^*_{t-1}<+\infty \;\;\mathcal{Q}^{t-1}-\mbox{q.s.}  \label{Nt_finite} \label{NtoPfinite}
\end{eqnarray}
Moreover, for all $P\in\mathcal{H}^T$,  Assertions (A2) and (B2) below hold true.\\
\noindent (A2): If $1/\alpha_{t-1}^P \in \mathcal{M}^{t-1}(P)$ and $1/\alpha_{t}^P$, $N_{t}^P$, $l_t^P$, $C_t$, $J_t(\cdot,0) \in \mathcal{M}^{t}(P)$, then $N_{t-1}^{P}\in \mathcal{M}^{t-1}(P)$.\\
\noindent (B2): If $1/\alpha_{t-1}^{P^*}\in\mathcal{M}^{t-1}(\widehat{P}_t)$ and $1/\alpha_t^{\widehat{P}_t}$, $N_{t}^{\widehat{P}_t}$, $l_t^{\widehat{P}_t}$, $C_t$, $J_t(\cdot,0)\in\mathcal{M}^t(\widehat{P}_t)$, then $N^*_{t-1}\in\mathcal{M}^{t-1}(P)$.

\label{lemma_N_finite}
\end{lemma}

\proof{Proof.}
Let $P:=q_1^P\otimes \cdots \otimes q_T^P\in\mathcal{H}^T$, $R:=q_1^R\otimes \cdots \otimes q_T^R\in\mathcal{H}^T$ and $0\leq t \leq T-1$. We define $N_t^{P,R} : \Omega^t \to \mathbb{N}\cup\{+\infty\}$ by
\begin{eqnarray*}
N_t^{P,R}(\w^t) :=\inf \left\{k\geq 1,\;  q_{t+1}^P\left (U_{t+1}^P(\w^t,\cdot, -k)\leq -i_t^R(\w^t) | \w^t \right)\geq 1-\frac{\alpha_t^R(\w^t)}{2}\right\}.
\end{eqnarray*}
The proof that $N_t^{P,R}$ is $\mathcal{B}_c(\Omega^t)$-measurable is completely similar to the proof that $N_t^P$ is $\mathcal{B}_c(\Omega^t)$-measurable in Lemma \ref{lemma_mes_N} and is thus omitted.\\

 \noindent\textit{Step 1: Case $1\leq t \leq T-1$.}\\ 
We prove that if there exists some $1\leq t\leq T-1$ such that $\widetilde{\Omega}^{t,P}$ is of $P^t$-full-measure for all $P\in\mathcal{H}^T$, then for all $P, R\in\mathcal{H}^T$, $N_{t-1}^{P,R}<+\infty\;\;P^{t-1}-\mbox{a.s.}$  and that Assertion (C1) below holds true. Recall that $P^{0}=\delta_{\{\w_0\}}$, see Section \ref{sett}.\\
(C1):  If $1/\alpha_{t-1}^R\in \mathcal{M}^{t-1}(P)$ and $1/\alpha_{t}^P$, $N_t^P$, $l_t^P$, $C_t$, $J_t(\cdot,0)\in \mathcal{M}^{t}(P)$ and 
$C_t$, $J_t(\cdot,0)\in \mathcal{M}^{t}(P^{t-1}\otimes q_t^R),$ then $N_{t-1}^{P,R}\in\mathcal{M}^{t-1}(P)$.\\
As $\widetilde{\Omega}^{t,P}$ is a $P^t$-full-measure set, there exists $\widehat{\Omega}^{t,P}\in\mathcal{B}_c(\Omega^t)$ such that $\widehat{\Omega}^{t,P}\subset \widetilde{\Omega}^{t,P}$ and  $P^t(\widehat{\Omega}^{t,P})=1$. Define 
$$\overline{\Omega}^{t,P}:= \widehat{\Omega}^{t,P}\cap \{\w^{t}\in\Omega^{t},\; C_{t-1}(\w^{t-1})<+\infty, \; J_{t-1}(\w^{t-1},0)<+\infty\}.$$
Then, $\overline{\Omega}^{t,P}\in\mathcal{B}_c(\Omega^t)$ and $P^t(\overline{\Omega}^{t,P})=1$, see Proposition \ref{CJ_t}. Let $\w^t\in\overline{\Omega}^{t,P}\subset \widetilde{\Omega}^{t,P}$. Lemma \ref{lemma_Assumption_true} shows that Assumptions  \ref{P^*} to 
\ref{pb_inequality}  hold true in the $P$-prior $(t+1)$-context (see Definitions \ref{def_one_prior} and \ref{def_ctxt}) and \eqref{v_v_q} in Proposition \ref{existence_uni} applies: 
\begin{eqnarray}
U_t^P(\w^t,x)=\mathcal{U}_t^P(\w^t,x)= \sup_{h\in\mathbb{R}^d}\mathbb{E}_{q_{t+1}^P(\cdot|\w^t)} U_{t+1}^P(\w^t,\cdot,x+h\Delta S_{t+1}(\w^t,\cdot)). \label{templemmeN2}
\end{eqnarray}
We can also apply Proposition \ref{as_little_v} in the $P$-prior $(t+1)$ context with $m=i_{t-1}^R(\w^{t-1})\geq 1$, see \eqref{It_P}. Indeed, $i_{t-1}^R(\w^{t-1})<+\infty$ using \eqref{ineq_ctp2} at time $t-1$ and $\alpha_{t-1}^R(\w^{t-1})>0$. We denote by $\widehat{N}_t^{P}(\w^t)$, the associated finite bound defined in \eqref{N_exp}. Recall that in the $P$-prior $(t+1)$ context, $\alpha^* = \alpha_t^P(\w^t)$, $n_0^* = N_t^P(\w^t)$, $c^*=c_t^P(\w^t)$ and $l^*=l_t^P(\w^t)$. When $\w^t\notin \overline{\Omega}^{t,P}$, we set $\widehat{N}_t^{P}(\w^t):=1$. Then, let $\overline{N}_t^{P}(\w^t):=\lceil \widehat{N}_t^{P}(\w^t) \rceil$. 
We get that $\overline{N}_t^{P}(\w^t)<+\infty$. \\ 
Now, we show that under the assumptions of (C1),  $\overline{N}_t^{P} \in \mathcal{M}^t(P)$. Lemma \ref{lemma_mes_N} shows that $\widehat{N}_t^P$ and thus $\overline{N}_t^P$ are $\mathcal{B}_c(\Omega^t)$-measurable. 
Moreover, if $1/\alpha_{t-1}^R$ and $c_{t-1}^R$ belong to $\mathcal{M}^{t-1}(P)$ and $1/\alpha_t^P$, $N_t^P$, $l_t^P$, $c_t^P$ to $\mathcal{M}^t(P)$, then $\overline{N}_t^P$ also belongs to $\mathcal{M}^t(P)$, see Lemma \ref{lemme_fdmt_W}. 
Thus, under the assumptions of (C1), it remains to show that $c_{t-1}^R \in\mathcal{M}^{t-1}(P)$ and $c_t^P \in \mathcal{M}^t(P).$ 
Lemma \ref{lemma_mes_N} shows that $c_t^P$ is $\mathcal{B}_c(\Omega^t)$-measurable. So, as $C_t$ and $J_t(\cdot,0)$ belong to $\mathcal{M}^t(P)$, \eqref{ineq_ctp2} and Lemma \ref{lemme_fdmt_W} imply that $c_t^P\in\mathcal{M}^t(P)$. 
Recall that $C_t$ and $J_t(\cdot,0)$ also belong to $\mathcal{M}^{t}(P^{t-1}\otimes q_t^R)$. Thus,  \eqref{ct^P} at $t-1$ and Lemma \ref{M_hered}  imply that $c_{t-1}^R \in \mathcal{M}^{t-1}(P)$.\\
Now, using $\overline{N}^P_{t}$, we construct a $P^{t-1}$-a.s. finite upper bound for $N_{t-1}^{P,R}$. 
For all $\w^t\in\overline{\Omega}^{t,P}$, $\overline{N}_t^{P}(\w^t)<+\infty$ and recalling \eqref{v_ineq_m} and \eqref{templemmeN2}, $
U_{t}^P(\w^t,-\overline{N}_t^P(\w^t))\leq -i_{t-1}^R(\w^{t-1}).$ So, as $\overline{\Omega}^{t,P}$ is a $P^t$-full-measure set, we get that
\begin{eqnarray}
A^t:=\left\{\w^t\in \Omega^t,\; U_{t}^P(\w^{t},-\overline{N}_t^P(\w^t))\leq -i_{t-1}^R(\w^{t-1}),\; \; \overline{N}_t^P(\w^t)<+\infty\right\}
\label{temp_N_retour}
\end{eqnarray}
is also a $P^t$-full-measure set. As $U_{t}^P$ is $\mathcal{B}_c(\Omega^t\times \mathbb{R})$-measurable and $i_{t-1}^R$ is $\mathcal{B}_c(\Omega^{t-1})$-measurable (see Proposition \ref{U_t_well} and Lemma \ref{lemma_mes_N}), we have that $A^t\in\mathcal{B}_c(\Omega^t)$. 
Let 
\begin{eqnarray}
A^{t-1}_1 :=\left\{\w^{t-1}\in\Omega^{t-1}, \;\; \int_{\Omega_{t}}\; 1_{A^t}(\w^{t-1},\w_t) q_t^P(d\w_t|\w^{t-1}) =1 \right\}. \label{set_fm_t-1_1}
\end{eqnarray}
Proposition \ref{univ_cvt} shows that $A^{t-1}_1\in\mathcal{B}_c(\Omega^{t-1})$. Moreover, $P^{t-1}(A^{t-1}_1)=1$ as $P^{t-1}\otimes q_t^P \left( A^t \right)=1$.
Let $\widetilde{N}_{t-1}^P : \Omega^{t-1} \to \mathbb{N}^*\cup \{+\infty\}$ be defined for all $\w^{t-1}\in\Omega^{t-1}$ by,
\begin{eqnarray*}
\widetilde{N}_{t-1}^P(\w^{t-1}):= \inf \left\{k\geq 1, \;q_t^P(\overline{N}_t^P(\w^{t-1},\cdot)\leq k | \w^{t-1})\geq 1- \frac{\alpha_{t-1}^R(\w^{t-1})}{2}\right\}.
\end{eqnarray*}
Fix $\w^{t-1}\in A^{t-1}_1$. Then,  $\lim_{x\to+\infty}q_t^P(\overline{N}_t^P(\w^{t-1},\cdot)\leq x | \w^{t-1})= q_t^P(\overline{N}_t^P(\w^{t-1},\cdot)< +\infty | \w^{t-1})=1$ using the monotone convergence theorem. 
So, $\widetilde{N}_{t-1}^P(\w^{t-1})<+\infty$ and  $q_t^P(\overline{N}_t^P(\w^{t-1},\cdot)\leq \widetilde{N}_{t-1}^P(\w^{t-1}) | \w^{t-1})\geq 1- {\alpha_{t-1}^R(\w^{t-1})}/{2}$. 
Using that $U_t^P(\w^t,\cdot)$ is nondecreasing, we get that 
\begin{eqnarray*}
q_t^P(U_t^P(\w^{t-1},\cdot,-\widetilde{N}_{t-1}^P(\w^{t-1}))\leq -i_{t-1}^R(\w^{t-1}) | \w^{t-1}) &\geq& 1- \frac{\alpha_{t-1}^R(\w^{t-1})}{2}. 
\end{eqnarray*}
So, as $P^{t-1}(A^{t-1}_1)=1$, we find by definition of $N_{t-1}^{P,R}$ that
\begin{eqnarray}
N_{t-1}^{P,R}\leq \widetilde{N}_{t-1}^P<+\infty \;\; P^{t-1}\mbox{-a.s}.
\label{NtleqNprime0}
\end{eqnarray}
Now, we construct a second upper bound for $N_{t-1}^{P,R}$ for which the estimation of the moment of order $r$ is easy for $r\geq 1$. Let $\widehat{N}_{t-1}^P : \Omega^{t-1} \to \mathbb{N}^*\cup\{+\infty\}$ be defined for all $\w^{t-1}\in\Omega^{t-1}$ by
$$\widehat{N}_{t-1}^P(\w^{t-1}) := 2\frac{\mathbb{E}_{q_{t}^P(\cdot|\w^{t-1})} \overline{N}_t^P(\w^{t-1},\cdot)}{\alpha_{t-1}^R(\w^{t-1})}.$$
Let $\w^{t-1}\in\Omega^{t-1}$. 
Applying Markov's inequality, we get that
\begin{eqnarray*}
q_t^P(\overline{N}_t^P(\w^{t-1},\cdot)\leq \widehat{N}_{t-1}^P(\w^{t-1}) | \w^{t-1})
&\geq & 1- \frac{\mathbb{E}_{q_{t+1}^P(\cdot|\w^{t-1})} \overline{N}_t^P(\w^{t-1},\cdot) }{\widehat{N}_{t-1}^P(\w^{t-1})}= 1-\frac{\alpha_{t-1}^R(\w^{t-1})}{2}.
\end{eqnarray*}
So, by definition of $\widetilde{N}_{t-1}^P$, we have that $\widetilde{N}_{t-1}^P(\w^{t-1})\leq \widehat{N}_{t-1}^P(\w^{t-1})$ and using also  \eqref{NtleqNprime0} that
\begin{eqnarray}
N_{t-1}^{P,R}\leq \widehat{N}_{t-1}^P \;\;P^{t-1}\mbox{-a.s}.
\label{temp_lemmaN_up1}
\end{eqnarray}
Let $r\geq 1$. Using Jensen's inequality, we have for all $\w^{t-1}\in\Omega^{t-1}$ that $$(\widehat{N}_{t-1}^P(\w^{t-1}))^r = \left(\mathbb{E}_{q_{t}^P(\cdot|\w^{t-1})} \frac{2\overline{N}_t^P(\w^{t-1},\cdot)}{\alpha_{t-1}^R(\w^{t-1})}\right)^r \leq  \mathbb{E}_{q_{t}^P(\cdot|\w^{t-1})}\left(\frac{2\overline{N}_t^P(\w^{t-1},\cdot)}{\alpha_{t-1}^R(\w^{t-1})}\right)^r.$$
Now, \eqref{temp_lemmaN_up1}, Fubini's theorem and Cauchy-Schwarz inequality show that 
\begin{eqnarray*}
\mathbb{E}_{P^{t-1}} \left(N_{t-1}^{P,R}\right)^r \leq 2^r \sqrt{\mathbb{E}_{P^t} (\overline{N}_t^P)^{2r}} \sqrt{\mathbb{E}_{P^{t-1}}\left(\frac{1}{\alpha_{t-1}^R}\right)^{2r}}.
\end{eqnarray*}
As a result, if $1/\alpha_{t-1}^R\in\mathcal{M}^{t-1}(P)$ and $\overline{N}_t^P\in\mathcal{M}^t(P)$, then $N_{t-1}^{P,R}\in\mathcal{M}^{t-1}(P)$. 
Recalling that $\overline{N}_t^P\in\mathcal{M}^t(P)$ under the assumptions of (C1), this shows assertion (C1).\\

\noindent\textit{Step 2: case $t=T$.}\\
We now prove that $N_{T-1}^{P,R}<+\infty\;\; P^{T-1}\mbox{-a.s}$ and that Assertion (C2) below holds true.\\
(C2): If $1/\alpha_{T-1}^R\in\mathcal{M}^{T-1}(P),$  $\underline{X}$, $1/|U(\cdot,\underline{X}(\cdot))+C(\cdot)|\in\mathcal{M}^{T}(P)$ and $C(\cdot)$, $U^-(\cdot,0) \in \mathcal{M}^{T}(P^{T-1}\otimes q_T^R)$, then $N_{T-1}^{P,R}\in\mathcal{M}^{T-1}(P)$.
\noindent Let
\begin{eqnarray*}
\overline{\Omega}^{T,P}:=\{C<+\infty\}\cap \{U(\cdot,\underline{X}(\cdot))<-C(\cdot)\}\cap\{C_{T-1}(\cdot)+J_{T-1}(\cdot,0)<+\infty\}.
\end{eqnarray*}
Assumptions \ref{AE} and \ref{nncst}, Remark \ref{joint_mes_U} and Proposition \ref{CJ_t} show that $\overline{\Omega}^{T,P}\in\mathcal{B}_c(\Omega^T)$ and is of $P$-full-measure. Let $\widetilde{N}_T^{P} : \Omega^T\to \mathbb{R}^*\cup\{+\infty\}$ be defined for all $\w^T\in\overline{\Omega}^{T,P}$ by
\begin{eqnarray*}
\widetilde{N}_T^{P}(\w^T)&:=&  \left(\frac{i_{T-1}^R(\w^{T-1})}{|U(\w^T,\underline{X}(\w^T))+C(\w^T)|} + 1\right)^{\frac{1}{\gamma}}|\underline{X}(\w^T)|
\end{eqnarray*}
and $\widetilde{N}_T^{P}(\w^T):=1$ otherwise. For all $\w^T\in\Omega^T$, we set $\overline{N}_T^{P}(\w^T):= \lceil \widetilde{N}_T^{P}(\w^T) \rceil$. We see easily that $\overline{N}_T^{P}$ is $\mathcal{B}_c(\Omega^T)$-measurable (recall that $i_{T-1}^R$ is $\mathcal{B}_c(\Omega^{T-1})$-measurable from Lemma \ref{lemma_mes_N}).\\ 
We show under the conditions of (C2) that $\overline{N}_{T}^P \in\mathcal{M}^{T}(P)$. Indeed,  $\overline{N}_{T}^P \in\mathcal{M}^{T}(P)$ if  $c_{T-1}^R$, $1/\alpha_{T-1}^R \in \mathcal{M}^{T-1}(P)$ and $\underline{X}$, $1/|U(\cdot,\underline{X}(\cdot)) + C(\cdot)| \in\mathcal{M}^T(P)$. Now as $C$ and $U^-(\cdot,0)$ belong to $\mathcal{M}^{T}(P^{T-1}\otimes q_T^R),$ $c_{T-1}^R\in\mathcal{M}^{T-1}(P)$, see Lemma \ref{M_hered} and \eqref{ct^P} 
(and also \eqref{J_t_eq} and \eqref{C_t_eq}).\\ 
Let $\w^T\in\overline{\Omega}^{T,P}$. Note that $c_{T-1}^R(\w^{T-1})<+\infty$ using \eqref{ineq_ctp2} at $T-1$. Moreover, $\alpha_{T-1}^R(\w^{T-1})>0$ and $U(\w^T,\underline{X}(\w^T))+C(\w^T) < 0$. So, $i_{T-1}^R(\w^{T-1})<+\infty$ and also $\overline{N}_T^{P}(\w^{T})<+\infty$. As $U(\w^T,\cdot)$ is nondecreasing and using \eqref{elas_gammaf} with $x=-|\underline{X}(\w^T)|=\underline{X}(\w^T)$ and $\lambda = \frac{\widetilde{N}_T^{P}(\w^T)}{|\underline{X}(\w^T)|} \geq 1$, we find that
\begin{eqnarray*}
U(\w^T,-\overline{N}_T^{P}(\w^T))\leq  U(\w^T,-\widetilde{N}_T^{P}(\w^T))&\leq & \frac{\left(\widetilde{N}_T^P(\w^T)\right)^\gamma}{|\underline{X}(\w^T)|^\gamma}\left(U(\w^T,\underline{X}(\w^T)) + C(\w^T) \right)\leq  -i_{T-1}^R(\w^{T-1}).
\end{eqnarray*}
So, we get that
$\overline{\Omega}^{t,P} \subset \left\{\w^T\in \Omega^T,\; U(\w^{T},-\overline{N}_T^{P}(\w^T))\leq -i_{T-1}^R(\w^{T-1})\; \mbox{and} \; \overline{N}_T^{P}(\w^T)<+\infty\right\}$
and this last set is of $P$-full measure. The rest of the proof is exactly as in Step 1 after \eqref{temp_N_retour}.\\

\noindent\textit{Step 3:}
We apply Steps 1 and 2 to $R=P$. Remark that for all $1\leq t \leq T$,  $\mathcal{M}^{t}(P^{t-1}\otimes q_t^R)=\mathcal{M}^{t}(P^t)$  and $N_{t}^P = N_{t}^{P,P}$ (see \eqref{eq_NtP}). 
Thus, \eqref{NtoPfiniteT} for $N_{T-1}^P$, \eqref{NtoPfinite} for $N_{t-1}^P$ as well as Assertions (A1) and (A2) hold true.\\

\noindent\textit{Step 4: \eqref{Nt_finiteT} for $N_{T-1}^*$, \eqref{Nt_finite} for $N_{t-1}^*$ and Assertions (B1) and (B2) hold true for all $1\leq t\leq T$ and $P\in\mathcal{H}^T$.}\\
Let $P:=q_1^P\otimes \cdots \otimes q_T^P \in\mathcal{H}^T$. Recall $P^*$ from \eqref{P^*_exp} and $\widehat{P}_t$ from \eqref{hatP_lemmaN}. Then, $P^*, \widehat{P}_t\in\mathcal{H}^T,$ see \eqref{setPT}. 
Let $1\leq t\leq T$. Steps 1 and 2 applied to $P=\widehat{P}_t$ and $R=P^*$ show that for all $\w^{t-1}$ in a $\widehat{P}^{t-1}_t$-full-measure set, $N_{t-1}^{\widehat{P}_t,P^*}(\w^{t-1})<+\infty$ and thus 
\begin{eqnarray*}
q_{t}^{\widehat{P}_t}\left(U_{t}^{\widehat{P}_t}(\w^{t-1},\cdot,-N_{t-1}^{\widehat{P}_t,P^*}(\w^{t-1}))\leq - i_{t-1}^{P^*}(\w^{t-1}) | \w^{t-1} \right)\geq 1-\frac{\alpha_{t-1}^{P^*}(\w^{t-1})}{2}.
\end{eqnarray*}
As $U_{t}\leq U_{t}^{\widehat{P}_t}$ (see Proposition \ref{U_t_well}) and $q_t^{\widehat{P}_t}=q_t^{P^*}$, we get that
\begin{eqnarray*}
q_{t}^{P^*}\left(U_{t}(\w^{t-1},\cdot,-N_{t-1}^{\widehat{P}_t,P^*}(\w^{t-1}))\leq - i_{t-1}^{P^*}(\w^{t-1}) | \w^{t-1} \right)\geq 1-\frac{\alpha_{t-1}^{P^*}(\w^{t-1})}{2}.
\end{eqnarray*}
So, we deduce from the definition of $N^*_{t-1}$ in \eqref{eq_Nt} that 
\begin{eqnarray}
\label{refouf}
0\leq N^*_{t-1} \leq N_{t-1}^{\widehat{P}_t,P^*}<+\infty\;\; \widehat{P}_t^{t-1}-\mbox{a.s.}
\end{eqnarray}
If $t=1$, then, $P^{0} = \delta_{\w_0} \ll \delta_{\w_0} =\widehat{P}_t^{0}.$ If $t>1$, Proposition \ref{domin_lcl} for $l=t-1$ shows that $P^{t-1}\ll \widehat{P}_t^{t-1}$. So, $N^*_{t-1} <+\infty\;\; P^{t-1}-\mbox{a.s.}$ 
As $P$ is arbitrary, \eqref{NtoPfiniteT} and \eqref{Nt_finite} hold true. \\If $t=1$, we have that $\mathcal{M}^{0}(\widehat{P}_t) = \mathcal{M}^{0}(P) = \mathbb{R}$. Now, if $t>1$, Lemma \ref{lemma_inc_M}, again with $l=t-1$, shows that $\mathcal{M}^{t-1}(\widehat{P}_t) \subset \mathcal{M}^{t-1}(P).$ 
For all $2\leq t\leq T$, as $q_t^{\widehat{P}_t}=q_t^{P^*}$, $\mathcal{M}^{t}(P^{t-1}\otimes q_t^R)=
\mathcal{M}^{t}(\widehat{P}_t^{t-1}\otimes q_t^{P^*})= 
\mathcal{M}^{t}(\widehat{P}_t)$. Thus, under the assumptions of (B2)  for $t$ (resp. (B1)  for $T$), 
 Assertion (C1)  for $t$ (resp. (C2) for $T$) shows $N_{t-1}^{\widehat{P}_t,P^*}\in \mathcal{M}^{t-1}(\widehat{P}_t) \subset \mathcal{M}^{t-1}(P).$
 So, \eqref{refouf} and Lemma \ref{lemme_fdmt_W} show that $N_{t-1}^{*}\in  \mathcal{M}^{t-1}(P)$ and 
 Assertion (B1) for $t$ (resp. (B2) for $T$) holds true. 
\Halmos \endproof


\end{document}